\documentclass[preprint]{revtex4}

\usepackage{epsfig}
\usepackage{amsmath}
\usepackage{amssymb}
\usepackage{verbatim}
\usepackage{psfrag}

\begin{document}

\begin{flushright}
{\tt SPIN-07/46, ITP-UU-07/60, Crete-07-13, UFIFT-QG-07-06}
\end{flushright}


\bigskip

\begin{center}
{\LARGE\bf  Two loop stress-energy tensor for} \\
\medskip
{\LARGE\bf  inflationary scalar electrodynamics}
\end{center}

\vskip 0.5cm

\begin{center}
{\large Tomislav Prokopec$^1$, Nicholas C. Tsamis$^2$
  and Richard P. Woodard$^3$}
\end{center}

\vskip .4cm

\begin{center}
{\it $^1$Institute for Theoretical Physics (ITF) \& Spinoza Institute,
Utrecht University,\\
Leuvenlaan 4, Postbus 80.195, 3508 TD Utrecht, The Netherlands}
\end{center}

\begin{center}
\it{$^2$Department of Physics, University of Crete,
        GR-710 03 Heraklion, Greece}
\end{center}

\begin{center}
\it{$^3$Department of Physics, University of Florida,
         Gainesville, FL 32611, USA}
\end{center}

\begin{center}
{\bf Abstract}
\end{center}

We calculate the expectation value of the coincident product of two
field strength tensors at two loop order in scalar electrodynamics 
on de Sitter background. The result agrees with the stochastic 
formulation which we have developed in a companion 
paper~\cite{Prokopec:2007ak} for the nonperturbative resummation of 
leading logarithms of the scale factor. When combined with a previous 
computation of scalar bilinears~\cite{Prokopec:2006ue}, our current 
result also gives the two loop stress-energy tensor for inflationary 
scalar electrodynamics. This shows a secular decrease in the vacuum 
energy which derives from the vacuum polarization induced by 
the inflationary production of charged scalars.

\section{Introduction}
\label{Introduction}

This is the third and final paper in a series with the goal of 
establishing a stochastic theory of inflation for scalar quantum 
electrodynamics (SQED). Stochastic SQED is outlined in 
Ref.~\cite{Prokopec:2007ak}. In this paper and in 
Ref.~\cite{Prokopec:2006ue} we compare the predictions of stochastic 
SQED with explicit and exact perturbative calculations. In particular, 
Ref.~\cite{Prokopec:2006ue} gives the mostly scalar contribution to 
the two-loop stress energy, and here we calculate the mostly photon 
contribution.

Stochastic inflationary
theory~\cite{Starobinsky:1986fx,Sasaki:1987gy,Starobinsky:1994bd,Martin:2005ir}
was developed primarily in order to capture the effects of small scale
stress-energy fluctuations (large momenta)
onto large scale fluctuations, which
are observed by current large scale structure and cosmological microwave
background radiation measurements.
Furthermore, stochastic inflation has also been used to study global structure
and dynamics~\cite{Nambu:1988je,Linde:1993xx}
of inflationary universe models.
 Even though corrections to stochastic inflationary theories
have not yet been systematically studied, these type of studies are feasible
within the {\it in-in} path integral formalism~\cite{Morikawa:1989xz}.
A phase space formulation of stochastic inflationary theory is given in
Ref.~\cite{Habib:1992ci}, where a quantum Liouville equation for
the Wigner function was obtained, which is the suitable
quantum generalization of the Fokker-Planck equation
for the (classical) density function.
More recently, a phase space formulation has been used
for computing the nongaussianity of cosmological
perturbations~\cite{Rigopoulos:2005ae,Rigopoulos:2005xx,Rigopoulos:2004ba,Rigopoulos:2004gr,Riotto:2008mv,Rigopoulos:2008kt} in various inflationary models.

We are interested in the stochastic theory of inflation mainly because
it captures correctly the leading logarithmic behavior (in the scale factor)
of quantum field theories during inflation.
In Ref.~\cite{Prokopec:2007ak} we have proven that a
stochastic theory of scalar electrodynamics (obtained by integrating
out the photon field and by subsequently stochasticizing the resulting
effective scalar field theory) correctly reproduces the
leading secular logarithms to all orders in perturbation theory.
The agreement is made explicit by a comparison between
the mostly scalar contribution to the two loop stress-energy tensor
in de Sitter inflation with the corresponding stochastic prediction.
The stochastic theory of inflation goes far beyond perturbation theory,
in that it provides an effective means for resummation
of the leading logarithm contributions~\cite{Prokopec:2007ak}.
Unlike full quantum field theories, which cannot be generally solved,
stochastic theories are classical, and hence they are relatively easily solved.
Indeed, solving a stochastic theory is tantamount to solving the corresponding
Fokker-Planck equation for the density function~\cite{Starobinsky:1994bd}.

One of the important application of stochastic theories is that they permit
to address the question of gravitational back-reaction
in accelerating space-times~\cite{Tsamis:1996qq,Tsamis:1996qm},
and can thus answer one of the fundamental unsolved problems of inflationary 
cosmology: how do quantum effects change the dynamics of space-time during 
inflation. More broadly, this will teach us something about the infrared
sector of quantum gravity and the cosmological constant problem.
Ultimately one would like to understand how quantum fluctuations of 
gravitons affect inflation. The hope is that this question can be 
answered using a stochastic theory of inflation~\cite{Tsamis:2005hd}.

Quite a lot is already known about quantum (radiative) effects in inflationary 
space-times. Perturbative analysis of scalar electrodynamics during de 
Sitter inflation shows that the photon acquires a mass during
inflation~\cite{Davis:2000zp,Prokopec:2002jn,Prokopec:2002uw,Prokopec:2003bx,Prokopec:2003iu,Prokopec:2003tm,Prokopec:2007ak},
which can be of relevance for generation of the cosmological magnetic
fields~\cite{Davis:2000zp,Dimopoulos:2001wx,Prokopec:2003bx,Prokopec:2003tm,ProkopecPuchwein:2004}.
The scalar field of SQED also acquires a mass during
inflation, but --- unlike the photon mass~\cite{Prokopec:2007ak} ---
the scalar mass remains perturbatively small~\cite{Prokopec:2007ak}.
Similarly, the fermions~\cite{Prokopec:2003qd,Garbrecht:2006jm} of Yukawa
theory acquire mass while the scalars remain 
light~\cite{Duffy:2005ue,Miao:2006pn}. In self-interacting scalar theories, 
the quantum loop effects during inflation can contribute to the stress-energy 
like a phantom field~\cite{Onemli:2002hr,Onemli:2004mb}, whose equation of 
state is even more negative than the pressure of a vacuum filled with a 
positive cosmological term. Although this generally betokens 
instability~\cite{Wise}, the phenomenon in this case is 
self-limiting~\cite{BOW,KO}. The quantum effects of gravitons on the 
dynamics of fermions and scalars during inflation have also been 
studied~\cite{Miao:2005am,Miao:2006gj,Kahya:2007bc,KW}, and it is now 
known that gravity induces a logarithmic correction to the fermionic wave 
function at one loop while no such correction occurs to a massless, 
minimally coupled scalar. The question of how large quantum corrections to 
the inflationary observables actually are, has been addressed by a number of
authors~\cite{Abramo:2001dc,GB,Weinberg:2005vy,Weinberg:2006ac,Boyanovsky:2005px,LU,WNF,Sloth:2006nu,Bilandzic:2007nb,Garriga:2007zk,van der Meulen:2007ah,TW},
but as yet no definite answer can be provided to that question.

This paper is organized as follows. In section II we introduce the relevant
scalar and gauge field propagators for de Sitter space and discuss some of 
their properties. In section III we compute the one loop expectation value 
of the SQED stress-energy tensor in de Sitter space. In section IV we
discuss some general properties of the stress-energy 
tensor and its constituents.
We also employ previous work~\cite{Prokopec:2006ue} to give the two loop
contributions from the scalar kinetic and potential energies. Section V is 
the central part of the paper, where we compute the two loop gauge field 
strength bilinear. In section VI we obtain the total, two loop
stress-energy tensor
and we discuss the result. Appendices A, B and C give identities and integrals
useful to readers who wish to reproduce the calculation.


\section{The propagators in de Sitter space}
\label{The propagators in de Sitter space}

 In this paper we approximate inflationary space-times by de Sitter space.
The de Sitter metric with flat spatial sections is,
\begin{equation}
 ds^2 = a^2(\eta) \left(-d\eta^2 + d\vec x^2\right)
\,,
\label{dS metric}
\end{equation}
where $a=a(\eta)$ is the scale factor,
which in de Sitter space has the form,
\begin{equation}
 a = - \frac{1}{H\eta} \,, \qquad (\eta < 0)
\label{scale factor}
\,.
\end{equation}
Here $H$ is the Hubble parameter and $\eta$ denotes conformal time.
Furthermore, it is convenient to define a de Sitter invariant function
$y=y(x;x')$ between two space-time points $x^\mu$ and ${x'}^\nu$. In
flat coordinates~(\ref{dS metric}),
\begin{equation}
 y = a a'H^2 \Delta x^2
\,, \qquad
 \Delta x^2 = -(|\eta-\eta'|-i\epsilon)^2 + \|\vec x - \vec x\,'\|^2
\label{y function}
\end{equation}
where $a \equiv a(\eta)$ and $a' \equiv a(\eta')$.
There is a simple relation between $y$ and the geodesic distance
$\ell=\ell(x;x')$ in de Sitter space,
\begin{equation}
 y = 4\sin^2 \left(\frac{H \ell}{2}\right)
\,.
\label{ell: geodesic distance}
\end{equation}

We work here in the Schwinger-Keldysh formalism \cite{JS,KTM,BM,LVK}, 
which is suitable for describing quantum field theory dynamics. In short,
the vertices of Feynman diagrams have two polarities: {\it plus} ($+$)
and {\it minus} ($-$). Because pairs of vertices have four possible 
polarities there are four propagators which we denote by $i\Delta_{++}$, 
$i\Delta_{+-}$, $i\Delta_{-+}$ and $i\Delta_{--}$, respectively. They are 
obtained from the Feynman propagator $i\Delta \equiv i\Delta_{++}$ by 
replacing $y\equiv y_{++}=y(x_+;x'_+)$ in Eq.~(\ref{y function}) 
by $y_{+-}$, $y_{-+}$ and $y_{--}$, respectively, defined by,
\begin{eqnarray}
 y_{+-} &=& a a'H^2 \Delta x^2_{+-}
\,,\qquad
  \Delta x^2_{+-} = -(\eta-\eta'+i\epsilon)^2 + \|\vec x - \vec x\,'\|^2
\label{y function:+-}
\\
 y_{-+} &=& a a'H^2 \Delta x^2_{-+}
\,,\qquad
  \Delta x^2_{-+} = -(\eta-\eta'-i\epsilon)^2 + \|\vec x - \vec x\,'\|^2
\label{y function:-+}
\\
 y_{--} &=& a a'H^2 \Delta x^2_{--}
\,,\qquad
  \Delta x^2_{--} = -(|\eta-\eta'|+i\epsilon)^2 + \|\vec x - \vec x\,'\|^2
\,.
\label{y function:--}
\end{eqnarray}

\subsection{The massless minimally coupled scalar propagator}
\label{The massless minimally coupled scalar propagator}

 The Feynman propagator for a minimally coupled massless scalar field
in de Sitter space obeys the differential equation,
\begin{equation}
  \sqrt{-g}\Box i\Delta(x;x') = i \delta^D(x-x')
\,,
\label{scalar eom}
\end{equation}
where $\Box = (-g)^{-1/2} \partial_\mu g^{\mu\nu}\sqrt{-g}\partial_\nu$
denotes the scalar d'Alembertian,
$g = {\rm det}[g_{\mu\nu}]$, and $\delta^D$ is the $D$-dimensional
Dirac delta function.
The addition of a mass term to (\ref{scalar eom}) results in a de Sitter
invariant propagator which was first constructed by Chernikov and 
Tagirov~\cite{Chernikov:1968zm}. However, there is no de Sitter invariant 
solution in the massless case~\cite{Allen:1987tz}.

The problem can be circumvented by adding de Sitter breaking contributions
to the propagator. A natural way of obtaining the form of the de Sitter
breaking term is to consider quasi-de Sitter space in which the 
de Sitter symmetry is mildly broken, such that
$\epsilon \equiv -\dot H/H^2 \ll 1$ and $\dot \epsilon = 0 $
(here $\dot H \equiv dH/dt$). One can show~\cite{Janssen:2007ht} that
the scalar propagator in such a space can be represented as a function
of the de Sitter invariant distance suitably rescaled
by powers of the two scale factors.
Remarkably, in the de Sitter limit when
$\epsilon \rightarrow 0$, one recovers
the massless minimally coupled scalar propagator
in $D$ space-time dimensions~\cite{Onemli:2004mb}
plus an infinite $\epsilon$-dependent constant.
Ignoring the constant, the propagator in de Sitter space
can be written in the form,
\begin{eqnarray}
  i \Delta(x;x^{\prime}) &=&  A(y)  + k  \ln(a a^{\prime})
\label{scalar propagator}
\\
 A(y) &=&  \frac{H^{D-2}}{(4\pi)^{\frac{D}2}}
  \left[- \frac{\Gamma(D \!-\! 1)}{\Gamma(\frac{D}2)}
                \pi\cot\Bigl(\frac{\pi}2 D\Bigr)
\right.
\label{scalar propagator:A}
\\
&& \hskip 1.2cm \left.
 - \sum_{n=-1}^{\infty}\! \frac1{n \!-\! \frac{D}2 \!+\! 2}
\frac{\Gamma(n \!+\!  \frac{D}2 \!+\! 1)}{\Gamma(n \!+\! 2)} \Bigl(\frac{y}4
\Bigr)^{n - \frac{D}2 +2}
+\,  \sum_{n=1}^{\infty}\!
\frac1{n} \frac{\Gamma(n \!+\! D \!-\! 1)}{\Gamma(n \!+\! \frac{D}2)}
\Bigl(\frac{y}4 \Bigr)^n
\right]
\nonumber\\
 k &=&   \frac{H^{D-2}}{(4\pi)^{\frac{D}2}}
           \frac{\Gamma(D \!-\! 1)}{\Gamma(\frac{D}2)}
\label{scalar propagator:k}
\,.
\end{eqnarray}

Note that the function $A=A(y)$ in Eq.~(\ref{scalar propagator:A}) 
obeys the equation,
\begin{equation}
  (4-y)yA''(y) + D(2-y)A'(y) = (D-1)k
\,.
\label{eom:A}
\end{equation}
This implies a relation which will have great importance for this
computation,
\begin{equation}
  (2 - y) A'(y) - k = -\frac{K'(y)}{D-2} \; , \label{AKrel}
\end{equation}
where we define $K(y)$ as,
\begin{equation}
  K(y) \equiv (4-y) y A'(y) + (2-y) k \; . \label{Kdef}
\end{equation}
The expansion of $K(y)$ around $y=0$ is,
\begin{eqnarray}
K(y) \!\!&=&\!\!
  - \frac{H^{D-2}}{(4\pi)^{D/2}}4\Gamma\left(\frac{D}{2}\right)
  \left\{
       \left(\frac{y}{4}\right)^{1-\frac{D}{2}}
   \!\!\! +  \frac{D-2}{2}\left(\frac{y}{4}\right)^{2-\frac{D}{2}}
    -\,\frac{\Gamma(D-1)}{2\Gamma\left(\frac{D}{2}\right)^2}
\right.\;
\nonumber\\
   \! &&\hskip .cm
   \! +\, \frac{D(D-2)}{8}\left(\frac{y}{4}\right)^{3-\frac{D}{2}}
    - \frac{\Gamma(D-1)}
        {\Gamma\left(\frac{D}{2}-1\right)\Gamma\left(\frac{D}{2}\!+\!1\right)}
                   \frac{y}{4}
\nonumber\\
   \!&&\!
    +\, \frac{D(D\!-\!2)(D\!+\!2)}{48}\left(\frac{y}{4}\right)^{4-\frac{D}{2}}
    -\frac{\Gamma(D)}
         {\Gamma\left(\frac{D}{2}-1\right)\Gamma\left(\frac{D}{2}\!+\!2\right)}
            \left(\frac{y}{4}\right)^2
\nonumber\\
   \!&&\!
     +\,\frac{D\!-\!4}{2}\sum_{n=3}^\infty\left(\frac{y}{4}\right)^n
                         \ln\left(\frac{y}{4}\right)
     +\,{\cal O}\left((D-4)^2\right)
      \bigg\}
\nonumber\\
   \!&=&\!
  - \frac{H^{D-2}}{(4\pi)^{D/2}}4\Gamma\left(\frac{D}{2}\right)
  \left\{
       \left(\frac{y}{4}\right)^{1-\frac{D}{2}}
     +\,\frac{D\!-\!4}{2}\frac{\ln\left(\frac{y}{4}\right)}
                              {1-{y}/{4}}
     +\,{\cal O}\left((D-4)^2\right)
 \right\}
\,. \;\;\;
\label{4-yyAp+2-yk}
\end{eqnarray}
Note that in $D=4$ the function $K(y)$ terminates at $y^{1-D/2}$.
In fact, up to the terms that are suppressed as $(D-4)$,
$K(y)$ is just $-2(D-2)$ times the conformal scalar propagator.
Another important relation is,
\begin{equation}
  (4-y) y K''(y) + D (2-y) K'(y) = (D-2) K(y)
\,.
\label{eom:K}
\end{equation}
Also note that dimensional regularization implies the following
coincidence limits ($y\rightarrow 0$),
\begin{eqnarray}
 A(0) &=&  - \frac{H^{D-2}}{(4\pi)^{\frac{D}2}}
              \cos\left(\frac{\pi D}{2}\right)
               \Gamma(D \!-\! 1)\Gamma\left(1-\frac{D}{2}\right)
\label{scalar propagator:A0}
\\
 A'(0) &=&  \frac{H^{D-2}}{4(4\pi)^{\frac{D}2}}
                    \frac{\Gamma(D)}{\Gamma\Big(\frac{D}{2}+1\Big)}
\label{scalar propagator:A'0}
\\
 2A'(0)-k &=& - \frac{H^{D-2}}{2(4\pi)^\frac{D}{2}}
                    \frac{\Gamma(D-1)}{\Gamma\Big(\frac{D}{2}+1\Big)}
                 = -\frac{K'(0)}{D-2} \,.
\label{scalar propagator:2A'0-k}
\end{eqnarray}

\subsection{The vector propagator}
\label{The vector propagator}

 The Lorentz gauge vector propagator in de Sitter space was discussed
by Allen and Jacobson~\cite{Allen:1985wd}, and a minor error in their
analysis was recently corrected~\cite{Tsamis:2006gj}. The propagator 
obeys the equation
\begin{equation}
 \sqrt{-g} \left(\Box g^{\rho\mu} - R^{\rho\mu}\right)
       i\Bigl[\mbox{}_{\mu} \Delta_{\nu}\Bigr](x;x')
         = \delta^\rho_\nu i\delta^D(x-x')
           + \sqrt{-g} \partial^\rho\partial'_\nu i\Delta(x;x')
\,,
\label{eom:vector propagator}
\end{equation}
where $\Box$ denotes the vector d'Alembertian, $R^{\rho\mu}$ is
the Ricci tensor and $i\Delta(x;x')$ represents the scalar 
propagator~(\ref{scalar propagator}). The transversality of 
Eq.~(\ref{eom:vector propagator}) follows from the Lorentz gauge condition
\begin{eqnarray}
\sqrt{-g}\nabla^\mu i\Bigl[\mbox{}_{\mu} \Delta_{\nu}\Bigr](x;x')
\;\equiv\;  \partial_\rho\,\, g^{\rho\mu}\,\sqrt{-g}\;
          i\Bigl[\mbox{}_{\mu} \Delta_{\nu}\Bigr](x;x')
 &=& 0
\nonumber\\
\sqrt{-g'}{\nabla'}^\nu i\Bigl[\mbox{}_{\mu} \Delta_{\nu}\Bigr](x;x')
\equiv {\partial'}_\sigma {g'}^{\sigma\nu}\sqrt{-g'}
   i\Bigl[\mbox{}_{\mu} \Delta_{\nu}\Bigr](x;x')
 &=& 0
\,,
\label{Lorentz gauge}
\end{eqnarray}
where $g^{\rho\mu} \equiv g^{\rho\mu} (x)$, ${g'}^{\sigma\nu} \equiv 
g^{\sigma\nu} (x')$, and $\nabla^{\mu}$ is the covariant derivative
operator.

On de Sitter background the photon propagator can be written in the 
following de Sitter invariant form
 \begin{equation}
 i\Bigl[\mbox{}_{\mu} \Delta_{\nu}\Bigr](x;x')
= B(y) \frac{\partial^2 y}{ \partial x^{\mu} \partial x^{\prime \nu}}
 + C(y) \frac{\partial y}{ \partial x^{\mu}}
 \frac{\partial y}{\partial x^{\prime \nu}}
 \; .
\label{photon propagator: Ansatz}
\end{equation}
We can use the Lorentz gauge condition (\ref{Lorentz gauge}) to express
$B(y)$ and $C(y)$ in terms of a single function $\gamma(y)$
 \begin{eqnarray}
B(y) & = & \frac1{4 (D\!-\!1) H^2} \Bigl\{ -(4 \!-\! y)y \gamma'(y)
- (D\!-\!1) (2 \!-\! y) \gamma(y) \Bigr\} \; ,
\label{B function}
\\
C(y) & = & \frac1{4 (D\!-\!1) H^2} \Bigl\{ (2 \!-\! y) \gamma'(y)
- (D\!-\!1) \gamma(y) \Bigr\} \; .
\label{C function}
\end{eqnarray}
The propagator equation will follow provided $\gamma(y)$ has the right
singularity at $y=0$ and obeys the differential equation,
\begin{equation}
(4 \!-\! y)y \gamma''(y) + (D\!+\!2) (2 \!-\! y) \gamma'(y)
-2(D\!-\!1) \gamma(y)
  = (D\!-\!1) \Bigl[ (2 \!-\!y) A'(y) - k\Bigr] \; .
\label{gamma function}
\end{equation}
The unique solution for $\gamma(y)$ is~\cite{Tsamis:2006gj}
\begin{eqnarray}
  \gamma(y) & = &
 \frac{D\!-\!1}{2(D\!-\!3)}
   \frac{H^{D-2}}{(4\pi)^\frac{D}{2}}
\nonumber\\
 &\times&\!\! \Bigg\{\sum_{n=0}^\infty
            \frac{(n\!+\!1)\Gamma(n\!+\!D\!-\!1)}
                 {\Gamma\left(n\!+\!\frac{D}{2}\!+\!1\right)}
 \bigg[\psi\Big(2\!-\!\frac{D}{2}\Big) - \psi\Big(\frac{D}{2}\!-\!1\Big)
       + \psi(n\!+\!D\!-\!1) - \psi(n\!+\!2) \bigg]
           \left(\frac{y}{4}\right)^n
\nonumber\\
 && \hskip 0.cm
 -\, \sum_{n=-1}^\infty \Big(n\!-\frac{D}{2}+\!3\Big)
           \frac{\Gamma\left(n\!+\!\frac{D}{2}\!+\!1\right)}
                 {\Gamma(n\!+\!3)}
\nonumber\\
 && \hskip 0.5cm
\times\bigg[\psi\Big(2\!-\!\frac{D}{2}\Big) - \psi\Big(\frac{D}{2}\!-\!1\Big)
       + \psi\Big(n\!+\!\frac{D}{2}\!+\!1\Big)
       - \psi\Big(n\!-\frac{D}{2}\!+\!4\Big) \bigg]
           \left(\frac{y}{4}\right)^{n-\frac{D}{2}+2}
  \Bigg\}
\,.
\label{gamma function:2}
\end{eqnarray}

An important simplification for the vector propagator is ``the outer leg 
identity'' 
\begin{equation}
\partial_{[\mu} i\Bigl[\mbox{}_{\nu]}\Delta_{\rho}\Bigr](x;x')
 = -\frac1{4 H^2} \frac{\partial y}{\partial x^{[\mu}}
     \frac{\partial^2 y}{ \partial x^{\nu]} \partial x^{\prime \rho}}
       \Bigl[ (2 \!-\! y) A'(y) - k\Bigr] 
 = \frac{K'(y)}{4 (D\!-\!2) H^2} \frac{\partial y}{\partial x^{[\mu}} 
      \frac{\partial^2 y}{\partial x^{\nu]} \partial x^{\prime \rho}}
\; .
\label{outer leg identity}
\end{equation}
Our convention throughout this paper is that square bracketed indices are 
anti-symmetrized:
 $H_{[\mu\nu]} \equiv (H_{\mu\nu} \!-\! H_{\nu\mu})/2$.
The outer leg identity can be easily obtained by making use of
Eqs.~(\ref{photon propagator: Ansatz}--\ref{gamma function}).
Note that Eq.~(\ref{outer leg identity}) allows us to express the 
integrals entirely in terms of the $A$-type 
propagator~(\ref{scalar propagator}--\ref{scalar propagator:k}),
without involving the complicated function $\gamma(y)$.
This is a very welcome simplification because the series for $A(y)$
terminates in $D\!=\!4$, whereas that for $\gamma(y)$ does
not.


\section{The one loop stress-energy tensor for SQED}
\label{The one loop stress-energy tensor for SQED}

\subsection{The SQED action}
\label{The SQED action}

 When expressed in terms of the renormalized fields, the action of
scalar electrodynamics is,
\begin{eqnarray}
 S_{SQED} &=& \int d^Dx {\cal L}_{SQED}
\label{action:renormalized}
\\
  {\cal L}_{SQED} &=& - \left(1+\delta Z_2\right)
            \left({\cal D}_\mu\varphi\right)^*\left({\cal D}_\nu\varphi\right)
                g^{\mu\nu}\sqrt{-g}
        -\frac14 (1 + \delta Z_3) F_{\mu\nu}F_{\rho\sigma}g^{\mu\rho}
                g^{\nu\sigma}\sqrt{-g} \\
& & \hspace{2cm} -\delta \xi R \varphi^*\varphi\sqrt{-g} - \frac14
\delta \lambda (\varphi^* \varphi)^2 \sqrt{-g} \; . \nonumber
\end{eqnarray}
Here ${\cal D}_\mu \equiv \partial_\mu + ie A_\mu$ is the covariant 
derivative operator, $A_\mu$ is the photon field, $\varphi$ is the scalar field,
$g = {\rm det}[g_{\mu\nu}]$ is the determinant of the metric
tensor $g_{\mu\nu}$, and $g^{\rho\sigma}$ is its inverse,
$g^{\mu\rho}g_{\rho\nu}=\delta^\mu_{\;\nu}$. Note that we have chosen the
renormalized values of the conformal coupling, the scalar mass and the
scalar self-coupling to vanish. There is no need for a mass counterterm 
because mass is multiplicatively renormalized in dimensional regularization.
However, full renormalization does require a conformal counterterm 
and a scalar self-coupling counterterm. The various one loop counterterms 
are~\cite{Prokopec:2002uw,Prokopec:2007ak,Prokopec:2006ue},
\begin{eqnarray}
 \delta Z_2 &=& -\frac{e^2 H^{D-4}}{(4\pi)^{D/2}}
               \frac{2(D-1)\Gamma\left(\frac{D}{2}-1\right)}{(D-3)(D-4)}
+ e^2 \delta Z_{\rm 2fin} + \mathcal{O}(e^4) \,, \label{delta Z2} \\
 \delta Z_3 &=& +\frac{e^2 H^{D-4}}{(4\pi)^{D/2}}
               \frac{2\Gamma\left(\frac{D}{2}-1\right)}{(D-1)(D-3)(D-4)}
+ e^2 \delta Z_{\rm 3fin} + \mathcal{O}(e^4) \,, \label{delta Z3} \\
 \delta \xi &=& +\frac{e^2 H^{D-4}}{(4\pi)^{\frac{D}2}} \frac{\Gamma(D\!-\!1)}{
2 (D\!-\!3) \Gamma(\frac{D}2 \!+\!1)} \Bigl[ -\psi(D\!-\!1) \!+\! \psi(2) \!+\! 
   \psi\Bigl(\frac{D}2 \!-\! 1\Bigr) \!-\! \psi\Bigl(2 \!-\! 
     \frac{D}2\Bigr)\Bigr] + \mathcal{O}(e^4) \,,\qquad
\label{delta xi} \\
 \delta \lambda &=& +\frac{e^4 H^{D-4}}{(4 \pi)^{\frac{D}2}}
  \frac{\Gamma(D\!+\!1)}{(D\!-\!3)^2 \Gamma(\frac{D}2\!+\!1)}
   \Biggl\{ \psi'(D\!-\!1) \!+\! \psi'(2) \!-\! \psi'\Bigl(\frac{D}2\!-\!1
\Bigr) \!-\! \psi'\Bigl(2\!-\!\frac{D}2\Bigr) 
\nonumber \\
& & \hspace{2cm} + \frac2{D\!-\!3} \Bigl[-\psi(D\!-\!1) \!+\! \psi(2) 
\!+\! \psi\Bigl(\frac{D}2 \!-\! 1\Bigr) \!-\! \psi\Bigl(2 \!-\! 
\frac{D}2\Bigr)\Bigr] \nonumber \\
& & \hspace{4cm} + \Bigl[ -\psi(D\!-\!1) \!+\! \psi(2) \!+\! 
   \psi\Bigl(\frac{D}2 \!-\! 1\Bigr) \!-\! \psi\Bigl(2 \!-\! 
     \frac{D}2\Bigr)\Bigr]^2 \Biggr\} + \mathcal{O}(e^6) \,. \qquad
\label{delta lambda}
\end{eqnarray}
Note that, while
the finite parts of the two field strength renormalizations are arbitrary,
the finite parts of $\delta \xi$ and $\delta \lambda$ are fixed (at one
loop order) by requiring that the scalar effective potential be as
flat as possible \cite{Prokopec:2007ak}.

 From the action~(\ref{action:renormalized}) one obtains the 
stress-energy tensor in the standard manner,
$T_{\alpha\beta} = -(2/\sqrt{-g})\delta S_{SQED}/\delta g^{\alpha\beta}$.
Its expectation value in the presence of Bunch-Davies vacuum $| \Omega
\rangle$ can be broken up into a part derived from the Maxwell field 
strength and a part derived from the scalar kinetic and potential 
energies,
\begin{eqnarray}
 \langle \Omega | T_{\alpha\beta}(x)|\Omega\rangle
   &=& \langle \Omega | T_{\alpha\beta}(x)|\Omega\rangle_{\rm Maxwell}
 + \langle \Omega | T_{\alpha\beta}(x)|\Omega\rangle_{\rm scalar}
\nonumber\\
\langle \Omega | T_{\alpha\beta}(x)|\Omega\rangle_{\rm Maxwell}
    &=& \left(\delta_\alpha^{\;\mu}\delta_\beta^{\;\rho}g^{\nu\sigma}
         - \frac14 g_{\alpha\beta}g^{\mu\rho}g^{\nu\sigma}
       \right) (1 + \delta Z_3)
         \langle\Omega | F_{\mu\nu}(x) F_{\rho\sigma}(x)|\Omega\rangle
\nonumber\\
  \langle \Omega | T_{\alpha\beta}(x)|\Omega\rangle_{\rm scalar}
   &=& \left(\delta_\alpha^{\;\mu}\delta_\beta^{\;\nu}
         - \frac12 g_{\alpha\beta}g^{\mu\nu}
       \right) 2(1+\delta Z_2)
       \langle\Omega |({\cal D}_\mu\varphi(x))^\dagger
                      ({\cal D}_\nu\varphi(x))|\Omega\rangle
\nonumber\\
   &+&2\delta \xi\left(G_{\alpha\beta}
                     + g_{\alpha\beta}g^{\mu\nu}\nabla_\mu\nabla_\nu
                    -  \nabla_\alpha\nabla_\beta
       \right)
       \langle\Omega |\varphi(x)^\dagger\varphi(x)|\Omega\rangle
\nonumber\\
   &-& \frac14 \delta \lambda g_{\alpha\beta}
       \langle\Omega |[\varphi(x)^\dagger\varphi(x)]^2|\Omega\rangle
\,.
\label{stressenergy:SQED}
\end{eqnarray}
Here $G_{\alpha\beta} \equiv R_{\alpha\beta} - \frac 12 
g_{\alpha\beta} R$ is the Einstein curvature tensor, and 
$\nabla_\mu$ denotes the covariant derivative. On de Sitter background
the various curvatures become,
\begin{equation}
 G_{\alpha\beta} = -\frac12 (D\!-\!1)(D\!-\!2) H^2 g_{\alpha\beta}
\qquad , \qquad R_{\alpha\beta} = (D\!-\!1) H^2 g_{\alpha\beta}
\qquad {\rm and} \qquad R = D(D\!-\!1) H^2 \,.
\label{curvature tensors}
\end{equation}

\subsection{One loop stress-energy tensor}
\label{One loop stress-energy tensor}

At one loop order the various counterterms are irrelevant and we have,
\begin{eqnarray}
\langle\Omega| T_{\alpha\beta}|\Omega\rangle_{\rm 1\; loop}
  &=& \Big(
           g_\alpha^{\;\mu} g_\beta^{\;\rho}g^{\nu\sigma}
        -  \frac14 g_{\alpha\beta} g^{\mu\rho} g^{\nu\sigma}
       \Big)
     \langle\Omega| F_{\mu\nu}(x) F_{\rho\sigma}(x)|\Omega\rangle
                        _{\rm 1\; loop}
\nonumber\\
 &+& \Big(g_{\alpha}^{\;\mu} g_\beta^{\;\nu}
          - \frac 12 g_{\alpha\beta}g^{\mu\nu}
     \Big)2\langle\Omega| \partial_\mu \varphi^*\partial_\nu \varphi
            |\Omega\rangle_{\rm 1\; loop}
\,.
\label{Tmunu:1 loop}
\end{eqnarray}
The electromagnetic contribution derives from the coincident product 
of two field strengths,
\begin{eqnarray}
  \langle\Omega| F_{\mu\nu}(x) F_{\rho\sigma}(x)|\Omega\rangle
                     _{\rm 1\; loop}
    &=& \Big(
            \partial_\mu\partial^\prime_\rho i[_\nu\Delta_\sigma](x;x^\prime)
        \Big)_{x=x^\prime}
     +  \Big(
            \partial_\nu\partial^\prime_\sigma i[_\mu\Delta_\rho](x;x^\prime)
        \Big)_{x=x^\prime}
\nonumber\\
   &-&  \Big(
            \partial_\nu\partial^\prime_\rho i[_\mu\Delta_\sigma](x;x^\prime)
        \Big)_{x=x^\prime}
     -  \Big(
            \partial_\mu\partial^\prime_\sigma i[_\nu\Delta_\rho](x;x^\prime)
        \Big)_{x=x^\prime}
\,,
\label{Tmunu:1 loop:Maxwell}
\end{eqnarray}
while the scalar contribution comes from the scalar kinetic energy,
\begin{eqnarray}
 \langle\Omega| \partial_\mu \varphi^*(x)\partial_\nu \varphi(x)
          |\Omega\rangle_{\rm 1\; loop}
   &=&  \Big(
            \partial_\mu\partial^\prime_\nu i\Delta(x;x^\prime)
        \Big)_{x=x^\prime}
\,.
\label{Tmunu:1 loop:scalar}
\end{eqnarray}

 Making use of the outer leg identity~(\ref{outer leg identity})
we find that the nonvanishing terms are
\begin{equation}
  \langle\Omega| F_{\mu\nu}(x) F_{\rho\sigma}(x)|\Omega\rangle
                     _{\rm 1\; loop}
=  \frac{1}{(D \!-\!2) H^2}
  \Big\{\Big(\partial'_\rho\partial_{[\mu}y\Big)
        \Big(\partial_{\nu]}\partial'_{\sigma}y\Big) \, K'(y)
 \Big\}_{x = x'}
   = \frac{4 H^2 K'(0)}{D\!-\!2} \, g_{\rho[\mu}g_{\nu]\sigma}
\,.\qquad
\label{Tmunu:1 loop:Maxwell:2}
\end{equation}
When this is inserted into Eq.~(\ref{scalar propagator:2A'0-k}) one arrives at,
\begin{eqnarray}
  \langle\Omega| F_{\mu\nu}(x) F_{\rho\sigma}(x)|\Omega\rangle
                     _{\rm 1\; loop}
    &=&\frac{H^D}{(4\pi)^\frac{D}{2}}
        \frac{\Gamma(D-1)}{\Gamma\Big(\frac{D}{2}+1\Big)}
  \big(g_{\mu\rho}g_{\nu\sigma}
    -  g_{\mu\sigma}g_{\nu\rho}
  \big)
\,,
\label{Tmunu:1 loop:Maxwell:3}
\end{eqnarray}
 Similarly, we make use of relation~(\ref{scalar propagator:A'0})
for the scalar propagator~(\ref{scalar propagator})
to find,
\begin{eqnarray}
 \langle\Omega| \partial_\mu \varphi^*(x)\partial_\nu \varphi(x)
          |\Omega\rangle_{\rm 1\; loop}
   &=& - \frac{H^D}{(4\pi)^\frac{D}{2}}
             \frac{\Gamma(D)}{2\Gamma\Big(\frac{D}{2}+1\Big)}
                  g_{\mu\nu}
\,.\qquad
\label{Tmunu:1 loop:scalar:2}
\end{eqnarray}

 The one-loop stress-energy tensor is easily obtained from
Eqs.~(\ref{Tmunu:1 loop}), (\ref{Tmunu:1 loop:Maxwell:3})
and~(\ref{Tmunu:1 loop:scalar:2}).
The electromagnetic contribution is
\begin{eqnarray}
\langle \Omega| T_{\alpha\beta}|\Omega\rangle_{\rm 1\; loop,\, Maxwell}
  &=& -\frac{H^D\Gamma(D)}{(4\pi)^\frac{D}{2}\Gamma\Big(\frac{D}{2}+1\Big)}
              \frac{D-4}{4}g_{\alpha\beta}
     \;\;\stackrel{D\rightarrow 4}{\longrightarrow}\;\; 0
\,,
\label{Tmunu:1 loop:Maxwell:5}
\end{eqnarray}
while the scalar contribution reads,
\begin{eqnarray}
\langle\Omega| T_{\alpha\beta}|\Omega\rangle_{\rm 1\; loop, scalar}
 &=& \frac{H^D}{(4\pi)^\frac{D}{2}}
     \frac{\Gamma(D)}{\Gamma\Big(\frac{D}{2}+1\Big)}
            \frac{D-2}{2}g_{\alpha\beta}
     \;\;\stackrel{D\rightarrow 4}{\longrightarrow}\;\;
   \frac{3H^4}{16\pi^2}g_{\alpha\beta}
\,.
\label{Tmunu:1 loop:scalar:4}
\end{eqnarray}
 Summing the two
contributions~(\ref{Tmunu:1 loop:Maxwell:5}--\ref{Tmunu:1 loop:scalar:4})
results in the final expression~(\ref{Tmunu:1 loop}),
\begin{eqnarray}
\langle\Omega| T_{\alpha\beta}|\Omega\rangle_{\rm 1\; loop}
    &=& \frac{H^D}{(4\pi)^\frac{D}{2}}
        \frac{\Gamma(D)}{2\Gamma\Big(\frac{D}{2}\Big)}
            g_{\alpha\beta}
     \;\;\stackrel{D\rightarrow 4}{\longrightarrow}\;\;
   \frac{3H^4}{16\pi^2}g_{\alpha\beta}
\,.
\label{Tmunu:1 loop:6}
\end{eqnarray}
These one loop contributions to the stress-energy tensor in de Sitter
space are of the form a constant times $g_{\alpha\beta}$, hence
they can be absorbed in the cosmological term.

\section{General properties}
\label{General properties}

The one loop stress-tensor and its constituents are de Sitter invariant.
Their tensor structures are comprised entirely from products of the
metric, and the coefficients of these products are constants. That is
not true of higher loop contributions. These receive de Sitter breaking
contributions from the undifferentiated scalar propagator 
(\ref{scalar propagator}) and from vertex integrations that reach back
to the finite initial time. Spatial homogeneity and isotropy is
preserved but tensor structures can involve the unit timelike vector 
$a \delta^0_{\mu}$, and coefficients can depend upon $\ln(a)$. Of course 
it is the temporal growth implied by factors of $\ln(a)$ that gives this 
computation its physical interest, and it was to reproduce and sum up 
the leading powers of these logarithms to all orders that the stochastic 
formulation was developed.

The expectation value of $T_{\mu\nu}$ at a general order derives from 
the expectation values of four composite operators: $F_{\mu\nu} 
F_{\rho\sigma}$, $(D_{\mu} \varphi)^* D_{\nu} \varphi$, $\varphi^* 
\varphi$ and $(\varphi^* \varphi)^2$. It is useful to parameterize 
these expectation values as follows:
\begin{eqnarray}
\langle \Omega \vert F_{\mu\nu} F_{\rho\sigma} \vert \Omega \rangle
& \equiv & E g_{\mu [\rho} g_{\sigma] \nu} + F a^2 
\delta^0_{[\mu} g_{\nu] [\sigma} \delta^0_{\rho]}
 \; ,  \label{EF}
 \\
\langle \Omega \vert (D_{\mu} \varphi)^* D_{\nu} \varphi \vert \Omega 
\rangle & \equiv & J g_{\mu \nu} + L a^2 \delta^0_{\mu} 
\delta^0_{\nu}
 \; , 
\label{JL}
 \\
\langle \Omega \vert \varphi^* \varphi \vert \Omega \rangle
 & \equiv & 
\frac{M}{H^2} 
\; , 
\label{M}
 \\
\delta \lambda \langle \Omega \vert (\varphi^* \varphi)^2 \vert \Omega 
\rangle
 & \equiv & 
N \; . \label{N}
\end{eqnarray}
The expectation value of the stress-tensor can be given a similar form,
\begin{equation}
\langle \Omega \vert T_{\mu\nu} \vert \Omega \rangle 
= p  g_{\mu\nu} + (\rho \!+\! p)  a^2 \delta^0_{\mu} \delta^0_{\nu} 
\; .
\label{Tmunu:form}
\end{equation}
The coefficient functions $E$, $F$, $J$, $L$, $M$, $N$, $p$ and 
$(\rho + p)$ all take the form of $H^D/(4\pi)^{\frac{D}2}$ times 
functions of the scale factor and the dimensionless loop counting 
parameter $e^2 H^{D-4}/(4\pi)^{\frac{D}2}$. In SQED there can be at 
most one additional factor of $\ln(a)$ for every additional 
loop~\cite{Prokopec:2007ak}, and it turns out that the temporal 
coefficient functions $F$, $L$ and $(\rho + p)$ are always down from this 
limit by at least one overall power of $\ln(a)$. Note that we have already 
encountered this pattern at one loop where there are no powers of 
$e^2 H^{D-4}/(4\pi)^{\frac{D}2}$, the coefficient functions $E$, $J$ and 
$p$ are constant, and the three temporal coefficient functions accordingly 
vanish.

In addition to factors of $\ln(a)$, the various coefficient functions 
generally depend in a complicated way upon inverse powers of $a$ that
redshift to zero at late times. This can be illustrated by the two loop 
result for the expectation value of the stress-energy tensor of a massless,
minimally coupled scalar with a quartic 
self-interaction~\cite{Onemli:2002hr,Onemli:2004mb}. We report this with
a minor change of the renormalization scheme to bring it into conformity
with our convention of not employing a mass counterterm,
\begin{eqnarray}
\Bigl(p\Bigr)_{\lambda \varphi^4} \!\!\!& = &\!
 - \frac{\delta \Lambda_{\rm fin}}{
8\pi G} + \frac{\lambda H^4}{(2 \pi)^4} \Biggl\{\!-\frac18 \ln^2(a) -
\Bigl(\frac{15}{32} \!+\! \frac{\gamma_E}{32}\Bigr) \ln(a) + \frac{37}{144}
+ \frac{\gamma_E}{48} 
-\frac1{24} \sum_{n=1}^{\infty} \frac{n^2 \!-\! 4}{(n \!+\! 
1)^2}\frac{1}{a^{n+1}} \Biggr\}
 \nonumber\\
& & \hspace{0cm} 
 +\, \mathcal{O}(\lambda^2) 
\; , \qquad \label{phi^4} \\
\Bigl(\rho + p\Bigr)_{\lambda \varphi^4} \!& = &\!
 \frac{\lambda H^4}{(2 \pi)^4} 
\Biggl\{-\frac1{12} \ln(a) - \frac{37}{288} - \frac{\gamma_E}{96} + \frac1{18} 
 \frac{1}{a^3} -\frac1{24} \sum_{n=1}^{\infty} \frac{n \!+\! 2}{n \!+\! 1} \, 
\frac{1}{a^{n+1}}\Biggr\} + \mathcal{O}(\lambda^2) \; . \label{2ndphi^4}
\end{eqnarray}
The larger number of infrared logarithms --- two for each extra loop in this
model, as opposed to one in SQED --- is simply explained by having four,
as opposed to two, undifferentiated scalars in the 
interaction~\cite{Prokopec:2007ak}. What concerns us at the moment is the 
terms which fall off like powers of $1/a$. These terms derive from the 
lower limits of conformal time integrations. Because they are separately 
conserved it has been conjectured that these terms can be absorbed into 
perturbative corrections of the initial state~\cite{Onemli:2004mb}. If 
so, they are an artifact of the initial state rather than a universal 
feature of inflationary particle production. Because these sorts of terms 
fall off at late times they play no role in the stochastic formalism we
are checking. Hence there is no point in struggling to retain these terms 
and we shall not bother to do so. This allows vast simplifications from 
partially integrating without concern for surface terms, which we have 
already done in evaluating the scalar coefficient functions $J$, $L$ and 
$M$ at two loop order~\cite{Prokopec:2006ue}, and which we shall be using
throughout this work 
{\it e.g.} when extracting d'Alembertians from the inner loop.

An all-orders relation between the scalar coefficient functions is implied
by the operator equation~\cite{Prokopec:2006ue},
\begin{equation}
\Box (\varphi^* \varphi) = 2 g^{\mu\nu} (D_{\mu} \varphi)^* D_{\nu}
\varphi + \frac{2 \delta \xi}{1 \!+\! \delta Z_2} \, R \varphi^* \varphi
+ \frac{\delta \lambda}{1 \!+\! \delta Z_2} \, (\varphi^* \varphi)^2 \; .
\end{equation}
Taking expectation values and substituting equations (\ref{JL}-\ref{M})
gives,
\begin{equation}
- M'' - (D\!-\!1) M' = 2D J - 2 L + \frac{2 D (D\!-\!1) \delta \xi}{1 \!+\!
\delta Z_2} \, M + \frac1{1 \!+\! \delta Z_2} \, N \; ,
\end{equation}
where a prime denotes differentiation with respect to $\ln(a)$. A slight
rearrangement produces a key result,
\begin{equation}
-2 D \Bigl[(1 \!+\! \delta Z_2) J + \delta \xi (D\!-\! 1) M\Bigr] = 
(1 \!+\! Z_2) \Bigl[M'' + (D\!-\!1) M'\Bigr] - 2 (1 \!+\! \delta Z_2) L + 
N \; . \label{keyres}
\end{equation}

It is now time to work out the Maxwell and scalar contributions to
$p$ and $(\rho + p)$ in terms of the various coefficient functions.
We find the Maxwell contributions from relations (\ref{stressenergy:SQED}) 
and (\ref{EF}),
\begin{eqnarray}
\Bigl(p\Bigr)_{\rm Maxwell} & = & (1 \!+\! \delta Z_3) \Bigl[
-\frac18 (D\!-\!1) (D\!-\!4) E + \frac18 (D\!-\!3) F\Bigr] \; , 
\label{pMaxwell} \\
\Bigl(\rho + p\Bigr)_{\rm Maxwell} & = & (1 \!+\! \delta Z_3) \, \frac14
(D\!-\!2) F \; . 
\end{eqnarray}
The scalar contribution to $(\rho + p)$ is almost as simple,
\begin{equation}
\Bigl(\rho + p\Bigr)_{\rm scalar} = (1 \!+\! \delta Z_2) \, 2 L +
\delta \xi \Bigl[-2 M'' + 2 M'\Bigr] \; . \label{rhoscalar}
\end{equation}
However, it is best to modify the scalar contribution to the pressure 
using Eqn. (\ref{keyres}),
\begin{eqnarray}
\lefteqn{\Bigl(p\Bigr)_{\rm scalar} \!\!\!\! = -(D\!-\!2) \Bigl[ (1 \!+\! 
\delta Z_2) J \!+\! \delta \xi (D\!-\!1) M\Bigr] 
+ (1 \!+\! \delta Z_2) L - 2 \delta \xi 
\Bigl[M'' \!+\! (D \!-\! 2) M'\Bigr] - \frac14 N \; , \qquad } \\
& & = (1 \!+\! \delta Z_2) \Bigl(\frac{D\!-\!2}{2 D}\Bigr) 
\Bigl[M'' \!+\! (D\!-\!1) M'\Bigr] \!-\! 2\delta \xi \Bigl[M'' \!+\! 
(D\!-\!2) M'\Bigr] \!+\! \frac2{D} (1 \!+\! \delta Z_2) L \!+\!
\Bigl(\frac{D\!-\!4}{4 D}\Bigr) N \; . \qquad \label{pscalar}
\end{eqnarray}
This final form (\ref{pscalar}) is quite significant because it precludes
$\ln(a)$ contributions from the scalar at two loop order. This is a key
prediction of the stochastic formalism~\cite{Prokopec:2007ak}.

Stress-energy conservation provides an important accuracy check,
\begin{equation}
\nabla^{\nu} \langle \Omega \vert T_{\mu\nu} \vert \Omega \rangle =
- H \, a \delta_{\mu}^0 \Bigl[ (\rho + p)' - p' + (D\!-\!1) (\rho + p)
\Bigr] = 0 \; .
\end{equation}
Recall that a prime denotes differentiation with respect to $\ln(a)$, 
and note that the powers of $1/a$ we ignore should be separately
conserved. The Maxwell and scalar contributions are not separately 
conserved but the Heisenberg equations of motion for SQED relate their
divergences to the expectation value of a field strength contracted
into a current,
\begin{equation}
\nabla^{\nu} \langle \Omega \vert T_{\mu\nu} \vert \Omega \rangle_{\rm
Maxwell} = - \nabla^{\nu} \langle \Omega \vert T_{\mu\nu} \vert \Omega 
\rangle_{\rm scalar} = -i e (1 \!+\! \delta Z_2) g^{\rho\sigma}
\Bigl\langle \Omega \Bigl\vert F_{\mu \rho} \Bigl[\varphi^* D_{\sigma}
\varphi \!-\! (D_{\sigma} \varphi)^* \varphi\Bigr] \Bigr\vert \Omega
\Bigr\rangle \; . \label{partial}
\end{equation}
Although we shall not give the derivation it is simple to evaluate
(\ref{partial}) at the order $e^2$ relevant to this analysis,
\begin{eqnarray}
\lefteqn{ -i e (1 \!+\! \delta Z_2) g^{\rho\sigma}
\Bigl\langle \Omega \Bigl\vert F_{\mu \rho} \Bigl[\varphi^* D_{\sigma}
\varphi \!-\! (D_{\sigma} \varphi)^* \varphi\Bigr] \Bigr\vert \Omega
\Bigr\rangle } \nonumber \\
& & \hspace{2cm} = -H \, a \delta^0_{\mu} \Biggl\{ \frac{e^2 H^{2D -4}}{
(4\pi)^D} \Gamma(D\!-\!1) \Bigl[ -\frac8{D\!-\!4} - 10 + 
\mathcal{O}\Bigl(a^{-1},D\!-\!4\Bigr) \Bigr] + \mathcal{O}(e^4)\Biggr\}
\; . \qquad \label{explicit}
\end{eqnarray}

This is probably the right point to discuss finiteness. Conventional
renormalization only serves to remove divergences from noncoincident
1PI (one particle irreducible) functions. Composite operators such as
$F_{\mu\nu}(x) F_{\rho\sigma}(x)$ and the others require additional, 
composite operator renormalization to remove their divergences. This
is evident from the divergence of expression (\ref{explicit}). One
renormalizes composite operators by adding a series of counter-operators 
which remove the divergences order-by-order in the loop 
expansion~\cite{Weinberg,Itzykson}. As with conventional renormalization,
there are ambiguities in finite parts, which are resolved by imposing
additional renormalization conditions. There is absolutely no reason
to burden an already formidable analysis by bothering with any of this.
The stochastic formalism which we are trying to check provides 
unambiguous predictions for the dimensionally regulated expectation
values of composite operators~\cite{Prokopec:2007ak}, so we may as well 
leave the regulator on and verify these predictions.

A related point concerns the stress-energy tensor. Unlike the other composite
operators, its expectation value is a 1PI function of an enlarged theory;
specifically, its represents matter contributions to the 1-graviton 1PI 
function in gravity + SQED. It must therefore be that divergences in the
expectation value of the stress-energy 
tensor can be absorbed with the addition 
of purely gravitational counterterms which all degenerate on de Sitter
background to changes in the bare cosmological constant. This was shown
at two loop order in $\lambda \varphi^4$ theory by explicit 
computation~\cite{Onemli:2002hr,Onemli:2004mb}, and the ability to
change the result by an additional, finite renormalization of the
cosmological constant is evident in expression (\ref{phi^4}). For SQED
this means three things:
\begin{itemize}
\item{Divergences in the temporal coefficient functions $F$ and $L$
must cancel when they are formed to give the full $(\rho + p)$;}
\item{The parts of $p$ which contain factors of $\ln(a)$ must be
finite; and}
\item{The parts of $p$ which are constant can diverge.}
\end{itemize}
The first two facts provide more accuracy checks; the third means that 
there is no point in working very hard to determine the constant 
contributions to $p$ because any constant part of $p$ can be subsumed 
into a renormalization of the cosmological constant.

It remains to summarize what is already known about the coefficient functions.
The Maxwell coefficient functions (\ref{EF}) take the form,
\begin{eqnarray}
E & = & \frac{H^D}{(4\pi)^{\frac{D}2}} \Biggl\{ \frac{2 \Gamma(D\!-\!1)}{
\Gamma(\frac{D}2 \!+\!1)} + \frac{e^2 H^{D-4}}{(4 \pi)^{\frac{D}2}} \Bigl[E_1
\ln(a) + E_2 + \mathcal{O}(a^{-1})\Bigr] + \mathcal{O}(e^4) \Biggr\} , \qquad
\label{Eexp} \\
F & = & \frac{H^D}{(4\pi)^{\frac{D}2}} \Biggl\{ 0 + \frac{e^2 H^{D-4}}{(4 
\pi)^{\frac{D}2}} \Bigl[F_1 + \mathcal{O}(a^{-1})\Bigr] + \mathcal{O}(e^4) 
\Biggr\} . \label{Fexp}
\end{eqnarray}
The point of this paper is to compute the $D$-dependent numbers $E_1$, 
$E_2$ and $F_1$! They give the order $e^2 H^{D-4}/(4\pi)^{\frac{D}2}$ 
electromagnetic contributions to $p$ and $(\rho + p)$,
\begin{eqnarray}
\Bigl( p \Bigr)_{\rm Maxwell} \!\!\!\!\! & = & \!\frac{H^D}{(4\pi)^{\frac{D}2}} 
\Biggl\{-\frac{(D \!-\! 4) \Gamma(D)}{4 \Gamma(\frac{D}2 \!+\! 1)} \!+\!
\frac{e^2 H^{D-4}}{(4\pi)^{\frac{D}2}} \Biggl[-\frac18 (D\!-\!1) (D\!-\!4) 
E_1 \ln(a) - \frac{2 \Gamma(D\!-\!1)}{D (D\!-\!2) (D\!-\!3)} \nonumber\\
& & \hspace{2cm} -\frac18 (D\!-\!1) (D\!-\!4) E_2 + \frac18 (D\!-\!3) F_1 + 
\mathcal{O}\Bigl(a^{-1},D\!-\!4\Bigr)\Biggr] + \mathcal{O}(e^4) \Biggr\} , 
\label{pmax} \\
\Bigl( \rho + p \Bigr)_{\rm Maxwell} \!\!\!\!\! & = & \!\frac{H^D}{(4\pi)^{
\frac{D}2}} \Biggl\{0 + \frac{e^2 H^{D-4}}{(4\pi)^{\frac{D}2}} \Bigl[\frac14
(D\!-\!2) F_1 + \mathcal{O}(a^{-1})\Bigr] + \mathcal{O}(e^4) \Biggr\} .
\label{rhomax}
\end{eqnarray}
Note that Eqn. (\ref{explicit}) implies a relation between the coefficients
$E_1$ and $F_1$,
\begin{equation}
\frac18 (D\!-\!1) (D\!-\!4) E_1 + \frac14 (D\!-\!1) (D\!-\!2) F_1
= -\frac{8 \Gamma(D\!-\!1)}{D\!-\!4} - 20 + \mathcal{O}(D\!-\!4)
\; . \label{E1F1rel}
\end{equation}
The stochastic formalism predicts~\cite{Prokopec:2007ak},
\begin{eqnarray}
E_1 & = & \frac{8 \Gamma^2(D\!-\!1)}{(D\!-\!3) \Gamma(\frac{D}2 \!+\! 1)
\Gamma(\frac{D}2)} \Bigl[-\psi\Bigl(2 \!-\! \frac{D}2\Bigr) + \psi\Bigl(
\frac{D}2 \!-\! 1\Bigr) - \psi(D\!-\!1) + \psi(2) \Bigr] \; , \qquad \\
& = & -\frac{16 \Gamma(D\!-\!1)}{D \!-\!4} + 16 + \mathcal{O}(D\!-\!4) \; .
\label{E1pre}
\end{eqnarray}
So we can use Eqn. (\ref{E1F1rel}) to infer what $F_1$ should be,
\begin{equation}
F_1 = -\frac{16 \Gamma(D\!-\!1)}{(D\!-\!1) (D \!-\!4)} + \mathcal{O}(D\!-\!4) 
\; . \label{F1pre}
\end{equation}
No comparable check involves $E_2$, and we see from (\ref{pmax}) that $E_2$
has no physical relevance because it can be absorbed into a renormalization
of the cosmological constant.

The coefficient functions (\ref{JL}) associated with the scalar kinetic 
energy are~\cite{Prokopec:2006ue},
\begin{eqnarray}
J & = & \frac{H^D}{(4\pi)^{\frac{D}2}} \Biggl\{- \frac{\Gamma(D)}{2
\Gamma(\frac{D}2 \!+\! 1)} + \delta \xi \times \frac{\Gamma(D)}{\Gamma(
\frac{D}2)} \Bigl[-2 \ln(a) + \pi \cot\Bigl(\frac{D\pi}2\Bigr)\Bigr] + 
\delta Z_2 \times \frac{\Gamma(D)}{2 \Gamma(\frac{D}2 \!+\! 1)} \nonumber \\
& & \hspace{3.5cm} + \frac{e^2 H^{D-4}}{(4 \pi)^{\frac{D}2}} \Bigl[
\frac{16 \Gamma(D\!-\!1)}{(D\!-\!1) (D\!-\!4)} \!-\! 17 \!+\! 2 \pi^2 \!+\!
\mathcal{O}\Bigl(a^{-1},D\!-\!4\Bigr)\Bigr] + \mathcal{O}(e^4) 
\Biggr\} , \qquad \\
L & = & \frac{H^D}{(4\pi)^{\frac{D}2}} \Biggl\{ 0 + \frac{e^2 H^{D-4}}{(4 
\pi)^{\frac{D}2}} \Bigl[\frac{10 \Gamma(D\!-\!1)}{(D\!-\!1) (D\!-\!4)}
+ \frac83 + \mathcal{O}\Bigl(a^{-1},D\!-\!4\Bigr)\Bigr] + \mathcal{O}(e^4) 
\Biggr\} .
\end{eqnarray}
Recall that $\delta Z_2$ and $\delta \xi$ are of order $e^2 H^{D-4}/(4\pi)^{
\frac{D}2}$ and are given in expressions (\ref{delta Z2}) and (\ref{delta xi}),
respectively. The coefficient function $M$ was defined in Eqn. (\ref{M}) and 
takes the form~\cite{Prokopec:2006ue},
\begin{eqnarray}
\lefteqn{M = \frac{H^D}{(4\pi)^{\frac{D}2}} \Biggl\{\frac{\Gamma(D\!-\!1)}{
\Gamma(\frac{D}2)} \Bigl[2 \ln(a) - \pi \cot\Bigl(\frac{\pi D}{2}\Bigr)\Bigr]
+ \delta Z_2 \times \frac{\Gamma(D\!-\!1)}{\Gamma(\frac{D}2)} \Bigl[-2 \ln(a) 
+ \pi \cot\Bigl(\frac{\pi D}{2}\Bigr)\Bigr] } \nonumber \\
& & \hspace{1cm} + \frac{e^2 H^{D-4}}{(4 \pi)^{\frac{D}2}} \Biggl[
\Bigl[\frac{-108 \Gamma(D\!-\!1)}{(D\!-\!1)^2 (D\!-\!4)} \!+\! 40 \!-\! 
\frac{16 \pi^2}3 \Bigr] \ln(a) + M_1 + \mathcal{O}\Bigl(a^{-1},D\!-\!4\Bigr)
\Biggr] + \mathcal{O}(e^4) \Biggr\} \; . \qquad
\end{eqnarray}
The constant $M_1$ has not been determined but it drops out because only
derivatives of $M$ enter expression (\ref{pscalar}) for the stress-energy 
tensor. 
Note that even the lowest order contribution to $M$ contains an infrared 
logarithm, so the $\ell$ loop result can have up to $\ell$ factors of 
$\ln(a)$, as opposed to only $\ell - 1$ in $J$. This is entirely in 
conformity with the rule that each additional loop brings at most one 
additional factor of $\ln(a)$~\cite{Prokopec:2007ak}. 
The final coefficient function~(\ref{N}) is,
\begin{equation}
N = \frac{H^D}{(4\pi)^{\frac{D}2}} \Biggl\{0 + \frac{\delta \lambda H^{D-4}}{
(4 \pi)^{\frac{D}2}} \frac{2 \Gamma^2(D\!-\!1)}{\Gamma^2(\frac{D}2)} \Bigl[2 
\ln(a) - \pi \cot\Bigl(\frac{\pi D}{2}\Bigr)\Bigr]^2 + \mathcal{O}(e^6)
\Biggr\} \; .
\end{equation}
Because $\delta \lambda$ in~(\ref{delta lambda})
is of order $[ e^2 H^{D-4}/(4\pi)^{\frac{D}2} ]^2$,
the function $N$ does not affect the two loop stress-energy tensor.

We infer the scalar contribution to the stress-energy tensor by substituting
these coefficient functions into expressions (\ref{pscalar}) and
(\ref{rhoscalar}),
\begin{eqnarray}
\Bigl( p \Bigr)_{\rm scalar} \!\!\!\! & = & \frac{H^D}{(4\pi)^{\frac{D}2}} 
\Biggl\{\frac{(D \!-\! 2) \Gamma(D)}{2 \Gamma(\frac{D}2 \!+\! 1)} + 
\frac{e^2 H^{D-4}}{(4\pi)^{\frac{D}2}} \Biggl[0 \times \ln(a) 
+ \frac{2 \Gamma(D \!-\! 1)}{(D\!-\!1) (D\!-\!4)} \nonumber\\
& & \hspace{6cm} + \frac{94}3 - 4\pi^2 + \mathcal{O}\Bigl(a^{-1},D\!-\!4
\Bigr)\Biggr] + \mathcal{O}(e^4) \Biggr\} , \qquad \label{pscal} \\
\Bigl( \rho + p \Bigr)_{\rm scalar} \!\!\!\! & = & \frac{H^D}{(4\pi)^{
\frac{D}2}} \Biggl\{0 + \frac{e^2 H^{D-4}}{(4\pi)^{\frac{D}2}} \Bigl[
\frac{8 \Gamma(D \!-\! 1)}{(D\!-\!1) (D\!-\!4)} + \frac{20}3 + \mathcal{O}
\Bigl(a^{-1},D\!-\!4\Bigr)\Bigr] + \mathcal{O}(e^4) \Biggr\} . \qquad
\label{rhoscal}
\end{eqnarray}
Note that there are no infrared logarithms in (\ref{pscal}) at this order.
Note also that (\ref{rhoscal}) obeys the partial conservation identity
(\ref{partial}-\ref{explicit}), which represents an additional check 
of the two loop scalar stress-energy calculation of
Ref.~\cite{Prokopec:2006ue}.

\section{The two loop gauge field strength bilinear}
\label{The two loop gauge field strength bilinear}

 Three diagrams contribute to the VEV at two loop order.
They are depicted in Fig.~1.
\begin{figure}
\centerline{\epsfig{file=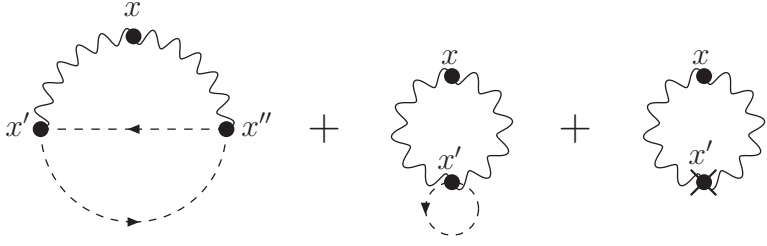}}
\caption{Two loop contributions to
$\langle \Omega \vert F_{\mu\nu}(x)
          F_{\rho\sigma}(x) \vert \Omega \rangle$.}
\end{figure}

\subsection{The counterterm diagram}
\label{The counterterm diagram}

 The diagram on the far right of Fig.~1 derives from the photon field
strength renormalization. Covariance and index symmetries dictate its form,
\begin{eqnarray}
{\cal I}_{\rm c.t.} = 4i \delta Z_3 \!\int\! d^Dx'
\Biggl\{ \partial_{[\mu}
i\Bigl[\mbox{}_{\nu]} \Delta_{\alpha}\Bigr]_{\scriptscriptstyle ++}
\!\!\!\!\!\!(x;x') \sqrt{-g'}
\Bigl[ g^{\prime \alpha\beta} g^{\prime \gamma\delta} \!-\!
g^{\prime \alpha \delta} g^{\prime \beta\gamma}\Bigr]
 \nabla'_{\gamma} \nabla'_{\delta}
\partial_{[\rho} i\Bigl[\mbox{}_{\sigma]} \Delta_{\beta}\Bigr]
_{ \scriptscriptstyle ++}\!\!\!\!\!\!(x;x') - ({\scriptscriptstyle +-})\Biggr\}
,\!\!\!
\nonumber\\
\label{I:counterterm}
 \end{eqnarray}
where $\nabla'_{\mu}$ stands for the covariant derivative operator
with respect to $x^{\prime \mu}$. To evaluate this diagram we
first interchange the order of covariant differentiation in the term
proportional to $g^{\prime \alpha\delta} g^{\prime \beta\gamma}$ and
then use the Lorentz gauge condition to get,
 \begin{eqnarray}
\sqrt{-g'}
 \Bigl[ g^{\prime \alpha\beta} g^{\prime \gamma\delta} \!-\!
g^{\prime \alpha \delta} g^{\prime \beta\gamma}\Bigr]
 \nabla'_{\gamma} \nabla'_{\delta}
 \partial_{[\rho} i\Bigl[\mbox{}_{\sigma]}
\Delta_{\beta}\Bigr]_{\scriptscriptstyle +\pm}\!\!\!\!\!\!(x;x')
= \sqrt{-g'} \Bigl[g^{\prime \alpha \beta}
\Box^{\prime}
- R^{\prime \alpha\beta}\Bigr]
\partial_{[\rho} i\Bigl[\mbox{}_{\sigma]}
\Delta_{\beta}\Bigr]_{\scriptscriptstyle +\pm}\!\!\!\!\!\!(x;x')
\; . \qquad
\end{eqnarray}
Now we use the appropriate Schwinger-Keldysh generalizations of the
propagator Eq.~(\ref{eom:vector propagator}),
\begin{eqnarray}
\sqrt{-g'} \Bigl[g^{\prime \alpha \beta} \Box'
 - R^{\prime \alpha\beta}\Bigr]
i\Bigl[\mbox{}_{\sigma} \Delta_{\beta}\Bigr]_{ \scriptscriptstyle ++}
\!\!\!\!\!\!(x;x')
&=&  i \delta^{\alpha}_{\sigma} \delta^D(x \!-\! x')
+ \sqrt{-g'} g^{\prime \alpha\beta}
\partial'_{\beta} \partial_{\sigma}
i\Delta_{\scriptscriptstyle ++}\!(x;x') \; ,
\qquad \label{+der} \\
\sqrt{-g'}
\Bigl[g^{\prime \alpha \beta} \Box'
- R^{\prime \alpha\beta}\Bigr]
i\Bigl[\mbox{}_{\sigma} \Delta_{\beta}\Bigr]_{ \scriptscriptstyle +-}
\!\!\!\!\!\!(x;x') &=& 0
\; .
\end{eqnarray}
The anti-symmetrized derivative with respect
to $x^{\rho}$ makes the second term in (\ref{+der}) drop out,
so the result for the far right diagram~(\ref{I:counterterm}) of Fig.~1 is
({\it cf.} Eqs.~(\ref{Tmunu:1 loop:Maxwell}),
(\ref{Tmunu:1 loop:Maxwell:2}) and~(\ref{Tmunu:1 loop:Maxwell:3})),
\begin{eqnarray}
\hskip -0.2cm
{\cal I}_{\rm c.t.} \!\!&=&\!\! -4 \delta Z_3 \!\! \int \!\! d^Dx'
 \partial_{[\mu}
i\Bigl[ \mbox{}_{\nu]} \Delta_{\alpha}\Bigr]_{\scriptscriptstyle ++}
\!\!\!\!\!\!(x;x') \partial_{[\rho} \delta^{\alpha}_{\sigma]}
 \delta^D(x \!-\! x')
\nonumber \\
\!\!&=&\!\! - \delta Z_3 \!
\lim_{x' \rightarrow x} \Biggl\{\! \partial_{\mu} \partial_{\rho}'
i\Bigl[\mbox{}_{\nu}\Delta_{\sigma}\Bigr]_{\scriptscriptstyle ++}\!\!\!\!\!\!
(x;x')
\!-\! \partial_{\nu} \partial_{\rho}'
i\Bigl[\mbox{}_{\mu}\Delta_{ \sigma}\Bigr]_{\scriptscriptstyle ++}
\!\!\!\!\!\!(x;x')
\!+\! \partial_{\nu} \partial_{\sigma}'
i\Bigl[\mbox{}_{\mu}\Delta_{\rho}\Bigr]_{\scriptscriptstyle ++}
\!\!\!\!\!\!(x;x')
\!-\! \partial_{\mu} \partial_{\sigma}'i\Bigl[\mbox{}_{\nu}
\Delta_{\rho}\Bigr]_{\scriptscriptstyle ++}\!\!\!\!\!\!(x;x')\! \Biggr\}
, \!\!\!
\nonumber
\\
\!\! &=&\!\! 2 \delta Z_3 H^2 \Bigl[2 A'(0) - k\Bigr] \Bigl(g_{\mu\rho} 
g_{\nu\sigma} - g_{\mu\sigma} g_{\nu\rho}\Bigr)
\nonumber\\
\!\! &=&\!\! -\delta Z_3 \times \langle \Omega \vert F_{\mu\nu}(x)
F_{\rho\sigma}(x) \vert \Omega \rangle_{\rm one\ loop} \; .
\label{Ict:res}
\end{eqnarray}
From expression (\ref{pMaxwell}) we see that the contribution of 
${\cal I}_{\rm c.t.}$ to the stress-energy tensor is exactly canceled by the
contribution from $\delta Z_3$ times the one loop result.

\subsection{The figure 8 diagram}
\label{The figure 8 diagram}

The middle diagram  of Fig.~1 is,
\begin{equation}
{\cal I}_{\infty}
 = -8i e^2 \int d^Dx' \sqrt{-g'}
 \Biggl\{ \partial_{[\mu} i\Bigl[\mbox{}_{\nu]} \Delta_{\alpha}
                          \Bigr]_{\scriptscriptstyle ++}\!\!\!\!\!\!(x;x')
\partial_{[ \rho} i\Bigl[\mbox{}_{\sigma]} \Delta_{\beta}
                   \Bigr]_{\scriptscriptstyle ++}\!\!\!\!\!\!(x;x')
 - (+-)\Biggr\} g^{\prime \alpha\beta} i\Delta(x';x') \; .
\end{equation}
Making use of the outer leg identity~(\ref{outer leg identity})
transforms this to,
\begin{equation}
{\cal I}_{\infty} =
 \frac{-i e^2}{2 (D\!-\!2)^2 H^4} \int d^Dx' \sqrt{-g'} 
 i\Delta(x';x') \Biggl\{ \Bigl[ K'(y_{\scriptscriptstyle ++})\Bigr]^2
     - (+-) \Biggr\} \frac{\partial y}{\partial x^{[\mu}}
\frac{\partial^2 y }{\partial x^{\nu]} \partial x^{\prime \alpha}}
  g^{\prime \alpha\beta}
\frac{\partial^2 y}{\partial x^{\prime \beta} \partial x^{[\sigma}}
 \frac{\partial y}{\partial x^{\rho]}}
 \; . 
\end{equation}
Now use the contraction identity~\cite{KW0,Tsamis:2006gj,Prokopec:2006ue},
\begin{equation}
\frac{\partial^2 y}{\partial x^{\nu} \partial x^{\prime \alpha}}
g^{\prime \alpha\beta}
\frac{\partial^2 y}{\partial x^{\prime \beta} \partial x^{\sigma}}
= 4 H^4 g_{\nu\sigma}
- H^2 \frac{\partial y}{\partial x^{\nu}}
     \frac{\partial y}{\partial x^{\sigma}}
\; .
\label{contraction identity}
\end{equation}
The final term drops out owing to the anti-symmetrizations,
so the middle diagram becomes
\begin{eqnarray}
{\cal I}_{\infty} &=& \frac{-2 i e^2}{(D\!-\!2)^2} \int d^Dx' \sqrt{-g'} 
i\Delta(x';x') \frac{\partial y}{\partial x^{[\mu}} g_{\nu ] [\sigma}
        \frac{\partial y}{\partial x^{\rho]}}
\Biggl\{ \Bigl[ K'(y_{\scriptscriptstyle ++})\Bigr]^2 - (+-) \Biggr\}
\nonumber\\
   &=& \frac{-2 i e^2}{(D\!-\!2)^2} \sum_{\pm}(\pm)\int d^Dx' 
\sqrt{-g'} i\Delta(x';x') \times \frac{\partial}{\partial x^{[\mu}}
  K(y_{\scriptscriptstyle +\pm}) g_{\nu ] [\sigma}
\frac{\partial}{\partial x^{\rho]}} K(y_{\scriptscriptstyle +\pm})
\, .
\label{diagram:figure 8}
\end{eqnarray}
From Eqs.~(\ref{scalar propagator:A}--\ref{scalar propagator:k})
and~(\ref{scalar propagator:A0})
we can read off the coincidence limit of the scalar propagator,
\begin{equation}
 i \Delta(x';x') = A(0) + 2 k \ln(a')
     = \frac{H^{D-2}}{(4\pi)^{\frac{D}2}}
           \frac{\Gamma(D \!-\! 1)}{\Gamma(\frac{D}2)}
  \left[-\cos\Bigl(\frac{\pi}2 D\Bigr)
            \Gamma\left(\frac{D}{2}\right)\Gamma\left(1-\frac{D}{2}\right)
             + 2\ln(a')
  \right]
\label{scalar propagator:coinc}
\; .
\end{equation}
We will see that the leading logarithm contribution $E_1$ derives entirely 
from the $2 k \ln(a')$ term in the integral~(\ref{diagram:figure 8}).


\subsection{The two leg diagram}
\label{The two leg diagram}

 The far left diagram on Fig.~1 is the most difficult in any gauge.
To simplify it we must derive new identities for contracting adjacent basis
tensors into one another. Let us first
make the following definitions for the three
 coordinate separations and their associated length functions,
\begin{eqnarray}
\Delta x^{\mu} \equiv x^\mu \!-\! {x'}^\mu & \Longrightarrow & y
\equiv y(x;x') = a a' H^2 \Delta x^2
\; ,
\\
\Delta x^{\prime \mu} \equiv {x'}^\mu \!-\! {x''}^\mu & \Longrightarrow & y'
\equiv
y(x';x'') = a' a'' H^2 \Delta x^{\prime 2}
 \; ,
\\
\Delta x^{\prime \prime \mu} \equiv x^\mu \!-\! {x''}^\mu
          & \Longrightarrow & y''
\equiv y(x;x'') = a a'' H^2 \Delta x^{\prime \prime 2} \; .
\end{eqnarray}
The three fundamental ``adjacent'' contractions we require are,
\begin{eqnarray}
\frac{\partial y}{\partial x^{\prime \rho}}
g^{\prime \rho\sigma}
 \frac{\partial y'}{\partial x^{\prime \sigma}}
& =  & H^2 \Bigl\{ 2 y + 2 y' - y y' - 2 y''\Bigr\}
 \; ,
\label{identity:1}
\\
 \frac{\partial^2 y}{\partial x^{\mu}
 \partial x^{\prime \rho}}
g^{\prime \rho\sigma}
 \frac{\partial y'}{\partial x^{\prime \sigma}}
& = & H^2 \Bigl\{ (2 \!-\! y') \frac{\partial y}{\partial x^{\mu}}
 - 2 \frac{\partial y''}{\partial x^{\mu}} \Bigr\}
\; ,
\label{identity:2}\\
 \frac{\partial^2 y}{\partial x^{\mu}
 \partial x^{\prime \rho}}
g^{\prime \rho\sigma}
\frac{\partial^2 y'}
     {\partial x^{\prime \sigma} \partial x^{\prime \prime \nu}}
 & = &
 -2 H^2 \frac{\partial^2 y''}
             {\partial x^{\mu} \partial x^{\prime\prime \nu}}
 - H^2 \frac{\partial y}{\partial x^{\mu}}
       \frac{\partial y'}{\partial x^{\prime\prime \nu}}
\label{identity:3}
\end{eqnarray}
For $x^{\prime\prime \mu} \rightarrow x^{\mu}$
these relations degenerate to
the well known identities~\cite{KW0,Tsamis:2006gj,Prokopec:2006ue},
\begin{eqnarray}
\frac{\partial y}{\partial x^{\prime \rho}}
g^{\prime \rho\sigma}
 \frac{\partial y}{\partial x^{\prime \sigma}}
   & =  & (4-y)y H^2
 \; ,
\label{identity:4}
\\
 \frac{\partial^2 y}{\partial x^{\mu}
 \partial x^{\prime \rho}}
g^{\prime \rho\sigma}
 \frac{\partial y}{\partial x^{\prime \sigma}}
& = & H^2 (2 \!-\! y) \frac{\partial y}{\partial x^{\mu}}
\; ,
\label{identity:5}
\\
 \frac{\partial^2 y}{\partial x^{\nu}\partial x^{\prime \alpha}}
   g^{\prime \alpha\beta}
\frac{\partial^2 y}
     {\partial x^{\prime \beta}\partial x^{\sigma}}
 & = &
  4 H^4 g_{\nu\sigma}
 - H^2 \frac{\partial y}{\partial x^{\nu}}
       \frac{\partial y}{\partial x^{\sigma}}
\; .
\label{identity:6}
\end{eqnarray}
Note that the last of these is just
the contraction identity~(\ref{contraction identity}).
The contractions on the ``other'' site $x''$ are trivial
 to obtain from these by interchanging
$x^{\prime \mu}$ and $x^{\prime\prime \mu}$,
 which interchanges $y$ and $y''$, while leaving $y'$ unchanged.
Two other useful identities are,
\begin{equation}
 \nabla_\nu\frac{\partial}{\partial x^\mu} y = H^2 (2-y)g_{\mu\nu}
\qquad {\rm and} \qquad
 \Box y = D(2-y)H^2
\,,
\label{identity:7}
\end{equation}
where $\nabla_\nu$ is the covariant derivative and $\Box$ is the
scalar d'Alembertian.

We are now ready to tackle the leftmost (two leg) diagram of Fig.~1,
\begin{eqnarray}
{\cal I}
& =& \lefteqn{8 e^2 \sum_{\pm \pm} (\pm) (\pm)
 \int d^Dx' \sqrt{-g'} \partial_{[\mu} i\Bigl[\mbox{}_{\nu]}
 \Delta_{\alpha}\Bigr]_{\scriptscriptstyle +\pm}\!\!\!\!\!(x;x')
\int d^Dx'' \sqrt{-g''} \partial_{[\rho} i\Bigl[\mbox{}_{\sigma]}
\Delta_{\gamma}\Bigr]_{\scriptscriptstyle +\pm}\!\!\!\!\!(x;x'') }
\nonumber \\
& & \hspace{-.3cm}
 \times g^{\prime \alpha\beta} g^{\prime\prime \gamma\delta}
\Biggl\{ \partial'_{\beta} i\Delta_{\scriptscriptstyle \pm\pm}(x';x'')
 \partial''_{\delta} i\Delta_{\scriptscriptstyle \pm\pm}(x';x'')
 - i\Delta_{\scriptscriptstyle \pm\pm}(x';x'')
 \partial'_{\beta} \partial''_{ \delta} i\Delta_{\scriptscriptstyle \pm\pm}
(x';x'') \Biggr\}
.\qquad
\label{I:nloc}
 \end{eqnarray}
The Lorentz gauge condition~(\ref{Lorentz gauge}) allows us to partially
integrate the $\partial'_{\beta}$
so that the inner loop consists of just a single term,
 \begin{eqnarray}
{\cal I}_{\rm}
&=& \lefteqn{
16 e^2 \sum_{\pm \pm} (\pm) (\pm)
\int d^Dx' \sqrt{-g'} \partial_{[\mu} i\Bigl[\mbox{}_{\nu]}
 \Delta_{\alpha}\Bigr]_{\scriptscriptstyle +\pm}
\!\!\!\!\!(x;x')
\int d^Dx'' \sqrt{-g''}
\partial_{[\rho} i\Bigl[\mbox{}_{\sigma]}
\Delta_{\gamma}\Bigr]_{\scriptscriptstyle +\pm}\!\!\!\!\!(x;x'')}
 \nonumber \\
 & & \hspace{5cm}
 \times g^{\prime \alpha\beta} g^{\prime\prime \gamma\delta}
\partial'_{\beta} i\Delta_{\scriptscriptstyle \pm\pm}(x';x'')
 \partial''_{\delta} i\Delta_{\scriptscriptstyle \pm\pm}(x';x'') 
 + {\cal O}(a^{-1}) \; . \qquad
\label{I:nloc:B}
\end{eqnarray}
The surface terms we have neglected can only involve the initial
time surface in the Schwinger-Keldysh formalism~\cite{Tsamis:1994ca}.
As explained in section III, these surface terms fall off like powers 
of $1/a$ and play no role in the check we are making of the stochastic
formalism. We shall consistently ignore them.

Expression (\ref{I:nloc:B}) is already a significant simplification 
because it means we do not have to worry about either the undifferentiated
logarithm term or the local contribution.
It also simplifies the tensor algebra.
The next step is to act with the derivatives on the two inner loop
propagators and expand,
\begin{eqnarray}
 \partial'_{\beta}
    i\Delta_{\scriptscriptstyle \pm\pm}(x';x'')
 \partial''_{\delta} i\Delta_{\scriptscriptstyle \pm\pm}(x';x'')
&=& A^{\prime 2}(y')
  \frac{\partial y'}{\partial x^{\prime \beta}}
 \frac{\partial y'}{\partial x^{\prime\prime \delta}} + k H a'
\delta^0_{\beta} A^{\prime}(y')
  \frac{\partial y'}{\partial x^{\prime\prime \delta}}
 \nonumber \\
& & \hspace{0cm}
+\; k H a'' \delta^0_{\delta} A^{\prime}(y')
\frac{\partial y'}{\partial x^{\prime \beta}}
+ k^2 H^2 a' a'' \delta^0_{\beta} \delta^0_{\delta}
 \; . \qquad
\label{4terms}
\end{eqnarray}
The term proportional to $k^2$ drops out in the polarity sum,
so we really only have three terms from the inner loop.

Let us define $I[f(y)]$ as the indefinite integral of $f(y)$ with 
respect to $y$. By making use of the identity,
\begin{equation}
  \frac{\partial}{\partial{x'}^\beta}
  \frac{\partial}{\partial{x''}^\delta} I^2[A'(y')^2]
   = \frac{\partial y'}{\partial{x'}^\beta}
  \frac{\partial y'}{\partial{x''}^\delta} A'(y')^2
   + \frac{\partial^2 y'}{\partial{x'}^\beta\partial{x''}^\delta}
            I[ A'(y')^2]
\,,
\label{second deriv}
\end{equation}
we can rewrite Eq.~(\ref{4terms}) as
\begin{eqnarray}
 \partial'_{\beta}
    i\Delta_{\scriptscriptstyle \pm\pm}(x';x'')
 \partial''_{\delta} i\Delta_{\scriptscriptstyle \pm\pm}(x';x'')
 &=& - \frac{\partial^2 y'}
            {\partial x^{\prime \beta}\partial x^{\prime\prime\delta}}
              I[A^{\prime 2}(y')]
  + \frac{\partial }{\partial x^{\prime \beta}}
     \frac{\partial}{\partial x^{\prime\prime \delta}}
        I^2[A^{\prime 2}(y')]
\nonumber\\
  && +\,   \frac{\partial }{\partial x^{\prime\prime \delta}}
     \left[k H a' \delta^0_{\beta} A^{\prime}(y') \right]
   + \frac{\partial }{\partial x^{\prime \beta}}
     \left[k H a'' \delta^0_{\delta} A^{\prime}(y')\right]
\nonumber\\
&&+\, \frac{\partial }{\partial x^{\prime \beta}}
\frac{\partial }{\partial x^{\prime\prime \delta}}
 \left[k^2 \ln(a')\ln(a'') \right]
 \; . \qquad
\label{4terms:B}
\end{eqnarray}
Observe that when Eq.~(\ref{4terms:B}) is inserted into Eq.~(\ref{I:nloc:B}),
all terms but the first vanish upon partial integration as a consequence 
of the Lorentz condition~(\ref{Lorentz gauge}), again up to $1/a$ surface
terms. With this Eq.~(\ref{I:nloc:B}) becomes
 \begin{eqnarray}
{\cal I}_{\rm}
&=&
\frac{- e^2}{(D\!-\!2)^2 H^4} \sum_{\pm \pm} (\pm) (\pm)
\int d^Dx' \sqrt{-g'} \, K'( y_{\scriptscriptstyle +\pm})
\int d^Dx'' \sqrt{-g''} \, K'(y''_{\scriptscriptstyle +\pm})
 \nonumber \\
 & & \hspace{2cm}
 \times
 \frac{\partial y}{\partial x^{[\mu}}
   \frac{\partial^2 y}{ \partial x^{\nu]} \partial x^{\prime \alpha}}
 g^{\prime \alpha\beta}
 \frac{\partial^2 y'}
            {\partial x^{\prime \beta}\partial x^{\prime\prime\delta}}
 g^{\prime\prime \delta\gamma}
 \frac{\partial y''}{\partial x^{[\rho}}
  \frac{\partial^2 y''}{\partial x^{\sigma]}
        \partial x^{\prime\prime\gamma}}
              \times I[A^{\prime 2}(y'_{\scriptscriptstyle \pm\pm})]
 + {\cal O}(a^{-1}) \; . \qquad
\label{I:nloc:C}
\end{eqnarray}
where we used the outer leg identity~(\ref{outer leg identity}) twice.

Note that one can add an integration constant to the first term
in~(\ref{4terms:B}):
$I[A^{\prime 2}(y')]\rightarrow I[A^{\prime 2}(y')] +c_I$,
provided the appropriate change is also made in the second term,
$I^2[A^{\prime 2}(y')]\rightarrow I^2[A^{\prime 2}(y')] +c_Iy'$.
We have just seen that $I^2[A^{\prime 2}(y')]$ in (\ref{4terms:B}) 
can be partially integrated to zero as a consequence of the Lorentz 
condition~(\ref{Lorentz gauge}), up to $1/a$ surface terms. Hence it
must be that any constant term in $I[A^{\prime 2}(y')]$ must fall 
off like powers of $1/a$.

Equation~(\ref{I:nloc:C}) can be further simplified by making use of the
identities~(\ref{identity:2}) and~(\ref{contraction identity}) and
the antisymmetry in $\mu$ and $\nu$ and in $\rho$ and $\sigma$
to obtain,
 \begin{eqnarray}
\lefteqn{{\cal I} =
 \frac{8 e^2 H^2}{(D\!-\!2)^2} \sum_{\pm \pm} (\pm) (\pm) \int d^Dx' \sqrt{-g'}
\int d^Dx'' \sqrt{-g''} } \nonumber \\
& & \hspace{5cm} \times
\frac{\partial}{\partial x^{[\mu}} K(y_{\scriptscriptstyle +\pm})
 g_{\nu][\sigma}
\frac{\partial}{\partial x^{\rho]}}
       K(y''_{\scriptscriptstyle +\pm}) \,
 I[A^{\prime 2}(y'_{\scriptscriptstyle \pm\pm})] + {\cal O}(a^{-1})
 \,.
\label{I:nloc:D}
 \end{eqnarray}
We shall often encounter integrals of this general form,
\begin{equation}
 \sum_{\pm \pm} (\pm) (\pm) \int d^Dx' \sqrt{-g'} \, \frac{\partial}{
   \partial x^{\mu}} K(y) \int d^Dx'' \sqrt{-g''} \, \frac{\partial}{
   \partial x^{\nu}} K(y'') \times F(y') \; . \label{genform}
\end{equation}
Our basic strategy for evaluating them is to extract d'Alembertians for 
the $y'$ dependent terms using the procedure described in Appendix A. In 
addition to d'Alembertians, this procedure will generally also produce a 
delta function and a finite remainder for which $D$ can be set to four,
\begin{equation}
F(y') = \frac{\Box'}{H^2} \Bigl[G(y')\Bigr] + {\rm const.} \times 
\delta^D(x' - x'') + R(y') \; . \label{genred}
\end{equation}
We partially integrate the $\Box'$ to act upon $K(y)$, ignoring the
order $1/a$ surface terms. To see the effect of acting $\Box'$ on $K(y)$,
note first that, up to delta function contributions arising from factors
of $y^{1-D/2}$,
\begin{equation}
 \frac{\Box'}{H^2} f(y) = (4-y)y f''(y) + D(2-y) f'(y)
\,.
\label{DAlembertian:F}
\end{equation}
In view of relation (\ref{eom:K}), and the coefficient of $y^{1-\frac{D}2}$
in the expansion (\ref{4-yyAp+2-yk}) of $K(y)$, we have,
\begin{equation}
\frac{\Box'}{H^2} K(y_{\scriptscriptstyle ++})
         = - \frac{2 (D \!-\! 2)}{a^D H^2} \, i\delta^D(x-x')
                        + (D\!-\!2) K(y) \qquad {\rm and} \qquad
\frac{\Box'}{H^2} K(y_{\scriptscriptstyle +-})
         = (D\!-\!2) K(y) \; .
\label{BoxBox IA}
\end{equation}
Hence the general reduction (\ref{genred}) will produce four sorts of terms:
\begin{enumerate}
\item{A less divergent, 2-vertex integral from the factor of $(D-2) K(y)$
in (\ref{BoxBox IA});}
\item{A potentially divergent, 1-vertex integral from the delta function
in (\ref{BoxBox IA});}
\item{A potentially divergent, 1-vertex integral from the delta function
in (\ref{genred}); and}
\item{A finite, 2-vertex integral from the remainder $R(y')$ in
(\ref{genred}).}
\end{enumerate}
Our strategy is to extract another d'Alembertian from the type-1 terms
and continue in this way until the limit $D=4$ can be taken in all 2-vertex 
integrals. Appendix C explains how to evaluate these. It turns out that the 
1-vertex integrals of type-3 can be usefully combined with the Figure 8 
diagram~(\ref{diagram:figure 8}) to cancel an overlapping divergence.

Note that the integral~(\ref{I:nloc:D}) is symmetric under the exchange 
of the inner legs, $x'\leftrightarrow x''$, under which $y\leftrightarrow y''$ 
and $y'\leftrightarrow y'$. Because the integral is quartically divergent,
it suffices to keep the three most divergent terms in the expansion of 
$I[A^{\prime 2}(y'_{\scriptscriptstyle \pm\pm})]$. From
\begin{eqnarray}
 A'(y) = \frac{H^{D-2}}{(4\pi)^{D/2}}\frac14
  \left\{
     -\sum_{n=-1}^\infty
           \frac{\Gamma\left(n+\frac{D}{2}+1\right)}{\Gamma\left(n+2\right)}
            \left(\frac{y}{4}\right)^{n-\frac{D}{2}+1}
    \!\!\! +\sum_{n=1}^\infty
           \frac{\Gamma\left(n+D-1\right)}{\Gamma\left(n+\frac{D}{2}\right)}
            \left(\frac{y}{4}\right)^{n-1}
  \right\}
\,\qquad
\label{A'}
\end{eqnarray}
we find
\begin{eqnarray}
 I[A'(y')^2] \!\!&=&\!\!
  - \frac{H^{2D-4}}{(4\pi)^D}
    \frac{\Gamma^2\left(\frac{D}{2}\right)}{4}
  \left\{
     \frac{1}{D-1}\left(\frac{y'}{4}\right)^{1-D}
   \!\!\! + \frac{D}{D-2} \left(\frac{y'}{4}\right)^{2-D}
\right.\;
\nonumber\\
 && \hskip 3cm
   +\, \frac{D(D\!+\!1)}{2(D-3)}\left(\frac{y'}{4}\right)^{3-D}
  - \frac{4\Gamma(D)}
          {\Gamma\left(\frac{D}{2}\right)
           \Gamma\left(\frac{D}{2}\!+\!1\right)(D-2)}
                  \left(\frac{y'}{4}\right)^{1-\frac{D}{2}}
\nonumber\\
 && \hskip 3cm
    +\,\frac{D(D\!+\!1)(D\!+\!2)}{6(D\!-\!4)}\left(\frac{y'}{4}\right)^{4-D}
    -\frac{D\!+\!6}{D\!-\!4}
    \frac{\Gamma(D+1)}
          {\Gamma\left(\frac{D}{2}\right)
           \Gamma\left(\frac{D}{2}\!+\!2\right)}
                  \left(\frac{y'}{4}\right)^{2-\frac{D}{2}}
\nonumber\\
 && \hskip 3cm
 +\,{\cal O}\left({y'}^{5-D},{y'}^{3-D/2},{y'}^1\right)
      \Bigg\}
\, . \qquad\;\;
\label{A'2}
\end{eqnarray}
Note that the terms of the order ${y'}^{4-D}$, ${y'}^{2-D/2}$
or higher in Eq.~(\ref{A'2}) are suppressed by $(D-4)^1$.
We are now ready to extract d'Alembertians from $I[A'(y')^2]$ defined 
in Eq.~(\ref{A'2}). Making use of relations in Appendix~A we obtain,
\begin{eqnarray}
 I[A'(y')^2] \!\!&=&\!\!
  - \frac{H^{2D-4}}{(4\pi)^D}
    \frac{\Gamma^2\left(\frac{D}{2}\right)}{4}
  \Bigg\{\frac{\Box'}{H^2}
  \Bigg[
      \frac{2}{(D-1)(D-2)^2}\left(\frac{y'}{4}\right)^{2-D}
\nonumber\\
  &&\hskip 3cm
      +\,\frac{2(D+1)}{(D-1)(D-3)(D-4)}
          \left(
                \left(\frac{y'}{4}\right)^{3-D}
               - \left(\frac{y'}{4}\right)^{1-\frac{D}{2}}
          \right)
\nonumber\\
  && \hskip 2cm
    +\,
             \frac{2(D+1)}{(D\!-\!4)(D\!-\!6)}
      \left(\frac{D}{2(D\!-\!3)}-\frac{4}{(D\!-\!1)(D\!-\!4)}
          \right)\left(\frac{y'}{4}\right)^{4-D}
\label{IA'2:Box1}
\\
  && \hskip 1cm
     -\, \frac{2}{D-4}
     \left(
          \frac{D(D-2)(D+1)}{2(D-1)(D-3)(D-4)}
          -  \frac{4}{D-2}\frac{\Gamma(D)}
                  {\Gamma\left(\frac{D}{2}\right)
                   \Gamma\left(\frac{D}{2}+1\right)}
      \right) \left(\frac{y'}{4}\right)^{2-\frac{D}{2}}
 \Bigg]
\nonumber\\
  && \hskip 0cm
 + \frac{2(D+1)}{(D-1)(D-3)(D-4)}
    \frac{(4\pi)^{D/2}}{\Gamma\left(\frac{D}{2}-1\right)(Ha')^D}
       i\delta^D(x'-x'')
\nonumber\\
 &&\hskip -0cm
 +\,\left[\frac{D(D\!+\!1)(D\!+\!2)}{6(D\!-\!4)}
    - \frac{6(D+1)}{D-6}\left(-\frac{4}{(D\!-\!1)(D\!-\!4)}
                          + \frac{D}{2(D\!-\!3)}
                      \right)
     \right]\left(\frac{y'}{4}\right)^{4-D}
\nonumber\\
 && \hskip -0.cm
  +\,\bigg[-\frac{D\!+\!6}{D\!-\!4}
                    \frac{\Gamma(D\!+\!1)}
                         {\Gamma\left(\frac{D}{2}\right)
                            \Gamma\left(\frac{D}{2}\!+\!2\right)}
\nonumber\\
 && \hskip 1.cm
        +\, \frac{D\!+\!2}{2}\left(\frac{D(D\!+\!1)(D\!-\!2)}
                              {2(D\!-\!1)(D\!-\!3)(D\!-\!4)}
                    - \frac{4}{D\!-\!2}\frac{\Gamma(D)}
                           {\Gamma\left(\frac{D}{2}\right)
                            \Gamma\left(\frac{D}{2}\!+\!1\right)}
                  \right)
      \bigg]\left(\frac{y'}{4}\right)^{\!2-\frac{D}{2}}\!
 \Bigg\}
\,.\qquad
\nonumber
\end{eqnarray}

When inserted into the integral~(\ref{I:nloc:D}) the last three lines 
in Eq.~(\ref{IA'2:Box1}) result in the finite, type-4 integral,
 \begin{eqnarray}
{\cal I}_{\rm fin, 1} &=&
- e^2\frac{ H^{2D-2}}{2(4\pi)^D}\Gamma^2\left(\frac{D}{2}\!-\!1\right)
\sum_{\pm \pm} (\pm)(\pm)\! \int d^Dx' {a'}^D\!
\int d^Dx'' {a''}^D\!
\nonumber\\
&&\times\,
\frac{\partial}{\partial x^{[\mu}}K(y_{\scriptscriptstyle+\pm})
 g_{\nu][\sigma}
\frac{\partial}{\partial x^{\rho]}}K(y''_{\scriptscriptstyle+\pm})
 \left[
10\ln\left(\frac{y'_{\pm\pm}}{4}\right) + {\cal O}\Bigl(a^{-1},D-4\Bigr)
\right]
,
\qquad\;
\label{I:nloc:E}
 \end{eqnarray}
where we dropped the order $D-4$ terms as well as an irrelevant constant.
The delta function in~(\ref{IA'2:Box1}) leads to the type-3 integral,
 \begin{eqnarray}
{\cal I}_{\rm sg, 1} \!=\!
 - ie^2\frac{H^{D-2}}{(4\pi)^{D/2}}
             \frac{\Gamma\left(\frac{D}{2}-1\right)(D+1)}
                  {(D\!-\!1)(D\!-\!3)(D\!-\!4)}
 \sum_{\pm}(\pm) \int d^Dx' {a'}^D
       \frac{\partial}{\partial x^{[\mu}}K(y_{\scriptscriptstyle+\pm})
 g_{\nu][\sigma}
\frac{\partial}{\partial x^{\rho]}}K(y_{\scriptscriptstyle+\pm})
\, .
\qquad\;
\label{I:sg:1}
\end{eqnarray}
The d'Alembertian terms Eq.~(\ref{IA'2:Box1}) produce a type-1
integral,
 \begin{eqnarray}
{\cal I}_1 &=&
  - e^2  \frac{H^{2D-2}}{(4\pi)^D}
    \Gamma\left(\frac{D}{2}\right)\Gamma\left(\frac{D}{2}\!-\!1\right)
   \sum_{\pm \pm} (\pm) (\pm) \int d^Dx' \sqrt{-g'}
\int d^Dx'' \sqrt{-g''}
\nonumber\\
&\times&\!
 \frac{\partial}{\partial x^{[\mu}}K(y_{\scriptscriptstyle +\pm})
 g_{\nu][\sigma}
\frac{\partial}{\partial x^{\rho]}}K(y''_{\scriptscriptstyle +\pm})
  \Bigg[
      \frac{2}{(D-1)(D-2)^2}\left(\frac{y'}{4}\right)^{2-D}
\nonumber\\
  &&\hskip 4.cm
      +\,\frac{2(D+1)}{(D-1)(D-3)(D-4)}
          \left(
                \left(\frac{y'}{4}\right)^{3-D}
               - \left(\frac{y'}{4}\right)^{1-\frac{D}{2}}
          \right)
\nonumber\\
  && \hskip 3.cm
 +\frac{2(D\!+\!1)}{(D\!-\!4)(D\!-\!6)}
    \left[-\frac{4}{(D\!-\!1)(D\!-\!4)} + \frac{D}{2(D\!-\!3)}\right]
           \left(\frac{y'}{4}\right)^{4-D}
\label{I:nloc:G}
\\
  && \hskip 2.cm
 -\,\frac{2}{(D\!-\!4)}\left[\frac{D(D\!-\!2)(D\!+\!1)}
                                {2(D\!-\!1)(D\!-\!3)(D\!-\!4)}
             - \frac{4}{D\!-\!2}\frac{\Gamma(D)}{\Gamma\left(\frac{D}{2}\right)
                                           \Gamma\left(\frac{D}{2}+1\right)}
                     \right] \left(\frac{y'}{4}\right)^{2-\frac{D}{2}}
   \Bigg]_{\pm\pm}
 ,
\nonumber
 \end{eqnarray}
and a type-2 integral,
 \begin{eqnarray}
{\cal I}_{\rm sg, 2} &=&
  ie^2\frac{H^{2D-4}}{(4\pi)^D}
    2\Gamma\left(\frac{D}{2}\right)\Gamma\left(\frac{D}{2}\!-\!1\right)
 \sum_{\pm}  (\pm) \int d^Dx' \sqrt{-g'}
\!\frac{\partial}{\partial x^{[\mu}}K(y_{\scriptscriptstyle +\pm})
   g_{\nu][\sigma}
\nonumber\\
&\times & \hspace{0cm}
\frac{\partial}{\partial x^{\rho]}}
  \Bigg[
      \frac{2}{(D-1)(D-2)^2}\left(\frac{y}{4}\right)^{2-D}
      +\,\frac{2(D+1)}{(D-1)(D-3)(D-4)}
          \left(
                \left(\frac{y}{4}\right)^{3-D}
               - \left(\frac{y}{4}\right)^{1-\frac{D}{2}}
          \right)
\nonumber\\
  && \hskip 0cm
+\frac{2(D\!+\!1)}{(D\!-\!4)(D\!-\!6)}
    \left[-\frac{4}{(D\!-\!1)(D\!-\!4)} + \frac{D}{2(D\!-\!3)}\right]
           \left(\frac{y}{4}\right)^{4-D}
\label{I:sg:2}
\\
  && \hskip -1.cm
 -\,\frac{2}{D\!-\!4}\left[\frac{D(D\!-\!2)(D\!+\!1)}
                                {2(D\!-\!1)(D\!-\!3)(D\!-\!4)}
                       - \frac{4}{D\!-\!2}
                           \frac{\Gamma(D)}{\Gamma\left(\frac{D}{2}\right)
                                             \Gamma\left(\frac{D}{2}+1\right)}
                     \right] \left(\frac{y}{4}\right)^{2-\frac{D}{2}}
 \Bigg]_{+\pm}
 \; . \!\!\!\!
\nonumber
 \end{eqnarray}

We have seen that the total integral~(\ref{I:nloc:D}) breaks up
to the sum,
\begin{equation}
{\cal I} = {\cal I}_{\rm fin,1}+{\cal I}_{\rm sg,1}
           + {\cal I}_{\rm sg,2}+{\cal I}_1 + {\cal O}(a^{-1})
\,.
\label{I broken}
\end{equation}
The next step is to extract another d'Alembertian from the $y'$
dependent terms of Eq.~(\ref{I:nloc:G}). The result is:
 \begin{eqnarray}
{\cal I}_1 &=&
  - e^2  \frac{H^{2D-2}}{(4\pi)^D}
   \Gamma\left(\frac{D}{2}\right)\Gamma\left(\frac{D}{2}\!-\!1\right)
\sum_{\pm \pm} (\pm) (\pm)\! \int\!\! d^Dx' \sqrt{-g'}
\!\int\!\! d^Dx'' \sqrt{-g''}
\nonumber\\
\!&\times&\!\frac{\partial}{\partial x^{[\mu}}K(y_{\scriptscriptstyle +\pm})
 g_{\nu][\sigma}
\frac{\partial}{\partial x^{\rho]}}K(y''_{\scriptscriptstyle+\pm})
 \Bigg\{\frac{\Box'}{H^2}
  \Bigg[
      \frac{4}{(D\!-\!1)(D\!-\!2)^2(D\!-\!3)(D\!-\!4)}
               \left(\left(\frac{y'}{4}\right)^{3-D}
                   \!- \left(\frac{y'}{4}\right)^{1-\frac{D}{2}}
              \right)
\nonumber\\
  &&\hskip 4.5cm
  +\,
  \frac{4}{(D\!-\!1)(D\!-\!4)^2(D\!-\!6)}
            \left(-\frac{4}{(D\!-\!2)^2}
               + \frac{D\!+\!1}{D\!-\!3}
            \right)\left(\frac{y'}{4}\right)^{4-D}
\nonumber\\
  &&\hskip 3.5cm
  -\, \frac{2}{(D\!-\!1)(D\!-\!3)(D\!-\!4)^2}
            \left(\frac{D}{D\!-\!2} - 2(D\!+\!1)
            \right)\left(\frac{y'}{4}\right)^{2-\frac{D}{2}}
 \Bigg]
\nonumber\\
  && \hskip 2.5cm
      + \frac{2}{(D\!-\!1)(D\!-\!2)(D\!-\!3)(D\!-\!4)}
        \frac{(4\pi)^{D/2}}{\Gamma\left(\frac{D}{2}\right)(a'H)^D}
          i\delta^D(x'-x'')
\nonumber\\
  && \hskip 1.5cm
    -\,\frac{20}{3(D\!-\!4)^2}+\frac{86}{9(D\!-\!4)}-\frac{23}{27}
     - 5 \ln\left(\frac{y'}{4}\right)
       + \frac{5}{3}\ln^2\left(\frac{y'}{4}\right) + {\cal O}(D-4)
  \Bigg\}_{\pm\pm}
 \; , \qquad
\label{I:nloc:H}
 \end{eqnarray}
where we expanded the terms in the last line in powers of $D-4$.
Just like ${\cal I}$ in Eq.~(\ref{I broken}), ${\cal I}_1$ can be 
decomposed into four sorts of integrals,
\begin{equation}
{\cal I}_1 = {\cal I}_{\rm fin,2}+{\cal I}_{\rm sg,3}
           + {\cal I}_{\rm sg,4}+{\cal I}_2 + {\cal O}(a^{-1})
\,.
\label{I1 broken}
\end{equation}
The contributions of type-4, type-3 and type-2 are, respectively,
 \begin{eqnarray}
{\cal I}_{\rm fin, 2} &=&
 - e^2 \frac{H^{2D-2}}{2(4\pi)^D}
    \Gamma\left(\frac{D}{2}\right)\Gamma\left(\frac{D}{2}\!-\!1\right)
\sum_{\pm \pm} (\pm) (\pm) \int d^Dx' {a'}^D
\int d^Dx'' {a''}^D
\\
&&\hskip 0.cm\times\,
\frac{\partial}{\partial x^{[\mu}}K(y_{\scriptscriptstyle+\pm})
 g_{\nu][\sigma}
\frac{\partial}{\partial x^{\rho]}}K(y''_{\scriptscriptstyle+\pm})
\label{I:fin:2}
 \left[-10\ln\left(\frac{y'_{\pm\pm}}{4}\right)
      + \frac{10}{3}\ln^2\left(\frac{y'_{\pm\pm}}{4}\right) + 
      {\cal O}\Bigl(a^{-1},D-4\Bigr)\right]
\, ,
\nonumber
\end{eqnarray}
 \begin{eqnarray}
{\cal I}_{\rm sg, 3} &=&
  - i e^2\frac{H^{D-2}}{(4\pi)^\frac{D}{2}}
    \frac{2\Gamma\left(\frac{D}{2}-1\right)}
         {(D\!-\!1)(D\!-\!2)(D\!-\!3)(D\!-\!4)}
 \sum_{\pm}  (\pm) \int d^Dx' \sqrt{-g'}
\nonumber\\
& & \hspace{0cm}
\times\,
  \frac{\partial}{\partial x^{[\mu}}K(y_{+\pm})
   g_{\nu][\sigma}
\frac{\partial}{\partial x^{\rho]}}K(y_{\scriptscriptstyle +\pm})
 \; , \qquad
\label{I:sg:3}
 \end{eqnarray}
and
 \begin{eqnarray}
{\cal I}_{\rm sg,4} &=&
   i e^2
  \frac{H^{2D-4}}{(4\pi)^{D}}
    4\Gamma^2\left(\frac{D}{2}\right)
 \sum_{\pm}  (\pm) \int d^Dx' \sqrt{-g'}
  \frac{\partial}{\partial x^{[\mu}}K(y_{+\pm})
   g_{\nu][\sigma}
\label{I:sg:4}
\\
&&\hskip -1.cm\times
\frac{\partial}{\partial x^{\rho]}}
    \Bigg\{\frac{4}{(D\!-\!1)(D\!-\!2)^2(D\!-\!3)(D\!-\!4)}
          \left(\left(\frac{y}{4}\right)^{3-D}\!\!
              - \left(\frac{y}{4}\right)^{1-\frac{D}{2}}
         \right)
\nonumber\\
&&\hskip 1.cm
 +\, \frac{4}{(D\!-\!1)(D\!-\!4)^2(D\!-\!6)}
            \left(-\frac{4}{(D\!-\!2)^2}
               + \frac{D\!+\!1}{D\!-\!3}
            \right)\left(\frac{y}{4}\right)^{4-D}
\nonumber\\
  &&\hskip 3.5cm
  -\, \frac{2}{(D\!-\!1)(D\!-\!3)(D\!-\!4)^2}
            \left(\frac{D}{D\!-\!2} - 2(D\!+\!1)
            \right)\left(\frac{y}{4}\right)^{2-\frac{D}{2}}
    \Bigg\}_{+\pm}
 \, . \!\!\!
\nonumber
\end{eqnarray}

The type-1 integral is,
 \begin{eqnarray}
{\cal I}_2 &=&
  - e^2  \frac{H^{2D-2}}{(4\pi)^D}
  2\Gamma^2\left(\frac{D}{2}\right)
\sum_{\pm \pm} (\pm) (\pm)\! \int\!\! d^Dx' \sqrt{-g'}
\!\int\!\! d^Dx'' \sqrt{-g''}
\frac{\partial}{\partial x^{[\mu}}K(y_{\scriptscriptstyle +\pm})
 g_{\nu][\sigma}
\frac{\partial}{\partial x^{\rho]}}K(y''_{\scriptscriptstyle+\pm})
\nonumber\\
&&\hskip 0cm
\times\,  \Bigg\{
      \frac{4}{(D\!-\!1)(D\!-\!2)^2(D\!-\!3)(D\!-\!4)}
        \left[
          \frac{2}{(D\!-\!4)(D\!-\!6)}
          \frac{\Box'}{H^2} \left(\frac{y'}{4}\right)^{4-D}
         + \frac{2}{(D\!-\!4)}
          \frac{\Box'}{H^2} \left(\frac{y'}{4}\right)^{2-\frac{D}{2}}
        \right]
\nonumber\\
  &&\hskip 1.5cm
+\frac{8}{3(D\!-\!4)^2}
-\frac{35}{9(D\!-\!4)}
+\frac{89}{27}
      + \frac{2}{3}\ln\left(\frac{y'}{4}\right)
      -\frac{2}{3}\ln^2\left(\frac{y'}{4}\right)
       + {\cal O}(D-4)
  \Bigg\}_{\pm\pm}
 \; . \qquad
\label{I:nloc:J}
 \end{eqnarray}
This time there is no type-3 integral,
\begin{equation}
 {\cal I}_2 = {\cal I}_{\rm sg,5} + {\cal I}_{\rm fin, 3}
            + {\cal I}_{\rm fin, 4} + {\cal O}(a^{-1})
\,.
\label{I2 broken}
\end{equation}
The contributions of type-4, type-1 (with $D=4$ taken) and type-3 are, 
respectively,
 \begin{eqnarray}
{\cal I}_{\rm fin, 3} &=&
 - e^2\frac{H^{2D-2}}{2(4\pi)^D}\Gamma^2\left(\frac{D}{2}\right)
 \sum_{\pm \pm} (\pm) (\pm) \int d^Dx' {a'}^D
\int d^Dx'' {a''}^D
\frac{\partial}{\partial x^{[\mu}}K(y_{\scriptscriptstyle+\pm})
 g_{\nu][\sigma}
\frac{\partial}{\partial x^{\rho]}}K(y''_{\scriptscriptstyle+\pm})
\nonumber
\\
&&\hskip 5cm\times\,
 \left[\frac{8}{3}\ln\left(\frac{y'_{\pm\pm}}{4}\right)
     - \frac{8}{3}\ln^2\left(\frac{y'_{\pm\pm}}{4}\right)
      + {\cal O}\Bigl(a^{-1},D-4\Bigr)
 \right]
\,
\qquad
\label{I:fin:3}
\end{eqnarray}
and
\begin{eqnarray}
{\cal I}_{\rm fin, 4} &=&
 - e^2\frac{H^{2D-2}}{2(4\pi)^D}\frac{D\!-\!2}{2}
                \Gamma^2\left(\frac{D}{2}\right)
\sum_{\pm\pm} (\pm)(\pm)\!\int\!\!d^Dx'{a'}^D \!\int\!\!d^Dx''{a''}^D
\frac{\partial}{\partial x^{[\mu}}K(y_{\scriptscriptstyle +\pm})
 g_{\nu][\sigma}
\frac{\partial}{\partial x^{\rho]}}K(y''_{\scriptscriptstyle +\pm})
\nonumber\\
&& \hskip 4.5cm
\times\,\left[
  \frac{4}{3} \ln\left(\frac{y'_{\pm\pm}}{4}\right)
 - \frac{2}{3} \ln^2\left(\frac{y'_{\pm\pm}}{4}\right)
  + {\cal O}\Bigl(a^{-1},D-4\Bigr)
 \right]
 \;,\qquad
\label{I:fin:4}
 \end{eqnarray}
and
 \begin{eqnarray}
{\cal I}_{\rm sg, 5} &=&
  ie^2 \frac{H^{2D-4}}{(4\pi)^D}
  4(D\!-\!2)\Gamma^2\left(\frac{D}{2}\right)
\sum_{\pm} (\pm)\! \int\!\! d^Dx' \sqrt{-g'}
\frac{\partial}{\partial x^{[\mu}}K(y_{\scriptscriptstyle +\pm})
 g_{\nu][\sigma}
\label{I:sg:5}
\\
&\times&\!\hskip -0cm
\frac{\partial}{\partial x^{\rho]}}
   \left\{ \frac{8}{(D\!-\!1)(D\!-\!2)^2(D\!-\!3)(D\!-\!4)^2}
     \left[\frac{1}{D\!-\!6}\left(\frac{y}{4}\right)^{4-D}
       + \left(\frac{y}{4}\right)^{2-\frac{D}{2}}
    \right]_{+\pm}
    \right\}
 \;.
\nonumber
 \end{eqnarray}

The finite, 2-vertex integrals ${\cal I}_{\rm fin, 1}$, ${\cal I}_{\rm fin, 2}$
${\cal I}_{\rm fin, 3}$ and ${\cal I}_{\rm fin, 4}$
in Eqs.~(\ref{I:nloc:E}), (\ref{I:fin:2}), (\ref{I:fin:3})
and~(\ref{I:fin:4}) can be summed to give
\begin{eqnarray}
{\cal I}_{\rm fin} &=& {\cal I}_{\rm fin, 1} + {\cal I}_{\rm fin, 2}
                    + {\cal I}_{\rm fin, 3} + {\cal I}_{\rm fin, 4}
\label{I:fin:D}
\\
 &=& -\, 2\frac{e^2 H^{2D-2}}{(4\pi)^D}\Gamma^2\left(\frac{D}{2}-1\right)
\sum_{\pm\pm} (\pm)(\pm)\!\int\!\!d^Dx'{a'}^D \!\int\!\!d^Dx''{a''}^D
\frac{\partial}{\partial x^{[\mu}}K(y_{\scriptscriptstyle +\pm})
 g_{\nu][\sigma}
\frac{\partial}{\partial x^{\rho]}}K(y''_{\scriptscriptstyle +\pm})
\nonumber\\
&& \hskip 5.5cm
\times\,\left[
\ln\left(\frac{y'_{\pm\pm}}{4}\right) + {\cal O}\Bigl(a^{-1},D-4\Bigr)
 \right]
\nonumber\\
 &=& -\, e^2\frac{H^{10}}{128\pi^8}
\sum_{\pm\pm} (\pm)(\pm)\!\int\!\!d^4x'{a'}^4 \!\int\!\!d^4x''{a''}^4
 \left(
      \frac{\partial}{\partial x^{[\mu}}\frac{1}{y_{\scriptscriptstyle +\pm}}
 \right)
 g_{\nu][\sigma}
\left(
  \frac{\partial}{\partial x^{\rho]}}\frac{1}{y''_{\scriptscriptstyle +\pm}}
\right)
 \ln\left(\frac{y'_{\pm\pm}}{4}\right)
 \nonumber\\
 &+& {\cal O}\Bigl(a^{-1},D-4\Bigr)
 \;.
\label{I:fin}
 \end{eqnarray}
In the last line we expanded around $D=4$ and made use of 
Eq.~(\ref{4-yyAp+2-yk}),
\begin{equation}
  K(y) \;\;\stackrel{D\rightarrow4}{\longrightarrow}\;\;
      -\frac{H^2}{\pi^2}\frac{1}{y}
\,.
\label{Ky in 4dim}
\end{equation}
In Appendix~C we evaluate the integral~(\ref{I:fin}). The 
result~(\ref{AppC:Imn:fin}), is a de Sitter invariant which does not 
contribute a leading logarithm.
 \begin{equation}
{\cal I}_{\rm fin} = \frac{H^D}{(4 \pi)^{\frac{D}2}} \times
\frac{e^2 H^{D-4}}{(4 \pi)^{\frac{D}2}} \Bigl\{-2 \times
g_{\mu[\rho}g_{\sigma]\nu} + \mathcal{O}\Bigl(a^{-1},D\!-\!4\Bigr)\Bigr\}
 \,.
\label{I:fin:result}
\end{equation}

We are now going to evaluate the one vertex integrals
${\cal I}_{\rm sg, n}$ ($n=1,2,..,5$). In order to do so, we shall
need the basic integrals
\begin{eqnarray}
  Y_n &=& \int d^D x' {a'}^D
         \left[
              \ln^n\left(\frac{y_{++}}{4}\right)
            - \ln^n\left(\frac{y_{+-}}{4}\right)
         \right]
\,,
\qquad  (n=1,2,3)
\label{Yn:def}
\\
  W_n &=& \int d^D x' {a'}^D\ln(a')
         \left[
              \ln^n\left(\frac{y_{++}}{4}\right)
            - \ln^n\left(\frac{y_{+-}}{4}\right)
         \right]
\qquad  (n=1,2)
\label{Wn:def}
\\
\Xi_\alpha &=&  \int d^D x' {a'}^D \left[
                          \left(\frac{y_{++}}{4}\right)^{-\alpha}
                        -  \left(\frac{y_{+-}}{4}\right)^{-\alpha}
                    \right]
\label{Xi:def}
\\
\Lambda_\alpha &=&  \int d^D x' {a'}^D \ln(a')\left[
                          \left(\frac{y_{++}}{4}\right)^{-\alpha}
                        -  \left(\frac{y_{+-}}{4}\right)^{-\alpha}
                    \right]
\label{Lambda:def}
\,.
\end{eqnarray}
The $Y_n$ and $W_n$ integrals are evaluated in Appendix B.
In the asymptotic regime when $a\rightarrow \infty$ these integrals are given by
Eqs.~(\ref{Y1b}), (\ref{Y2:total:finite}), (\ref{Y3b})
and by~(\ref{W1:asym}--\ref{W2:asym}). The $\Xi_\alpha$ integrals can be found in
Appendix B3,
Eqs.~(\ref{xi:3D/2-6}--\ref{xi:D-1}),
while the $\Lambda_\alpha$ integrals ($\alpha = D-4, D-3, D-2, D-1$)
are in Appendix B4,
Eqs.~(\ref{Lambda:D-4}--\ref{Lambda:D-1}).

Recall that the single leg 
integral~(\ref{diagram:figure 8}--\ref{scalar propagator:coinc}) is,
\begin{eqnarray}
{\cal I}_{\infty} &=& {\cal I}_{\infty,\rm sg}
                         + {\cal I}_{\infty,\rm fin}
\nonumber\\
 {\cal I}_{\infty,\rm sg}
  &=& ie^2\frac{H^{D-2}}{(4\pi)^{\frac{D}2}}
        \frac{2\cos\Bigl(\frac{\pi}2 D\Bigr)\Gamma(D \!-\! 2)
            \Gamma\left(1\!-\!\frac{D}{2}\right)}{D\!-\!2}
\sum_{\pm}(\pm)\!\! \int \!\! d^Dx' \sqrt{-g'}
\frac{\partial}{\partial x^{[\mu}} K(y_{\scriptscriptstyle +\pm})
 g_{\nu ] [\sigma}
\frac{\partial}{\partial x^{\rho]}} K(y_{\scriptscriptstyle +\pm})
\nonumber\\
{\cal I}_{\infty,\rm fin}
  &=&-ie^2 \frac{H^{D-2}}{(4\pi)^{\frac{D}2}}
           \frac{4\Gamma(D \!-\! 2)}{(D\!-\!2)\Gamma(\frac{D}2)}
\sum_{\pm}(\pm)\!\! \int \!\! d^Dx' \sqrt{-g'}
          \ln(a')
\frac{\partial}{\partial x^{[\mu}} K(y_{\scriptscriptstyle +\pm})
 g_{\nu ] [\sigma}
\frac{\partial}{\partial x^{\rho]}} K(y_{\scriptscriptstyle +\pm})
\,,\qquad\;
\label{diagram:figure 8:2}
\end{eqnarray}
 The singular part of this integral ${\cal I}_{\infty,\rm sg}$
can be combined with the singular one leg integrals
${\cal I}_{\rm sg, 1}$ and ${\cal I}_{\rm sg, 3}$
in Eqs.~(\ref{I:sg:1}) and~(\ref{I:sg:3}) to yield,
\begin{eqnarray}
 {\cal I}_{\rm sg, A} &=&  {\cal I}_{\infty,\rm sg}
                    + {\cal I}_{\rm sg, 1} +  {\cal I}_{\rm sg, 3}
\nonumber\\
  &=&
 ie^2 \frac{H^{D-2}}{(4\pi)^{\frac{D}{2}}}\frac{1}{D\!-\!2}
\left\{
      2\cos\Bigl(\frac{\pi}2 D\Bigr)\Gamma(D \!-\! 2)
            \Gamma\left(1\!-\!\frac{D}{2}\right)
    - \frac{D\Gamma\left(\frac{D}{2}-1\right)}
                  {(D\!-\!3)(D\!-\!4)}
\right\}
\nonumber\\
&&\times\, \sum_{\pm}(\pm) \int d^Dx' {a'}^D
  \frac{\partial}{\partial x^{[\mu}}
       K\left(y_{\scriptscriptstyle+\pm}\right)
  g_{\nu][\sigma}
\frac{\partial}{\partial x^{\rho]}}
              K\left(y_{\scriptscriptstyle+\pm}\right)
 ,\!\!\!\!\!\!\!\!\!
\label{I:sg:A}
\end{eqnarray}
where $K\left(y_{\scriptscriptstyle+\pm}\right)$ is given by
Eq.~(\ref{4-yyAp+2-yk}). The terms in curly parentheses can be expanded in
 powers of $D-4$. The result is,
\begin{equation}
 5
 + (D\!-\!4)\left(-4-\frac{\pi^2}{6}-\frac52\gamma_E\right)
 +{\cal O}\left((D\!-\!4)^2\right)
\,.
\label{expansion:ovelap}
\end{equation}
This means that the most divergent contributions from the two left diagrams
in Fig. 1 cancel. This is an example of a general phenomenon
of cancellation of overlapping divergences and simplifies our calculation
considerably. 

A further simplification is facilitated by the identity,
\begin{equation}
\frac{\partial y}{\partial x^{\mu}} \frac{\partial y}{\partial x^{\rho}} f(y)
 = - H^2 g_{\mu\rho} (2 \!-\! y) I[f](y) + \nabla_{\mu} \nabla_{\rho}
I^2[f](y) \; ,
\label{identity:double-der}
\end{equation}
where $I[f](y)$ stands for the indefinite integral with respect to $y$.
By making use of this identity and Eq.~(\ref{expansion:ovelap})
we can express the integral in~(\ref{I:sg:A})
in terms of two relatively straightforward integrals as follows,
\begin{eqnarray}
 {\cal I}_{\rm sg,A}
 &=& -ie^2\frac{H^{D}}{(4\pi)^{\frac{D}2}}
   \frac{1}{D\!-\!2}
\left\{
      2\cos\Bigl(\frac{\pi}2 D\Bigr)\Gamma(D \!-\! 2)
            \Gamma\left(1\!-\!\frac{D}{2}\right)
    - \frac{D\Gamma\left(\frac{D}{2}-1\right)}
                  {(D\!-\!3)(D\!-\!4)}
\right\}
\nonumber\\
&&\times\,
\bigg[
 g_{\mu [\rho} g_{\sigma] \nu}
  \sum_\pm(\pm)
        \int d^Dx' \sqrt{-g'} \, (2 \!-\! y_{\scriptscriptstyle +\pm})
  I\Bigl[ \big(K^{\prime}\big)^2\Bigr](y_{\scriptscriptstyle +\pm})
\nonumber \\
&& \hskip 0.5cm
  -\, \frac{\nabla_{[\mu } g_{\nu] [\sigma} \nabla_{\rho]}}{H^2}
\sum_{\pm}(\pm)\int d^Dx' \sqrt{-g'} \,
 I^2\Bigl[\big(K^{\prime}\big)^2\Bigr](y_{\scriptscriptstyle +\pm})
\bigg]
. \qquad
 \label{twoints}
\end{eqnarray}

 In order to accurately extract both the divergent
and finite contributions from this integral, we shall keep
the D dimensional form of the terms in $K$ up to $y^{4-D/2}$ and $y^2$;
in higher order terms $y^n$ $(n\geq 3)$ we set $D=4$, since these
yield contributions that are suppressed linearly in $D-4$.
Making use of Eq.~(\ref{4-yyAp+2-yk}) we find,
\begin{eqnarray}
 K'(y)^2 &=& \frac{H^{2D-4}}{(4\pi)^D}\frac{(D-2)^2}{4}
                \Gamma^2\left(\frac{D}{2}\right)
  \Bigg\{
         \left(\frac{y}{4}\right)^{-D}
       + (D\!-\!4) \left(\frac{y}{4}\right)^{1-D}
       + \frac{D^2\!-\!7D\!+\!8}{2}\left(\frac{y}{4}\right)^{2-D}
\nonumber\\
 &&    +\, \frac{4\Gamma(D-2)}{\Gamma\left(\frac{D}{2}\!-\!1\right)
                              \Gamma\left(\frac{D}{2}\!+\!1\right)}
                                \left(\frac{y}{4}\right)^{-\frac{D}{2}}
       + \frac{D(D\!-\!2)(D\!-\!7)}{6}\left(\frac{y}{4}\right)^{3-D}
\nonumber\\
 &&    +\, (D\!+\!8)\frac{\Gamma(D-1)}{\Gamma\left(\frac{D}{2}\!-\!1\right)
                                      \Gamma\left(\frac{D}{2}\!+\!2\right)}
                                \left(\frac{y}{4}\right)^{1-\frac{D}{2}}
       +{\cal O}\left(y^{4-D},y^{2-D/2},y^{0},\right)
  \Bigg\}
\,,
\label{K'2}
\end{eqnarray}
from which it follows,
\begin{eqnarray}
 (2\!-\!y)I\Big[{K'}^2\Big](y)
     \!&=&\! -\frac{H^{2D-4}}{(4\pi)^D}4(D\!-\!2)^2
                \Gamma^2\!\left(\frac{D}{2}\right) \!\!
  \Bigg\{
         \frac{1}{2(D\!-\!1)}\left(\frac{y}{4}\right)^{1-D}
       \!+\! \left[\frac{D\!-\!4}{2(D\!-\!2)}-\frac{1}{D\!-\!1}\right] \!
                              \left(\frac{y}{4}\right)^{2-D}
\nonumber\\
 && \hskip 2.3cm
    +\, \left[\frac{D^2\!-\!7D\!+\!8}{4(D\!-\!3)}-\frac{D\!-\!4}{D\!-\!2}
         \right]\left(\frac{y}{4}\right)^{3-D}
     + \frac{2\Gamma(D-2)}{\Gamma\left(\frac{D}{2}\right)
                              \Gamma\left(\frac{D}{2}\!+\!1\right)}
                                \left(\frac{y}{4}\right)^{1-\frac{D}{2}}
\nonumber\\
 && \hskip 0.25cm
    +\, \left[\frac{D(D\!-\!2)(D\!-\!7)}{12(D\!-\!4)}
            -\frac{D^2\!-\!7D\!+\!8}{2(D\!-\!3)}
        \right]\left(\frac{y}{4}\right)^{4-D}
\label{2-y:I:K'2}
\\
 && \hskip -2cm
   +\, \left[
            \frac{(D\!-\!2)^2(D\!+\!8)}{2(D\!-\!4)}-2(D\!+\!2)
        \right]
                \frac{\Gamma(D-2)}{\Gamma\left(\frac{D}{2}\right)
                                      \Gamma\left(\frac{D}{2}\!+\!2\right)}
                                \left(\frac{y}{4}\right)^{2-\frac{D}{2}}
       +{\cal O}\left(y^{5-D},y^{3-D/2},y^{1},\right)
  \Bigg\}
\,,
\nonumber
\end{eqnarray}
and
\begin{eqnarray}
 I^2\Big[{K'}^2\Big](y) \!&=&\! \frac{H^{2D-4}}{(4\pi)^D}4(D\!-\!2)^2
                \Gamma^2\!\left(\frac{D}{2}\right)
  \Bigg\{
         \frac{1}{(D\!-\!1)(D\!-\!2)}\left(\frac{y}{4}\right)^{2-D}
       \!+ \frac{(D\!-\!4)}{(D\!-\!2)(D\!-\!3)}
                              \left(\frac{y}{4}\right)^{3-D}
\nonumber\\
 && \hskip 3.5cm
    +\, \frac{D^2\!-\!7D\!+\!8}{2(D\!-\!3)(D\!-\!4)}
                           \left(\frac{y}{4}\right)^{4-D}
     + \frac{8\Gamma(D-2)}{(D\!-\!4)\Gamma\left(\frac{D}{2}\right)
                              \Gamma\left(\frac{D}{2}\!+\!1\right)}
                                \left(\frac{y}{4}\right)^{2-\frac{D}{2}}
\nonumber
\\
 && \hskip 4.5cm
       +\,{\cal O}\left(y^{5-D},y^{3-D/2},y^{1},\right)
  \Bigg\}
\,,
\label{I2:K'2}
\end{eqnarray}
When these expressions~(\ref{2-y:I:K'2}--\ref{I2:K'2}) are inserted into
Eq.~(\ref{twoints}) and one makes use of
the integrals~(\ref{xi:3D/2-6}--\ref{xi:D-1}) evaluated in
Appendix~B3, one arrives at the following expression
for ${\cal I}_{\rm sg,A}$~(\ref{twoints}),
\begin{eqnarray}
 {\cal I}_{\rm sg,A}
 &=& ie^2\frac{H^{3D-4}}{(4\pi)^\frac{3D}{2}}
              8\Gamma^2\left(\frac{D}{2}\right)
\left\{
      \cos\Bigl(\frac{\pi}2 D\Bigr)\Gamma(D \!-\! 1)
            \Gamma\left(1\!-\!\frac{D}{2}\right)
    - \frac{2\Gamma\left(\frac{D}{2}+1\right)}
                  {(D\!-\!3)(D\!-\!4)}
\right\}
\nonumber\\
&\times&
\Bigg\{
 g_{\mu [\rho} g_{\sigma] \nu}
  \Bigg[
         \frac{1}{2(D\!-\!1)}\Xi_{D-1}
       \!+ \left(\frac{D\!-\!4}{2(D\!-\!2)}-\frac{1}{D\!-\!1}\right)
                              \Xi_{D-2}
    + \left(\frac{D^2\!-\!7D\!+\!8}{4(D\!-\!3)}-\frac{D\!-\!4}{D\!-\!2}
         \right)\Xi_{D-3}
\nonumber\\
 &&\hskip 2cm
    +\, \frac{2\Gamma(D-2)}{\Gamma\left(\frac{D}{2}\right)
                              \Gamma\left(\frac{D}{2}\!+\!1\right)}
                               \Xi_{\frac{D}{2}-1}
    + \left(\frac{D(D\!-\!2)(D\!-\!7)}{12(D\!-\!4)}
          - \frac{D^2\!-\!7D\!+\!8}{2(D\!-\!3)}
      \right)\Xi_{D-4}
\nonumber\\
 && \hskip 2cm
   +\, \left(
            \frac{(D\!-\!2)^2(D\!+\!8)}{2(D\!-\!4)}-2(D\!+\!2)
        \right)
                \frac{\Gamma(D-2)}{\Gamma\left(\frac{D}{2}\right)
                                      \Gamma\left(\frac{D}{2}\!+\!2\right)}
                               \Xi_{\frac{D}{2}-2}
  \Bigg]
\nonumber \\
&&+\, \frac{\nabla_{[\mu } g_{\nu] [\sigma} \nabla_{\rho]}}{H^2}
  \Bigg[
         \frac{1}{(D\!-\!1)(D\!-\!2)}\Xi_{D-2}
       \!+ \frac{D\!-\!4}{(D\!-\!2)(D\!-\!3)}
                              \Xi_{D-3}
\nonumber\\
 && \hskip 2.5cm
    +\, \frac{D^2\!-\!7D\!+\!8}{2(D\!-\!3)(D\!-\!4)}
                              \Xi_{D-4}
     + \frac{8\Gamma(D-2)}{(D\!-\!4)\Gamma\left(\frac{D}{2}\right)
                              \Gamma\left(\frac{D}{2}\!+\!1\right)}
                              \Xi_{\frac{D}{2}-2}
  \Bigg]
\Bigg\}
. \qquad
 \label{IsgA:4}
\end{eqnarray}
Recalling that the operator,
\begin{equation}
\frac{\nabla_{[\mu } g_{\nu] [\sigma} \nabla _{\rho]}}{H^2}
  = a^2\delta_{[\mu}^0g_{\nu][\sigma}\delta_{\rho]}^0
             a^2\frac{d^2}{da^2}
  -  g_{\mu[\rho}g_{\sigma]\nu}a\frac{d}{da}
\,,
\label{DmDr:operator}
\end{equation}
Eq.~(\ref{IsgA:4}) can be evaluated and expanded in powers of $D-4$.
Here we provide an intermediate result, which also shows
the terms linear in $(D-4)$,
\begin{eqnarray}
 {\cal I}_{\rm sg,A}
 &=& - e^2\frac{H^{2D-4}}{(4\pi)^D}
     8 \frac{\Gamma^3\left(\frac{D}{2}\right)}{\Gamma(D)}
       \left[\cos\left(\frac{\pi D}{2}\right)\Gamma(D\!-\!1)\Gamma\left(1\!-\!\frac{D}{2}\right)
           -\frac{2\Gamma\left(\frac{D}{2}\!+\!1\right)}{(D\!-\!3)(D\!-\!4)}
       \right]
\nonumber\\
&\times&
  \Bigg\{
  g_{\mu[\rho} g_{\sigma]\nu}
  \Bigg[
      \frac{\Gamma(D)}{\Gamma^2\left(\frac{D}{2}\right)}
       \frac{D\!-\!6}{2(D\!-\!3)(D\!-\!4)}
\nonumber\\
&& \hskip 2cm  -\, 6
     + (D\!-\!4)\left(-\frac{1}{2}\ln^2(a)+\frac{13}{6}\ln(a)
                    + \frac{19}{12}+\frac{\pi^2}{12}
                \right)
   \Bigg]
\nonumber\\
&& +\, a^2\delta_{[\mu}^0g_{\nu][\sigma}\delta_{\rho]}^0
         (D\!-\!4)\left(\ln(a)-\frac{5}{3}\right)
   \Bigg\}
    +{\cal O}\Bigl(a^{-1},(D\!-\!4)^2\Bigr)
\,.
 \label{IsgA:5}
\end{eqnarray}
The final result is,
\begin{equation}
 {\cal I}_{\rm sg,A}
 = \frac{H^D}{(4\pi)^{\frac{D}2}} \times \frac{e^2 H^{D-4}}{(4\pi)^\frac{D}2}
     \Biggl[ \frac{20 \Gamma(D \!-\!1)}{D \!-\! 4} - 72 - \frac43 \pi^2\Biggr]
     g_{\mu[\rho} g_{\sigma]\nu}
    +{\cal O}\Bigl(a^{-1},D\!-\!4\Bigr)
\,.
 \label{IsgA:6}
\end{equation}

The second part of the singular contribution to the one leg integral
is the sum of Eqs.~(\ref{I:sg:2}), (\ref{I:sg:4}) and~(\ref{I:sg:5}),
\begin{eqnarray}
 {\cal I}_{\rm sg, B} &=& {\cal I}_{\rm sg, 2} + {\cal I}_{\rm sg, 4}
                  + {\cal I}_{\rm sg, 5}
 \nonumber\\
  &=& ie^2
  \frac{H^{2D-4}}{(4\pi)^D}
    2\Gamma\left(\frac{D}{2}\right)\Gamma\left(\frac{D}{2}\!-\!1\right)
 \sum_{\pm}  (\pm) \int d^Dx' \sqrt{-g'}
\!\frac{\partial}{\partial x^{[\mu}}K(y_{\scriptscriptstyle +\pm})
   g_{\nu][\sigma}
\label{I:sg:B}
\\
&\times & \hspace{0cm}
\frac{\partial}{\partial x^{\rho]}}
  \Bigg\{
      \frac{2}{(D\!-\!1)(D\!-\!2)^2}\left(\frac{y}{4}\right)^{2-D}
      +\,\frac{2D}{(D\!-\!2)(D\!-\!3)(D\!-\!4)}
          \left(
                \left(\frac{y}{4}\right)^{3-D}
               - \left(\frac{y}{4}\right)^{1-\frac{D}{2}}
          \right)
\nonumber\\
  &&  \hskip 1.0cm +\,\frac{D(D\!+\!2)}{(D\!-\!2)(D\!-\!4)(D\!-\!6)}
                         \left(\frac{y}{4}\right)^{4-D}
\nonumber\\
  && \hskip 1.cm
 +\,
   \left[
       - \frac{D}{(D\!-\!3)(D\!-\!4)}
       + \frac{8}{(D\!-\!2)(D\!-\!4)}
        \frac{\Gamma(D)}{\Gamma\left(\frac{D}{2}\right)
                                           \Gamma\left(\frac{D}{2}+1\right)}
                     \right] \left(\frac{y}{4}\right)^{2-\frac{D}{2}}
 \Bigg\}_{+\pm}
 \; , \!\!\!\!
\nonumber
\,.
\end{eqnarray}
This can be expressed in terms of the $\Omega_{\alpha,\beta}$ integrals
evaluated in Appendix~B5 (as usually,  the expression below is
anti-symmetrized in both pairs of indices
$[\mu ,\nu]$ and $[\rho, \sigma]$),
\begin{eqnarray}
 {\cal I}_{\rm sg, B} &=& -ie^2
  \frac{H^{3D-6}}{(4\pi)^{\frac{3D}{2}}}\,
    8\Gamma^2\left(\frac{D}{2}\right)\Gamma\left(\frac{D}{2}\!-\!1\right)
   g_{\nu[\sigma}
\label{I:sg:B:2}
\\
&\times & \hspace{0cm}
  \Bigg\{
      \frac{2}{(D\!-\!1)(D\!-\!2)^2}
         \Bigg[\Omega_{D\!-\!2,\frac{D}{2}\!-\!1}
            + \frac{D\!-\!2}{2}\Omega_{D\!-\!2,\frac{D}{2}\!-\!2}
            + \frac{D(D\!-\!2)}{8}\Omega_{D\!-\!2,\frac{D}{2}\!-\!3}
\nonumber\\
&-& \frac{\Gamma(D\!-\!1)}{\Gamma\left(\frac{D}{2}\!-\!1\right)
                                     \Gamma\left(\frac{D}{2}\!+\!1\right)}
                  \Omega_{D\!-\!2,-\!1}
            + \frac{D(D\!-\!2)(D\!+\!2)}{48}\Omega_{D\!-\!2,\frac{D}{2}\!-\!4}
            - \frac{\Gamma(D)}{\Gamma\left(\frac{D}{2}\!-\!1\right)
                                         \Gamma\left(\frac{D}{2}\!+\!2\right)}
                  \Omega_{D\!-\!2,-\!2}
         \Bigg]
\nonumber\\
&&      +\,\frac{2D}{(D\!-\!2)(D\!-\!3)(D\!-\!4)}
          \Bigg[
               \left(\Omega_{D\!-\!3,\frac{D}{2}-\!1}
              - \Omega_{\frac{D}{2}-\!1,\frac{D}{2}-\!1}\right)
            + \frac{D\!-\!2}{2}
                  \left(\Omega_{D\!-\!3,\frac{D}{2}\!-\!2}
                       - \Omega_{\frac{D}{2}\!-\!1,\frac{D}{2}\!-\!2}
                    \right)
\nonumber\\
&& \hskip 1cm
        +\, \frac{D(D\!-\!2)}{8}
                    \left(\Omega_{D\!-\!3,\frac{D}{2}\!-\!3}
                       -  \Omega_{\frac{D}{2}\!-\!3,\frac{D}{2}\!-\!1}
                     \right)
        - \frac{\Gamma(D\!-\!1)}{\Gamma\left(\frac{D}{2}\!-\!1\right)
                                     \Gamma\left(\frac{D}{2}\!+\!1\right)}
                    \left(
                          \Omega_{D\!-\!3,-\!1}
                        - \Omega_{\frac{D}{2}\!-\!1,-\!1}
                     \right)
          \Bigg]
\nonumber\\
  &&  \hskip 0.cm +\,\frac{D(D\!+\!2)}{(D\!-\!2)(D\!-\!4)(D\!-\!6)}
   \left[
         \Omega_{D\!-\!4,\frac{D}{2}\!-\!1}
       + \frac{(D\!-\!2)}{2}\Omega_{D\!-\!4,\frac{D}{2}-\!2}
   \right]
\nonumber\\
  &+&\! \hskip 0.cm
   \left[
     -\frac{D}{(D\!-\!3)(D\!-\!4)}
       + \frac{8}{(D\!-\!2)(D\!-\!4)}
        \frac{\Gamma(D)}{\Gamma\left(\frac{D}{2}\right)
                                           \Gamma\left(\frac{D}{2}+1\right)}
   \right]
   \left[
         \Omega_{\frac{D}{2}\!-\!1,\frac{D}{2}\!-\!2}
       + \frac{D\!-\!2}{2}\Omega_{\frac{D}{2}\!-\!2,\frac{D}{2}-\!2}
   \right]
 \Bigg\}_{\rho]\mu}
\nonumber
\,.
\end{eqnarray}
The result is,
\begin{eqnarray}
 {\cal I}_{\rm sg, B} &=& e^2\frac{H^{2D-4}}{(4\pi)^D}\,
   \frac{8\Gamma^3\left(\frac{D}{2}\right)\Gamma\left(\frac{D}{2}\!-\!1\right)}
        {\Gamma(D)}
\nonumber\\
&\times&\!\!\bigg\{
    g_{\nu[\sigma}g_{\rho]\mu}
        \bigg[
        \frac{\Gamma(D)}{\Gamma\left(\frac{D}{2}\right)\Gamma\left(\frac{D}{2}\!-\!1\right)}
        \frac{1}{(3D\!-\!10)(D\!-\!3)(D\!-\!4)}
        \bigg(\!\!-\!\frac{D^2\!-\!5D\!+\!2}{2(D\!-\!1)(D\!-\!2)}
               \!+\! \frac{D(2D\!-\!7)}{(D\!-\!3)(D\!-\!4)}\bigg)
\nonumber\\
&&   +\,\frac{1}{2(D\!-\!4)} + 3
         +(D\!-\!4)\left(-3\ln^3(a) + \Big(\!-\frac{67}{6}+\frac{9\pi^2}{2}\Big)\ln(a)-\frac{191}{9}+\frac{5\pi^2}{48}-18\zeta(3)\right)
       \bigg]
\nonumber\\
&&  +\,    a^2\delta_{[\rho}^0 g_{\sigma][\nu}\delta_{\mu]}^0
       (D\!-\!4)\left(
                \!-4\ln^2(a)+\frac{50}{3}\ln(a)-\frac{145}{6}+\frac{13\pi^2}{6}
                \right)
\bigg\} + {\cal O}\Bigl(a^{-1},(D\!-\!4)^2\Bigr)
.\quad
\label{I:sg:B:3}
\end{eqnarray}
Upon expanding this in powers of $D-4$ one obtains,
\begin{equation}
 {\cal I}_{\rm sg, B} = \frac{H^D}{(4\pi)^{\frac{D}2}} \times \frac{e^2
H^{D-4}}{(4\pi)^\frac{D}2} \Biggl[ \frac{8 \Gamma(D\!-\!1)}{(D\!-\!4)^2}
   - \frac{40 \Gamma(D\!-\!1)}{(D\!-\!1) (D\!-\!4)} + \frac{67}3 - \frac23
   \pi^2\Biggr] g_{\nu[\sigma}g_{\rho]\mu}
   + {\cal O}\Bigl(a^{-1},D\!-\!4\Bigr)
\,.
\label{I:sg:B:4}
\end{equation}

\bigskip

 What remains to calculate is the integral~(\ref{diagram:figure 8:2}).
This integral can be expressed in terms of the
$\Lambda_\alpha$ integrals~(\ref{Lambda:alpha}).
Firstly, we make use of Eqs.~(\ref{identity:double-der})
to write~(\ref{diagram:figure 8:2}) as
\begin{eqnarray}
{\cal I}_{\infty,\rm fin}
  &=&ie^2 \frac{H^D}{(4\pi)^{\frac{D}2}}
           \frac{4\Gamma(D \!-\! 2)}{(D\!-\!2)\Gamma(\frac{D}2)}
\sum_{\pm}(\pm)\int d^Dx' \sqrt{-g'}
          \ln(a')
\nonumber\\
&&\times\,
\bigg[
 g_{\mu [\rho} g_{\sigma] \nu}
  \sum_\pm(\pm)
        \int d^Dx' \sqrt{-g'}(2 \!-\! y_{\scriptscriptstyle +\pm})
  I\Bigl[ \big(K^{\prime}\big)^2\Bigr](y_{\scriptscriptstyle +\pm})
\nonumber \\
&& \hskip 0.5cm
  -\, \frac{\nabla_{[\mu } g_{\nu] [\sigma} \nabla_{\rho]}}{H^2}
\sum_{\pm}(\pm)\int d^Dx' \sqrt{-g'}
 I^2\Bigl[\big(K^{\prime}\big)^2\Bigr](y_{\scriptscriptstyle +\pm})
\bigg]
. \qquad
\label{diagram:figure 8:3}
\end{eqnarray}
Secondly, we make use of Eqs.~(\ref{2-y:I:K'2}--\ref{I2:K'2})
and the definition of $\Lambda_\alpha$ integrals~(\ref{Lambda:alpha})
to obtain ({\it cf.} Eqs.~(\ref{IsgA:4})),
\begin{eqnarray}
{\cal I}_{\infty,\rm fin}
  &=& - ie^2 \frac{H^{3D-4}}{(4\pi)^{\frac{3D}2}}
             16\Gamma(D \!-\!1)\Gamma(\frac{D}2)
\nonumber\\
&\times&
\Bigg\{
 g_{\mu [\rho} g_{\sigma] \nu}
  \Bigg[
         \frac{1}{2(D\!-\!1)}\Lambda_{D-1}
       \!+ \left(\frac{D\!-\!4}{2(D\!-\!2)}-\frac{1}{D\!-\!1}\right)
                              \Lambda_{D-2}
    + \left(\frac{D^2\!-\!7D\!+\!8}{4(D\!-\!3)}-\frac{D\!-\!4}{D\!-\!2}
         \right)\Lambda_{D-3}
\nonumber\\
 &&\hskip 2cm
    +\, \frac{2\Gamma(D-2)}{\Gamma\left(\frac{D}{2}\right)
                              \Gamma\left(\frac{D}{2}\!+\!1\right)}
                               \Lambda_{\frac{D}{2}-1}
    + \left(-\frac{D^2\!-\!7D\!+\!8}{2(D\!-\!3)}
             + \frac{D(D\!-\!2)(D\!-\!7)}{12(D\!-\!4)}\right)
                               \Lambda_{D-4}
\nonumber\\
 && \hskip 2cm
   +\, \left(
            \frac{(D\!-\!2)^2(D\!+\!8)}{2(D\!-\!4)}-2(D\!+\!2)
        \right)
                \frac{\Gamma(D-2)}{\Gamma\left(\frac{D}{2}\right)
                                      \Gamma\left(\frac{D}{2}\!+\!2\right)}
                               \Lambda_{\frac{D}{2}-2}
  \Bigg]
\nonumber \\
&+& \frac{\nabla_{[\mu } g_{\nu] [\sigma} \nabla_{\rho]}}{H^2}
  \Bigg[
         \frac{1}{(D\!-\!1)(D\!-\!2)}\Lambda_{D-2}
       \!+ \frac{D\!-\!4}{(D\!-\!2)(D\!-\!3)}
                              \Lambda_{D-3}
\nonumber\\
 && \hskip 2.cm
    +\, \frac{D^2\!-\!7D\!+\!8}{2(D\!-\!3)(D\!-\!4)}
                              \Lambda_{D-4}
     + \frac{8\Gamma(D-2)}{(D\!-\!4)\Gamma\left(\frac{D}{2}\right)
                              \Gamma\left(\frac{D}{2}\!+\!1\right)}
                              \Lambda_{\frac{D}{2}-2}
  \Bigg]
\Bigg\}
. \qquad
\label{diagram:figure 8:4}
\end{eqnarray}
 This can be evaluated by making use of Eq.~(\ref{DmDr:operator})
and Eqs.~(\ref{Lambda:D-4}--\ref{Lambda:D-1}). Keeping the terms up
to order $(D-4)$ we have,
\begin{eqnarray}
{\cal I}_{\infty,\rm fin}
  &=& e^2 \frac{H^{2D-4}}{(4\pi)^{D}}
                   \frac{16\Gamma^2(\frac{D}2)}{D \!-\!1}
\nonumber\\
\!&\times& \!\!
\Bigg\{
 g_{\mu [\rho} g_{\sigma] \nu} \!
  \Bigg[\frac{\Gamma(D)}{\Gamma^2\left(\frac{D}2\right)}\frac{1}{(D\!-\!3)(D\!-\!4)}
  \left(\frac{D\!-\!6}{2}\ln(a) - \frac{2D\!-\!3}{(D\!-\!1)(D\!-\!2)}\right)
               + \left(-6\ln(a) +1 - \pi^2 \right)
\nonumber\\
&& + (D\!-\!4)\left( \frac{7}{12}\ln^3(a)
                  + \frac{13}{12}\ln^2(a)
                  + \left(\frac{25}{24}-\frac{2\pi^2}{3}\right)\ln(a)
                  + \frac{1309}{432}-\frac{7\pi^2}{9} + 4\zeta(3)
              \right)
  \Bigg]
\nonumber\\
\!&+&\! a^2\delta_{[\mu}^0  g_{\nu] [\sigma}\delta_{\rho]}^0
  \Bigg[- \frac{\Gamma(D)}{\Gamma^2\left(\frac{D}2\right)}
           \frac{1}{(D\!-\!1)(D\!-\!3)(D\!-\!4)} - 2
\nonumber\\
&&\hskip 2cm
         + (D\!-\!4)\left(
                   \frac{5}{4}\ln^2(a)
                  - \frac{53}{12}\ln(a)
                  + \frac{119}{18} - \frac{2\pi^2}{3}
              \right) 
  \Bigg]
\Bigg\} + {\cal O}\Bigl(a^{-1},D\!-\!4\Bigr)
.
\label{diagram:figure 8:5}
\end{eqnarray}
Upon rearranging the divergent contributions and dropping the
${\cal O}(D\!-\!4)$ terms one obtains,
\begin{eqnarray}
{\cal I}_{\infty,\rm fin}
 \! &=&\!  \frac{H^D}{(4 \pi)^{\frac{D}2}} \times \frac{e^2 H^{D-4}}{(4\pi)^{
\frac{D}2}} \Biggl\{ \Biggl[ \Bigl(-\frac{16 \Gamma(D\!-\!1)}{D \!-\!4} + 16
\Bigr) \ln(a) -\frac{40 \Gamma(D\!-\!1)}{(D \!-\!1) (D\!-\!4)} + \frac{104}3 
- \frac{16}3 \pi^2 \Biggr] g_{\mu [\rho} g_{\sigma] \nu} \nonumber \\
& & \hspace{5.5cm} 
- \frac{16 \Gamma(D\!-\!1)}{(D\!-\!1) (D\!-\!4)} \times a^2\delta_{[\mu}^0  
g_{\nu] [\sigma}\delta_{\rho]}^0 \Bigg\} + {\cal O}\Bigl(a^{-1},D\!-\!4\Bigr)
. \qquad \label{diagram:figure 8:6}
\end{eqnarray}

Summing the integrals ~(\ref{Ict:res}), (\ref{I:fin:result}), (\ref{IsgA:6}),
(\ref{I:sg:B:4}) and (\ref{diagram:figure 8:6}) we finally get,
\begin{eqnarray}
\lefteqn{\left\langle\Omega| F_{\mu\nu}(x)
                   F_{\rho\sigma}(x)|\Omega\right\rangle_{\rm two\; loop}
= {\cal I}_{\rm c.t.} + {\cal I}_{\rm fin} + {\cal I}_{\rm sg, A} + 
{\cal I}_{\rm sg, B} +  {\cal I}_{\infty,\rm fin} + {\cal O}(a^{-1})} 
\nonumber \\
& & = -\delta Z_3 \times \left\langle\Omega| F_{\mu\nu}(x) F_{\rho\sigma}(x)|
\Omega\right\rangle_{\rm one\; loop} + \frac{H^D}{(4\pi)^{\frac{D}2}} \times 
\frac{e^2 H^{D-4}}{(4 \pi)^{\frac{D}2}} \nonumber \\
& & \times \Biggl\{ \Biggl[ \Bigl[- \frac{16 \Gamma(D\!-\!1)}{D \!-\! 4} + 
16 \Bigr] \ln(a) + \frac{8 \Gamma(D\!-\!1)}{(D \!-\!4)^2} -\frac{20
\Gamma(D\!-\!1)}{(D\!-\!1) (D\!-\!4)} - \frac{11}3 - \frac{22}3 \pi^2\Biggr] 
g_{\mu [\rho} g_{\sigma] \nu} \nonumber \\
& & \hspace{6cm} - \frac{16 \Gamma(D \!-\!1)}{(D\!-\!1) (D\!-\!4)} \times
a^2 \delta_{[\mu}^0  g_{\nu] [\sigma}\delta_{\rho]}^0 \Biggr\}
+ {\cal O}\Bigl(a^{-1},D\!-\!4\Bigr)\,. \qquad \label{final log contribution}
\end{eqnarray}
This corresponds to the following values for the $D$-dependent quantities
$E_1$, $E_2$ and $F_1$ defined in expressions (\ref{EF}) and 
(\ref{Eexp}-\ref{Fexp}),
\begin{eqnarray}
E_1 & = & -\frac{16 \Gamma(D\!-\!1)}{D\!-\!4} + 16 + {\cal O}(D\!-\!4) \; , 
\label{E1actual} \\
E_2 & = & -\delta Z_3 \times \frac{(4\pi)^{\frac{D}2}}{e^2 H^{D-4}} \times
\frac{2 \Gamma(D\!-\!1)}{\Gamma(\frac{D}2 \!+\! 1)} \nonumber \\
& & \hspace{3cm} + \frac{8 \Gamma(D\!-\!1)}{(D\!-\!4)^2} -\frac{20
\Gamma(D\!-\!1)}{ (D\!-\!1) (D\!-\!4)} - \frac{11}3 - \frac{22}3 \pi^2
+ {\cal O}(D\!-\!4) \; , \qquad \label{E2actual} \\
F_1 & = & -\frac{16 \Gamma(D\!-\!1)}{(D\!-\!1) (D\!-\!4)} + {\cal O}(D\!-\!4) 
\; . \label{F1actual}
\end{eqnarray}
Our result for $E_1$ agrees with the stochastic prediction 
(\ref{E1pre})~\cite{Prokopec:2007ak}, and our result for $F_1$ agrees with 
the prediction (\ref{F1pre}) based upon conservation.

\section{Discussion}
\label{Discussion}

By substituting (\ref{E1actual}-\ref{F1actual}) into expressions 
(\ref{pmax}-\ref{rhomax}) we obtain two loop results for the field
strength contributions to the pressure $p$ and to $(\rho + p)$,
\begin{eqnarray}
\Bigl( p \Bigr)_{\rm Maxwell} \!\!\!\! & = & \frac{H^D}{(4\pi)^{\frac{D}2}} 
\Biggl\{-\frac{(D \!-\! 4) \Gamma(D)}{4 \Gamma(\frac{D}2 \!+\! 1)} + 
\frac{e^2 H^{D-4}}{(4\pi)^{\frac{D}2}} \Biggl[12 \ln(a) \nonumber \\
& & \hspace{4cm} - \frac{11 \Gamma(D\!-\!1)}{(D\!-\!1) (D \!-\! 4)} - \frac13
+ {\cal O}\Bigl(a^{-1},D\!-\!4\Bigr) \Biggr] + \mathcal{O}(e^4) \Biggr\} , 
\label{pmaxfin} \\
\Bigl( \rho + p \Bigr)_{\rm Maxwell} \!\!\!\! & = & \frac{H^D}{(4\pi)^{
\frac{D}2}} \Biggl\{0 + \frac{e^2 H^{D-4}}{(4\pi)^{\frac{D}2}} \Biggl[
-\frac{8 \Gamma(D\!-\!1)}{(D\!-\!1) (D\!-\!4)} - \frac83 
+ \mathcal{O}\Bigl(a^{-1},D\!-\!4\Bigr) \Biggr] + \mathcal{O}(e^4) \Biggr\} .
\qquad \label{rhomaxfin}
\end{eqnarray}
The $\ln(a)$ term in (\ref{pmaxfin}) agrees exactly with the stochastic
prediction~\cite{Prokopec:2007ak}. Together with the two loop results for 
the expectation values of $(D_{\mu} \varphi)^* D_{\nu} \varphi$ and 
$\varphi^* \varphi$~\cite{Prokopec:2006ue}, this constitutes an impressive
verification of the stochastic formalism.

Note that we have not bothered to work out the terms which fall off like
powers of $1/a$. Just as for the stress-energy tensor of $\lambda \varphi^4$
 ---
expressions (\ref{phi^4}-\ref{2ndphi^4}) --- these terms are separately
conserved and play no role at late times. It has been conjectured that
the $1/a$ contributions can be subsumed into a perturbative correction of 
the initial state~\cite{Onemli:2004mb}. Such a correction is in any case
inevitable because free vacuum cannot be a very realistic
initial state of an interacting theory.

We have also not bothered to renormalize the composite operators
$F_{\mu\nu} F_{\rho\sigma}$, $(D_{\mu} \varphi)^* D_{\nu} \varphi$ and
$\varphi^* \varphi$. Recall that conventional renormalization only 
removes divergences from noncoincident 1PI functions. One would need
to additionally subtract a series of counter-operators from $F_{\mu\nu}
F_{\rho\sigma}$ and the others to remove their divergences, and there 
would be the usual ambiguities about the finite part~\cite{Weinberg,Itzykson}.
This is neither necessary nor desirable. The stochastic formalism which it 
was our purpose to check makes unique predictions for the leading logarithm
contributions to the dimensionally regulated expectation values of 
the three composite operators~\cite{Prokopec:2007ak}, and these predictions
agree in each case.

Accuracy is always an issue in such an intricate computation. Our result
essentially consists of three numbers: $E_1$, $E_2$ and $F_1$, which were
defined in expressions (\ref{EF}) and (\ref{Eexp}-\ref{Fexp}) and reported
in (\ref{E1actual}-\ref{F1actual}). An obvious check on $E_1$ is that it
agrees with the stochastic formalism for which a compelling and independent 
theoretical justification exists~\cite{Prokopec:2007ak}. A powerful check
on $F_1$ is provided by partial conservation within the electromagnetic
sector (\ref{partial}-\ref{explicit}). No direct check exists for $E_2$
but it was of course computed using the same reduction strategy and many 
of the same integrals that produced correct results for $E_1$ and $F_1$.

Combining the field strength contributions (\ref{pmaxfin}-\ref{rhomaxfin})
with the corresponding scalar results (\ref{pscal}-\ref{rhoscal}) gives 
the total for the stress-energy tensor of SQED~(\ref{Tmunu:form}),
\begin{eqnarray}
\Bigl( p \Bigr)_{\rm SQED} \!\!\!\! & = & \frac{H^D}{(4\pi)^{\frac{D}2}} 
\Biggl\{\frac{\Gamma(D)}{2 \Gamma(\frac{D}2)} + \frac{e^2 H^{D-4}}{
(4\pi)^{\frac{D}2}} \Biggl[12 \ln(a) \nonumber \\
& & \hspace{3.2cm} - \frac{3 \Gamma(D\!-\!1)}{(D \!-\! 4)} 
+ 33 - 4 \pi^2 + {\cal O}\Bigl(a^{-1},D\!-\!4\Bigr) \Biggr] + 
\mathcal{O}(e^4) \Biggr\} , \qquad \label{ptotal} \\
\Bigl( \rho + p \Bigr)_{\rm SQED} \!\!\!\! & = & \frac{H^D}{(4\pi)^{
\frac{D}2}} \Biggl\{0 + \frac{e^2 H^{D-4}}{(4\pi)^{\frac{D}2}} \Biggl[ 4
+ \mathcal{O}\Bigl(a^{-1},D\!-\!4\Bigr) \Biggr] + \mathcal{O}(e^4) \Biggr\} .
\qquad \label{rhototal}
\end{eqnarray}
Note that the divergent field strength and scalar contributions to
$(\rho + p)$ have canceled. This is required by the fact that the
expectation value of $T_{\mu\nu}$ gives the matter contributions to
 the 1-graviton 1PI function in gravity + SQED. Hence the expectation
value of $T_{\mu\nu}$ can be renormalized by purely gravitational
counterterms, and all of these degenerate to constants times $g_{\mu\nu}$
on de Sitter background. Because the expectation value of $T_{\mu\nu}$
takes the form $p g_{\mu\nu} + (\rho + p) a^2 \delta^0_{\mu} \delta^0_{\nu}$,
we see that expression (\ref{ptotal}) for $p$ can contain divergent
constants, but the $\ln(a)$ terms in $p$ and all of $(\rho + p)$ must be
finite. Because the constant contribution to the pressure can be absorbed
into a renormalization of the cosmological constant, this term has no
physical significance --- which is reassuring because it is the least well
checked.

Although this exercise was undertaken to check the stochastic formalism,
our results (\ref{ptotal}-\ref{rhototal}) have considerable physical
interest in their own right. After many e-foldings the $\ln(a)$ 
contribution to the pressure (\ref{ptotal}) must dominate and we see that 
SQED induces a growing, {\it negative} vacuum energy. The physical 
interpretation is that the vacuum becomes polarized by the 
inflationary production of charged scalars. Just as a dielectric slab will 
be drawn into the region between the plates of a charged capacitor, so the 
production of additional scalars seems to be favored by the electric fields 
of those that came before.

It is natural to wonder how far this progresses, both at higher orders in 
the loop expansion and in time. Recall that the two loop $\ln(a)$ derived
entirely from the field strength contributions. That is not true at higher
orders, however, only the field strength contributions act to decrease the 
vacuum energy~\cite{Prokopec:2007ak}. At $\ell$ loop order one gets $\ell-1$
factors of $e^2/(4\pi)^2$, and there can be up to $\ell-1$ factors of 
$\ln(a)$~\cite{Prokopec:2007ak}. The leading logarithms at all loops 
become order one after $\ln(a) \sim (4\pi)^2/e^2$ e-foldings. At this point
perturbation theory breaks down and one must employ a nonperturbative
resummation scheme to evolve further. It was to solve this sort of problem 
that the stochastic formulation of SQED was developed, and the answer is 
known~\cite{Prokopec:2007ak},
\begin{equation}
\lim_{t \rightarrow \infty} p(t) \approx 0.6551 \times \frac{3 H^4}{8 \pi^2}
\approx \frac{\Lambda}{8\pi G} \times 0.2085 \times G H^2 \; . \label{pshift}
\end{equation}
Here $\Lambda = 3 H^2$ is the bare cosmological constant and $G$ is Newton's
constant. 

Although (\ref{pshift}) is a nonperturbatively large decrease in the 
vacuum energy, it is suppressed by $G H^2$ relative to the vacuum energy 
of the bare cosmological constant. The largest value of $G H^2$ consistent 
with the current upper limit on the tensor-to-scalar ratio for anisotropies 
in the cosmic microwave background is about $10^{-12}$~\cite{WMAP}, so 
(\ref{pshift}) does not represent a significant decrease of the vacuum 
energy. On the other hand, it is an enormous amount of vacuum energy by 
current scales. Further, the shift is dynamic; it was caused by inflationary 
particle production and it would presumably dissipate, on some time scale, 
after the end of inflation. This may well have important consequences for 
cosmology~\cite{Prokopec:2003bx}.

\section*{Acknowledgements}

T.P. acknowledges financial support by FOM grant FOM-07.0583
and by Utrecht University.  N.T acknowledges financial support by
the European Social Fund and National Resources
$\Upsilon\Pi{\rm E}\Pi\Theta$-Pythagoras II-2103,
by European Union grants MRTN-CT-2004-512194 and FP-6-12679.
R.P.W. was partially supported by NSF grants PHY-0244714 and PHY-0653085,
and by the Institute for Fundamental Theory at the University of Florida.

\section*{\large \bf Appendix A: Extracting d'Alembertians}
\label{Appendix A}

 Here we list the results of extracting the d'Alembertians from powers
of $y/4$ which reduce the degree of divergence by 2.

 When a d'Alembertian acts on a nonsingular function $F=F(y)$
 one obtains~(\ref{DAlembertian:F}),
 \begin{equation}
  \frac{\Box}{H^2}F(y) = (4-y)yF''(y) + D(2-y)F'(y)
  \,.
\label{A:DAlembertian:F}
 \end{equation}
(A nonsingular function $F$ of $y=y(x;x')$ is a function which,
when expanded in powers of $y$, does not contain the power $y^{1-D/2}$.)
 Equation~(\ref{A:DAlembertian:F}) is helpful for establishing the relations
\begin{eqnarray}
 \left(\frac{y}{4}\right)^{-\alpha}
  &=& -\frac{1}{(\alpha-1)\left(\frac{D}{2}-\alpha\right)}
      \frac{\Box}{H^2}\left(\frac{y}{4}\right)^{1-\alpha}
   + \frac{D-\alpha}{\frac{D}{2}-\alpha}
      \left(\frac{y}{4}\right)^{1-\alpha}
\quad (\alpha \neq D/2)\qquad
\label{A1:DAlembertian:alpha}
\\
   \frac{\Box}{H^2} \left(\frac{y}{4}\right)^{1-\frac{D}{2}}
  &=& \frac{(4\pi)^{D/2}}{\Gamma\left(\frac{D}{2}-1\right)(Ha)^D}
       i\delta^D(x-x')
   + \frac{D(D-2)}{4}\left(\frac{y}{4}\right)^{1-\frac{D}{2}}
\label{A1:DAlembertian:D2-1}
 \end{eqnarray}
 For example for $\alpha = 3D/2-3,3D/2-4,3D/2-5,
D, D-1, D-2, D-3, D-4, (D/2) - 1, (D/2)-2$
these give,
\begin{eqnarray}
 \left(\frac{y}{4}\right)^{3-\frac{3D}{2}}
 &=& \left[\frac{2}{(3D-8)(D-3)}\frac{\Box}{H^2} + \frac{D-6}{2(D-3)}\right]
                           \left(\frac{y}{4}\right)^{4-\frac{3D}{2}}
\label{A1:DAlembertian:3D2m3}
\\
 \left(\frac{y}{4}\right)^{4-\frac{3D}{2}}
  &=& \left[\frac{2}{(3D-10)(D-4)}\frac{\Box}{H^2} + \frac{D-8}{2(D-4)}\right]
                           \left(\frac{y}{4}\right)^{5-\frac{3D}{2}}
\label{A1:DAlembertian:3D2m4}
\\
  & & \hspace{-1.5cm} + \frac{2}{(3D\!-\!10)(D\!-\!4)}
               \left[-\frac{\Box}{H^2} + \frac{D(D\!-\!2)}{4}\right] 
                           \left(\frac{y}{4}\right)^{1-\frac{D}{2}}
    + \frac{2}{(3D\!-\!10)(D\!-\!4)}
       \frac{(4\pi)^\frac{D}{2}}{\Gamma\left(\frac{D}{2}\!-\!1\right)}
            \frac{i\delta^D(x\!-\!x')}{(aH)^D}
\nonumber\\
 \left(\frac{y}{4}\right)^{5-\frac{3D}{2}}
  &=& \left[\frac{2}{3(D-4)(D-5)}\frac{\Box}{H^2} + \frac{D-10}{2(D-5)}\right]
                           \left(\frac{y}{4}\right)^{6-\frac{3D}{2}}
\label{A1:DAlembertian:3D2m5}
\\
 \left(\frac{y}{4}\right)^{-D}
  &=& \frac{2}{D(D-1)}\frac{\Box}{H^2}\left(\frac{y}{4}\right)^{1-D}
\label{A1:DAlembertian:D}
\end{eqnarray}
\begin{eqnarray}
 \left(\frac{y}{4}\right)^{1-D}
  &=& \frac{2}{(D-2)^2}\frac{\Box}{H^2}\left(\frac{y}{4}\right)^{2-D}
     - \frac{2}{D-2}\left(\frac{y}{4}\right)^{2-D}
\label{A1:DAlembertian:Dm1}
\\
 \left(\frac{y}{4}\right)^{2-D}
  &=& \left[\frac{2}{(D-3)(D-4)}\frac{\Box}{H^2} - \frac{4}{D-4}\right]
                   \left(\frac{y}{4}\right)^{3-D}
\label{A1:DAlembertian:Dm2}
\\
  &+& \frac{2}{(D\!-\!3)(D\!-\!4)}
               \left[-\frac{\Box}{H^2} + \frac{D(D\!-\!2)}{4}\right]
                           \left(\frac{y}{4}\right)^{1-\frac{D}{2}}
    + \frac{2}{(D\!-\!3)(D\!-\!4)}
       \frac{(4\pi)^\frac{D}{2}}{\Gamma\left(\frac{D}{2}\!-\!1\right)}
            \frac{i\delta^D(x-x')}{(aH)^D}
\nonumber
\\
 \left(\frac{y}{4}\right)^{3-D}
  &=& \frac{2}{(D-4)(D-6)}\frac{\Box}{H^2}\left(\frac{y}{4}\right)^{4-D}
     - \frac{6}{D-6}\left(\frac{y}{4}\right)^{4-D}
\label{A1:DAlembertian:Dm3}
\\
 \left(\frac{y}{4}\right)^{4-D}
  &=& \frac{2}{(D-5)(D-8)}\frac{\Box}{H^2}\left(\frac{y}{4}\right)^{5-D}
     - \frac{8}{D-8}\left(\frac{y}{4}\right)^{5-D}
\label{A1:DAlembertian:Dm4}
\end{eqnarray}
\begin{eqnarray}
 \left(\frac{y}{4}\right)^{1-\frac{D}{2}}
  &=& -\frac{2}{D-4}\frac{\Box}{H^2}\left(\frac{y}{4}\right)^{2-\frac{D}{2}}
     + \frac{D+2}{2}\left(\frac{y}{4}\right)^{2-\frac{D}{2}}
\label{A1:DAlembertian:D2m1}
\\
 \left(\frac{y}{4}\right)^{2-\frac{D}{2}}
  &=& -\frac{1}{D-6}\frac{\Box}{H^2}\left(\frac{y}{4}\right)^{3-\frac{D}{2}}
     + \frac{D+4}{4}\left(\frac{y}{4}\right)^{3-\frac{D}{2}}
\,.
 \end{eqnarray}

 When the d'Alembertian acts on a power of $y$ one gets
\begin{eqnarray}
 \frac{\Box}{H^2}\left(\frac{y}{4}\right)^{-\beta}
 &=& -\beta\left(\frac{D}{2}-\beta-1\right)\left(\frac{y}{4}\right)^{-\beta-1}
   + \beta(D-\beta-1) \left(\frac{y}{4}\right)^{-\beta}
\quad
\label{A1:DAlembertian:1}
\end{eqnarray}
which when applied to
 $\beta = (D/2)-1, (D/2)-2, (D/2)-3, (D/2)-4,-2,-1$ yields
\begin{eqnarray}
 0 &=&
 - \frac{\Box}{H^2} \left(\frac{y}{4}\right)^{1-\frac{D}{2}}
+ \frac{(4\pi)^{D/2}}{\Gamma\left(\frac{D}{2}-1\right)(Ha)^D}
       i\delta^D(x-x')
   + \frac{D(D-2)}{4}\left(\frac{y}{4}\right)^{1-\frac{D}{2}}
\qquad\;
\label{A1:DAlembertian:D2m1:b}
\\
   \frac{\Box}{H^2} \left(\frac{y}{4}\right)^{2-\frac{D}{2}}
  &=&  - \frac{D-4}{2}\left(\frac{y}{4}\right)^{1-\frac{D}{2}}
      + \frac{(D-4)(D+2)}{4}\left(\frac{y}{4}\right)^{2-\frac{D}{2}}
\label{A1:DAlembertian:D2m2:b}
\\
   \frac{\Box}{H^2} \left(\frac{y}{4}\right)^{3-\frac{D}{2}}
  &=&  - (D-6)\left(\frac{y}{4}\right)^{2-\frac{D}{2}}
      + \frac{(D-6)(D+4)}{4}\left(\frac{y}{4}\right)^{3-\frac{D}{2}}
\label{A1:DAlembertian:D2m3:b}
\\
   \frac{\Box}{H^2} \left(\frac{y}{4}\right)^{4-\frac{D}{2}}
  &=&  - \frac{3(D-8)}{2}\left(\frac{y}{4}\right)^{3-\frac{D}{2}}
      + \frac{(D-8)(D+6)}{4}\left(\frac{y}{4}\right)^{4-\frac{D}{2}}
\\
   \frac{\Box}{H^2} \left(\frac{y}{4}\right)^{2}
  &=&  - 2(D+1)\left(\frac{y}{4}\right)^{2}
      + (D+2)\left(\frac{y}{4}\right)
\label{A1:DAlembertian:2:b}
\\
   \frac{\Box}{H^2} \left(\frac{y}{4}\right)
  &=&  - D\left(\frac{y}{4}\right)
      + \frac{D}{2}
\label{A1:DAlembertian:1:b}
\,.
\end{eqnarray}

 Next we make use of the identity~(\ref{identity:7}),
\begin{equation}
 \nabla_\mu \nabla_\rho y = (2-y) H^2 g_{\mu\rho}
\label{DDy}
\end{equation}
to obtain ($\alpha+\beta+1 \neq D/2$),
\begin{eqnarray}
 \partial_\mu\left(\frac{y}{4}\right)^{-\alpha}
 \partial_\rho\left(\frac{y}{4}\right)^{-\beta}
    &=& \frac{\alpha\beta}{(\alpha+\beta)(\alpha+\beta+1)}
        \nabla_\rho \nabla_{\mu}
             \left[\left(\frac{y}{4}\right)^{-(\alpha+\beta)}\right]
\label{DyalphaDybeta}
\\
 &+& \frac{\alpha\beta H^2g_{\mu\rho}}{D-2(\alpha+\beta+1)}
       \left[
           - \frac{1}{(\alpha+\beta)(\alpha+\beta+1)}
             \frac{\Box}{H^2}\left(\frac{y}{4}\right)^{-(\alpha+\beta)}
           + \left(\frac{y}{4}\right)^{-(\alpha+\beta)}
      \right]
\,.
\nonumber
\end{eqnarray}
The relevant cases are
$(\alpha,\beta) = (D/2-1,D-2), (D/2-1,D-3), (D/2-1,D-4), (D/2-1,D-5)
      (D/2-1,D/2-1), (D/2-1,D/2-2), (D/2-1,D/2-3)$ and $(D/2-1,-1)$.
 From Eq.~(\ref{DyalphaDybeta}) we have,
\begin{eqnarray}
 \partial_\mu\left(\frac{y}{4}\right)^{1-\frac{D}{2}}
 \partial_\rho\left(\frac{y}{4}\right)^{2-D}
    &=& \frac{2(D-2)}{3(3D-4)}
       \nabla_\rho \nabla_{\mu}
             \left[\left(\frac{y}{4}\right)^{3-\frac32D}\right]
\label{Dy1-D2Dy2-D}
\\
 &+& H^2g_{\mu\rho}
       \left[
            \frac{1}{3(3D-4)}
             \frac{\Box}{H^2}\left(\frac{y}{4}\right)^{3-\frac32D}
           - \frac{D-2}{4} \left(\frac{y}{4}\right)^{3-\frac32D}
      \right]
\nonumber
\\
 \partial_\mu\left(\frac{y}{4}\right)^{1-\frac{D}{2}}
 \partial_\rho\left(\frac{y}{4}\right)^{3-D}
    &=& \frac{2(D-3)}{3(3D-8)}
        \nabla_\rho \nabla_{\mu}
             \left[\left(\frac{y}{4}\right)^{4-\frac32D}\right]
\label{Dy1-D2Dy3-D}
\\
 &+& H^2g_{\mu\rho}
       \left[
           \frac{1}{3(3D-8)}
             \frac{\Box}{H^2}\left(\frac{y}{4}\right)^{4-\frac32D}
           - \frac{D-2}{4} \left(\frac{y}{4}\right)^{4-\frac32D}
      \right]
\nonumber
\\
 \partial_\mu\left(\frac{y}{4}\right)^{1-\frac{D}{2}}
 \partial_\rho\left(\frac{y}{4}\right)^{4-D}
    &=& \frac{2(D-2)(D-4)}{(3D-10)(3D-8)}
        \nabla_\rho \nabla_{\mu}
             \left[\left(\frac{y}{4}\right)^{5-\frac32D}\right]
\label{Dy1-D2Dy4-D}
\\
 &+& (D\!-\!2) H^2g_{\mu\rho}
       \left[
            \frac{1}{(3D\!-\!10)(3D\!-\!8)}
             \frac{\Box}{H^2}\left(\frac{y}{4}\right)^{5-\frac32D}
           - \frac14\left(\frac{y}{4}\right)^{5-\frac32D}
      \right]
\nonumber
\\
 \partial_\mu\left(\frac{y}{4}\right)^{1-\frac{D}{2}}
 \partial_\rho\left(\frac{y}{4}\right)^{5-D}
    &=& \frac{2(D-2)(D-5)}{3(3D-10)(D-4)}
        \nabla_\rho \nabla_{\mu}
             \left[\left(\frac{y}{4}\right)^{6-\frac32D}\right]
\label{Dy1-D2Dy5-D}
\\
 &+& (D\!-\!2) H^2g_{\mu\rho}
       \left[
            \frac{1}{3(3D\!-\!10)(D\!-\!4)}
             \frac{\Box}{H^2}\left(\frac{y}{4}\right)^{6-\frac32D}
           - \frac14\left(\frac{y}{4}\right)^{6-\frac32D}
      \right]
\nonumber
\end{eqnarray}
\begin{eqnarray}
 \partial_\mu\left(\frac{y}{4}\right)^{1-\frac{D}{2}}
 \partial_\rho\left(\frac{y}{4}\right)^{1-\frac{D}{2}}
    &=& \frac{D-2}{4(D-1)}
        \nabla_\rho \nabla_{\mu}
             \left[\left(\frac{y}{4}\right)^{2-D}\right]
\label{Dy1-D2Dy1-D2}
\\
 &+& H^2g_{\mu\rho}
       \left[
            \frac{1}{4(D-1)}
             \frac{\Box}{H^2}\left(\frac{y}{4}\right)^{2-D}
           - \frac{D-2}{4}\left(\frac{y}{4}\right)^{2-D}
      \right]
\nonumber
\\
 \partial_\mu\left(\frac{y}{4}\right)^{1-\frac{D}{2}}
 \partial_\rho\left(\frac{y}{4}\right)^{2-\frac{D}{2}}
    &=& \frac{D-4}{4(D-3)}
        \nabla_\rho \nabla_{\mu}
             \left[\left(\frac{y}{4}\right)^{3-D}\right]
\label{Dy1-D2Dy2-D2}
\\
 &+& H^2g_{\mu\rho}
       \left[
            \frac{1}{4(D-3)}
             \frac{\Box}{H^2}\left(\frac{y}{4}\right)^{3-D}
           - \frac{D-2}{4}\left(\frac{y}{4}\right)^{3-D}
      \right]
\nonumber
\\
 \partial_\mu\left(\frac{y}{4}\right)^{1-\frac{D}{2}}
 \partial_\rho\left(\frac{y}{4}\right)^{3-\frac{D}{2}}
    &=& \frac{(D-2)(D-6)}{4(D-3)(D-4)}
        \nabla_\rho \nabla_{\mu}
             \left[\left(\frac{y}{4}\right)^{4-D}\right]
\label{Dy1-D2Dy3-D2}
\\
 &+& \frac{(D-2)H^2g_{\mu\rho}}{4}
       \left[
            \frac{1}{(D-3)(D-4)}
             \frac{\Box}{H^2}\left(\frac{y}{4}\right)^{4-D}
           - \left(\frac{y}{4}\right)^{4-D}
      \right]
\nonumber
\end{eqnarray}
\begin{eqnarray}
 \partial_\mu\left(\frac{y}{4}\right)^{1-\frac{D}{2}}
 \partial_\rho\left(\frac{y}{4}\right)
    &=& -\frac{2}{D-4}
        \nabla_\rho \nabla_{\mu}
             \left[\left(\frac{y}{4}\right)^{2-\frac{D}{2}}\right]
\label{Dy1-D2Dy-1}
\\
 &+& H^2g_{\mu\rho}
       \left[
            \frac{1}{D-4}
             \frac{\Box}{H^2}\left(\frac{y}{4}\right)^{2-\frac{D}{2}}
           - \frac{D-2}{4}\left(\frac{y}{4}\right)^{2-\frac{D}{2}}
      \right]
\nonumber
\,.
\end{eqnarray}

\section*{\large \bf
           Appendix B: Some useful integrals}
\label{Appendix Integrals}

\subsection*{B1. The $Y$ integrals}
\label{The $Y$ integrals}

 Here we evaluate the basic integrals of the form,
\begin{equation}
  Y_n = \int d^D x' {a'}^D
         \left[
              \ln^n\left(\frac{y_{++}}{4}\right)
            - \ln^n\left(\frac{y_{+-}}{4}\right)
         \right]
\,,
\qquad  (n=0,1,2,3)
\label{Yn}
\,.
\end{equation}
 In order to perform the integrations, the following relations are
needed,
\begin{eqnarray}
\ln(\mu^2\Delta x^2_{++}) - \ln(\mu^2\Delta x^2_{+-})
              &=& 2\pi i \theta(\Delta\eta)\theta(\Delta\eta^2- r^2)
\label{ln:++ +-}
\\
\ln^2(\mu^2\Delta x^2_{++}) - \ln^2(\mu^2\Delta x^2_{+-})
              &=& 4\pi i \theta(\Delta\eta)\theta(\Delta\eta^2- r^2)
                   \ln\left|\mu^2(\Delta\eta^2 - r^2)\right|
\label{ln2:++ +-}
\\
\ln^3(\mu^2\Delta x^2_{++}) - \ln^3(\mu^2\Delta x^2_{+-})
              &=& 6\pi i \theta(\Delta\eta)\theta(\Delta\eta^2- r^2)
                   \Big[
                        \ln^2\left|\mu^2(\Delta\eta^2 - r^2)\right|
                     -  \frac{\pi^2}{3}
                   \Big]
\,,\qquad
\label{ln3:++ +-}
\end{eqnarray}
where $r = \|\vec x - \vec x^{\,\prime}\|$,
$\mu^2 = aa'H^2/4$
and we made use of
\begin{eqnarray}
 \Delta x^2_{++} &=& - (|\eta-\eta^\prime|-i\epsilon)^2
                       + \|\vec x - \vec x^{\,\prime}\|^2
\label{Delta x2:++}
\\
 \Delta x^2_{+-} &=& - (\eta-\eta^\prime+i\epsilon)^2
                       + \|\vec x - \vec x^{\,\prime}\|^2
\,.
\label{Delta x2:+-}
\end{eqnarray}
We shall also need the polylogarithm functions, defined as,
\begin{equation}
  {\rm Li}_n(z) \equiv {\rm PolyLog}[n,z] = \sum_{i=1}^\infty \frac{z^i}{i^n}
\qquad (n=2,3)
\,.
\label{polylog}
\end{equation}
Note that the natural logarithm is
a special case of the Li function, $\ln(1-z) = - {\rm Li}_1(z)$.
Moreover, we shall make use of the properties of polylogarithms,
\begin{eqnarray}
  {\rm Li}_2(z) &=&  - {\rm Li}_2(1-z)
                 +  \ln\left(\frac{1}{z}\right)\ln(1-z)
                 + \frac{\pi^2}{6}
\label{polylog:prop:1}
\\
  {\rm Li}_2(z) + \ln(z)\ln(1-z)
          &=&  {\rm Li}_2\left(1-\frac{1}{z}\right)
                 +  \frac12\ln^2(z)
                 + \frac{\pi^2}{6}
\quad (z\notin [-\infty,0])
\label{polylog:prop:2}
\\
  {\rm Li}_3\left(1-\frac1z\right)
  &=&  \sum_{n=0}^\infty \frac{\left(1-\frac1z\right)^n}{n^3}
\label{polylog:prop:3}\\
                  &=& \zeta(3) - \frac{\pi^2}{6}\frac1z
                   + \left(\frac34-\frac{\pi^2}{12}-\frac12\ln\left(\frac{1}{z}\right)\right)\frac{1}{z^2}
                   +{\cal O}\left(\frac{\ln(z)}{z^3}\right)
                     \quad (|z|\gg 1)
\,.
\nonumber
\end{eqnarray}

 Of course,
\begin{equation}
  Y_0 = 0
\,.
\end{equation}
 The simplest nontrivial integral ($n=1$) in Eq.~(\ref{Yn}) is,
\begin{eqnarray}
  Y_1 &=& \int_{\eta_0}^\eta d\eta' {a'}^D  \int_0^{\Delta\eta}d^{D-1}x'
         \left[
               2\pi i
         \right]
\nonumber\\
      &=& i\frac{(4\pi)^\frac{D}{2}}{H^D}
           \frac{\Gamma\left(\frac{D}{2}\right)}{\Gamma\left(D\right)}
            \int_{1}^a \frac{da'}{a'}
            \left(1-\frac{a'}{a}\right)^{D-1}
\,,
\label{Y1}
\end{eqnarray}
where $\eta_0$ denotes an initial conformal time for which we choose
$a(\eta_0)=1$ and we made use of the surface area
of the $D-2$ dimensional sphere,
\begin{equation}
  S^{D-2} = \frac{2\pi^\frac{D-1}{2}}{\Gamma\left(\frac{D-1}{2}\right)}
          = \frac{2(4\pi)^{\frac{D}{2}-1}\Gamma\left(\frac{D}{2}\right)}
                 {\Gamma\left(D\!-\!1\right)}
\,.
\end{equation}
The integral~(\ref{Y1}) can be expressed in terms of a hypergeometric
function. Since we are ultimately interested in expanding in powers of $D-4$,
it is more convenient to expand the integrand in powers of $D-4$,
\begin{eqnarray}
 \left(1-\frac{a'}{a}\right)^{D-1}
   &=& \left(1-\frac{a'}{a}\right)^{3}
   + (D-4)\left(1-\frac{a'}{a}\right)^{3}\ln\left(1-\frac{a'}{a}\right)
\nonumber\\
  &+& \frac12(D-4)^2\left(1-\frac{a'}{a}\right)^{3}
                      \ln^2\left(1-\frac{a'}{a}\right)
   + {\cal O}((D-4)^3)
\label{expand D-1}
\end{eqnarray}
and then to integrate. The result is,
\begin{eqnarray}
  Y_1 &=& i\frac{(4\pi)^\frac{D}{2}}{H^D}
           \frac{\Gamma\left(\frac{D}{2}\right)}{\Gamma\left(D\right)}
    \Bigg\{
           \left[
                 \ln(a)
              - \frac{11}{6} + \frac{3}{a}
              - \frac{3}{2a^2} + \frac{1}{3a^3}
           \right]
\label{Y1a}\\
 &+&\!(D\!-\!4)\left[
            \left(\frac{49}{36} \!-\! \frac{\pi^2}{6}
            \!-\! \frac{11}{6a} \!+\! \frac{7}{12a^2} \!-\! \frac{1}{9a^3}
             \right)
         + \left(\!-\frac{11}{6} \!+\! \frac{3}{a}
            \!-\! \frac{3}{2a^2} \!+\! \frac{1}{3a^3}
            \right)\ln\left(1\!-\!\frac{1}{a}\right)
         +{\rm Li}_2\left(\frac{1}{a}\right)
       \right]
\nonumber\\
 &+&\!\frac{(D\!-\!4)^2}{2}
     \bigg[
           \left(-\frac{251}{108} \!+\! \frac{49}{18a}
           \!-\! \frac{17}{36a^2}\!+\! \frac{2}{27a^3}
             \right)
         - \ln(a)\ln^2\left(1\!-\!\frac{1}{a}\right)
\nonumber\\
       &&\hskip 1.5cm
        +\, \left(\frac{49}{18} \!-\!\frac{\pi^2}{3} \!-\! \frac{11}{3a}
           \!+\! \frac{7}{6a^2}\!-\! \frac{2}{9a^3}
             \right)\ln\left(1\!-\!\frac{1}{a}\right)
         + \left(-\frac{11}{6} \!+\!\frac{3}{a}
           \!-\! \frac{3}{2a^2}\!+\! \frac{1}{3a^3}
             \right)\ln^2\left(1\!-\!\frac{1}{a}\right)
\nonumber\\
       &&\hskip 1.5cm
         +\, 2{\rm Li}_2\left(\frac{1}{a}\right)\ln\left(1-\frac{1}{a}\right)
         + 2{\rm Li}_3\left(1-\frac{1}{a}\right)
       \bigg]
     \Bigg\}
\,.
\nonumber
 \end{eqnarray}

The next integral in Eq.~(\ref{Yn}) is $Y_2$. With the help of
 Eqs.~(\ref{ln2:++ +-}), (\ref{expand D-1}) and~(\ref{J1}) we can write
 ($x=r/\Delta \eta,v=a'/a,u=1-v$),
\begin{eqnarray}
  Y_2 &=& i\frac{(4\pi)^\frac{D}{2}}{H^D}
           \frac{\Gamma\left(\frac{D}{2}\right)}{\Gamma\left(D-1\right)}
             2
           \int_{1}^a \frac{da'}{a'}
            \left(1-\frac{a'}{a}\right)^{D-1}
            \int_0^1 x^{D-2} dx
             \left[
                  \ln\left(\frac{aa'H^2\Delta\eta^2}{4}\right)
                + \ln(1-x^2)
             \right]
\nonumber\\
 &=& i\frac{(4\pi)^\frac{D}{2}}{H^D}
           \frac{\Gamma\left(\frac{D}{2}\right)}{\Gamma\left(D\right)}
             2
           \int_{1}^a \frac{da'}{a'}
            \left(1-\frac{a'}{a}\right)^{3}
             \bigg\{
                \left[
                     2\ln\left(1-\frac{a'}{a}\right)
                   - \ln\left(\frac{a'}{a}\right)
                   - \frac83
                \right]
\nonumber\\
     &&\hskip .cm
    +\,(D\!-\!4)
       \left[
           2\ln^2\left(1-\frac{a'}{a}\right)
           - \ln\left(1-\frac{a'}{a}\right)\ln\left(\frac{a'}{a}\right)
           - \frac83\ln\left(1-\frac{a'}{a}\right)
           + \left(\frac{20}{9}-\frac{\pi^2}{4}\right)
      \right]
\nonumber\\
     &&\hskip .cm
                 +\,{\cal O}\left((D\!-\!4)^2\right)
             \bigg\}
\nonumber\\
 &=& i\frac{(4\pi)^\frac{D}{2}}{H^D}
           \frac{\Gamma\left(\frac{D}{2}\right)}{\Gamma\left(D\right)}
             2
           \int_{0}^{1-\frac{1}{a}} \frac{du}{1-u}u^3
             \bigg\{
                \left[2\ln(u)-\ln(1-u)-\frac83\right]
\label{Y2:u}
\\
     &&\hskip .cm
    +\,(D\!-\!4)
       \left[2\ln^2(u)-\ln(u)\ln(1-u) - \frac83\ln(u)
           + \left(\frac{20}{9}-\frac{\pi^2}{4}\right)
      \right]
                 +\,{\cal O}\left((D\!-\!4)^2\right)
   \bigg\}
\,.
\nonumber
\end{eqnarray}
where we introduced a new variable, $u=1-a'/a$.
These integrals can be integrated to give,
\begin{eqnarray}
 Y_2  &=& i\frac{(4\pi)^\frac{D}{2}}{H^D}\frac{\Gamma\left(\frac{D}{2}\right)}
                                               {\Gamma(D)}
      \Bigg\{
      \Bigg[
             \ln^2(a)
        +  \left(
              - \frac{16}{3} \!+\! \frac{6}{a} \!-\! \frac{3}{a^2} \!+\!
            \frac{2}{3a^3} \right)\ln(a)
        +  \left(
               \frac{21}{2} \!-\! \frac{2\pi^2}{3} \!-\! \frac{52}{3a}
              \!+\! \frac{53}{6a^2} \!-\! \frac{2}{a^3}
            \right)
\nonumber\\
 && \hskip 2.5cm  +\,  \left(
              - \frac{22}{3} + \frac{12}{a}
              - \frac{6}{a^2} + \frac{4}{3a^3}
            \right)\ln\left(1-\frac1a\right)
        + 4{\rm Li}_2\left(\frac{1}{a}\right)
        \Bigg]
\nonumber\\
 &+& (D\!-\!4)
        \Bigg[\left(
                 \frac{40}{9}-\frac{\pi^2}{2}
                - \frac{11}{3a}+\frac{7}{6a^2}
                - \frac{2}{9a^3}
                 +\left(
                   - \frac{13}{3}+\frac{6}{a}
                    - \frac{3}{a^2} + \frac{2}{3a^3}
                \right)\ln\left(1-\frac1a\right)
            \right)\ln(a)
\nonumber\\
   &&        +\,   \left(
                - \frac{313}{18}+\frac{43\pi^2}{36}
                + \frac{461}{18a}-\frac{3\pi^2}{2a}
                - \frac{94}{9a^2}+\frac{3\pi^2}{4a^2}
                + \frac{20}{9a^3}-\frac{\pi^2}{6a^3}
            \right)
\nonumber\\
    &&   +\, \left(
                 \frac{287}{18} -\frac{4\pi^2}{3} - \frac{74}{3a}
                + \frac{67}{6a^2} - \frac{22}{9a^3}
         \right)\ln\left(1-\frac1a\right)
\nonumber\\
   &&   +\, \left(- 4\ln(a)  - \frac{22}{3}+\frac{12}{a}
              - \frac{6}{a^2} + \frac{4}{3a^3}
            \right)\ln^2\left(1-\frac1a\right)
         + 2\ln(a){\rm Li}_2\left(\frac1a\right)
\label{Y2:total}
\\
   &&    -\, \frac53{\rm Li}_2\left(\frac1a\right)
         +8\ln\left(1-\frac1a\right){\rm Li}_2\left(\frac1a\right)
         - 2\zeta(3) +2{\rm Li}_3\left(\frac1a\right)
         +  8{\rm Li}_3\left(1-\frac1a\right)
        \Bigg]
       \Bigg\}
\,.
\nonumber
\end{eqnarray}

 The last integral in Eq.~(\ref{Yn}) is $Y_3$. Similarly as in the
case of $Y_2$, we take account of Eq.~(\ref{ln3:++ +-}) and get,
\begin{eqnarray}
  Y_3 &=& i\frac{(4\pi)^\frac{D}{2}}{H^D}\frac{\Gamma\left(\frac{D}{2}\right)}
                                               {\Gamma(D-1)}3
\int_{1}^a da' {a'}^2
            \left(\frac{1}{a'}-\frac{1}{a}\right)^3
            \int_0^1 x^2 dx
             \bigg[
                  \ln^2\left(\frac{aa'H^2\Delta\eta^2}{4}\right)
\nonumber\\
&&\hskip 3.5cm
           +\, 2 \ln\left(\frac{aa'H^2\Delta\eta^2}{4}\right)\ln(1-x^2)
                + \ln^2(1-x^2)
                - \frac{\pi^2}{3}
                +{\cal O}(D-4)
             \bigg]
\nonumber\\
    &=& i\frac{(4\pi)^\frac{D}{2}}{H^D}\frac{\Gamma\left(\frac{D}{2}\right)}
                                               {\Gamma(D)} 3
     \int_{1}^a \frac{da'}{a'}
            \left(1-\frac{a'}{a}\right)^3
             \bigg[
                 4\ln^2\left(1-\frac{a'}{a}\right)
        - 4\ln\left(\frac{a'}{a}\right)\ln\left(1-\frac{a'}{a}\right)
          + \ln^2\left(\frac{a'}{a}\right)
\nonumber\\
&&\hskip 5.5cm
          -\,\frac{16}{3}\left(
                2\ln\left(1-\frac{a'}{a}\right)
                - \ln\left(\frac{a'}{a}\right)
                         \right)
                + \frac{104}{9} - \frac{2\pi^2}{3}
             \bigg]
\nonumber\\
    &=& i\frac{(4\pi)^\frac{D}{2}}{H^D}\frac{\Gamma\left(\frac{D}{2}\right)}
                                               {\Gamma(D)}3
     \int_{0}^{1-\frac{1}{a}} \frac{du}{1-u}u^3
             \bigg[4\ln^2(u)- 4\ln(u)\ln(1-u) + \ln^2(1-u)
\nonumber\\
&&\hskip 5.2cm
          -\frac{16}{3}\left(
                2\ln(u) - \ln\left(1-u\right)
                         \right)
                + \frac{104}{9} - \frac{2\pi^2}{3}
             \bigg]
\label{Y3}
\end{eqnarray}
Upon performing the integrals in~(\ref{Y3}) we finally obtain,
\begin{eqnarray}
  Y_3 &=& i\frac{(4\pi)^\frac{D}{2}}{H^D}\frac{\Gamma\left(\frac{D}{2}\right)}
                                               {\Gamma(D)}
      \bigg\{
             \ln^3(a)
        +  \left[
              - 8 + \frac{9}{a} - \frac{9}{2a^2} + \frac{1}{a^3}
            \right]\ln^2(a)
\nonumber\\
  &&  +\,  \left[
                \frac{104}{3} - 2\pi^2
              - \frac{52}{a} + \frac{53}{2a^2} - \frac{6}{a^3}
            \right]\ln(a)
\nonumber\\
  &&    +\,  \left[
              - 12\zeta(3) - \frac{2495}{36} + \frac{16\pi^2}{3}
              + \frac{115}{a} - \frac{6\pi^2}{a}
              - \frac{237}{4a^2} + \frac{3\pi^2}{a^2}
              + \frac{122}{9a^3} - \frac{2\pi^2}{3a^3}
            \right]
\nonumber\\
  &&    +\,  \left[
              - 22 + \frac{36}{a} - \frac{18}{a^2} + \frac{4}{a^3}
            \right]\ln\left(1-\frac{1}{a}\right)\ln(a)
        - 12 \ln^2\left(1-\frac{1}{a}\right)\ln(a)
        + 12 {\rm Li}_2\left(\frac{1}{a}\right)\ln(a)
\nonumber\\
 && \hskip 0cm +\,
         \left[
              63 - 4\pi^2
              - \frac{104}{a} + \frac{53}{a^2}
              - \frac{12}{a^3} + 24 {\rm Li}_2\left(\frac{1}{a}\right)
         \right]\ln\left(1-\frac1a\right)
\label{Y3a}
\\
 && \hskip 0cm
     +\, \left[
              - 22
              + \frac{36}{a} - \frac{18}{a^2}
              + \frac{4}{a^3}
         \right]\ln^2\left(1-\frac1a\right)
      - 10 {\rm Li}_2\left(\frac1a\right)
      + 24 {\rm Li}_3\left(1-\frac1a\right)
      + 12 {\rm Li}_3\left(\frac1a\right)
       \bigg\}
\,.
\nonumber
\end{eqnarray}

\subsection*{B2. The $W$ integrals}
\label{The $W$ integrals}

 The following class of basic integrals we need
(these integrals we calculate only up to order $(D-4)^2$) is of the form,
\begin{eqnarray}
 W_n = \int d^Dx' {a'}^D\ln(a')
      \left[
           \ln^n\left(\frac{y_{++}}{4}\right)
          - \ln^n\left(\frac{y_{+-}}{4}\right)
       \right]
\qquad (n=0,1,2)
\label{Wn}
\end{eqnarray}
The trivial integral is $W_0=0$. The first nontrivial integral can be written
as
\begin{eqnarray}
 W_1 =  \int d^Dx' {a'}^D\ln(a')
      \left[2i\pi\theta(\Delta \eta)\theta(\Delta\eta^2-r^2)
       \right]
\,.
\label{W1}
\end{eqnarray}
After the radial integration is performed, this integral can be reduced to,
\begin{eqnarray}
 W_1 &=&  i\frac{(4\pi)^\frac{D}{2}}{H^D}
        \frac{\Gamma\left(\frac{D}{2}\right)}{\Gamma(D)}
                 \int_{1}^{a} \frac{da'}{a'}
           \left(1-\frac{a'}{a}\right)^3\ln(a')
               \left[1+(D\!-\!4)\ln\left(1-\frac{a'}{a}\right)\right]
      + {\cal O}\left((D-4)^2\right)
\label{W1:2}
\\
&=&  i\frac{(4\pi)^\frac{D}{2}}{H^D}
        \frac{\Gamma\left(\frac{D}{2}\right)}{\Gamma(D)}
                 \int_{\frac{1}{a}}^{1} \frac{dv}{v}
           \left(1-v\right)^3\left[\ln(v)+\ln(a)\right]
               \left[1+(D\!-\!4)\ln\left(1-v\right)\right]
      + {\cal O}\left((D-4)^2\right)
\,.\qquad\;
\nonumber
\end{eqnarray}
This can be integrated to give
\begin{eqnarray}
 W_1 &=&  i\frac{(4\pi)^\frac{D}{2}}{H^D}
        \frac{\Gamma\left(\frac{D}{2}\right)}{\Gamma(D)}
 \Bigg\{
         \left[\frac{1}{2}\ln^2(a)-\frac{11}{6}\ln(a)
             + \frac{85}{36}
             - \frac{3}{a}+\frac{3}{4a^2}
             - \frac{1}{9a^3}
         \right]
\nonumber\\
 &+& (D\!-\!4)
      \bigg[\left(\frac{49}{36} - \frac{\pi^2}{6}
           \right)\ln(a)
          + \left(-\frac{395}{108}+\frac{11\pi^2}{36}
               + \frac{151}{36a} - \frac{11}{18a^2}
               + \frac{2}{27a^3}
           \right)
\nonumber\\
&& \hskip 1.2cm
        +\, \left(
               \frac{85}{36}
               - \frac{3}{a} + \frac{3}{4a^2}
               - \frac{1}{9a^3}
           \right)\ln\left(1\!-\!\frac{1}{a}\right)
\nonumber\\
 && \hskip 1.2cm
         - \frac{11}{6}{\rm Li}_2\left(\frac{1}{a}\right)
         - {\rm Li}_3\left(\frac{1}{a}\right)
         + \zeta(3)
      \bigg]
   +{\cal O}\left((D-4)^2\right)
\,.
\label{W1:3}
\end{eqnarray}

 The second integral $W_2$ in~(\ref{Wn}) we only need to leading order in
the $D-4$ expansion. We begin with,
\begin{eqnarray}
 W_2 =  \int d^Dx' {a'}^D\ln(a')
      \left[4i\pi\theta(\Delta \eta)\theta(\Delta\eta^2-r^2)
          \ln\left(\frac{aa'H^2(\Delta\eta^2-r^2)}{4}\right)
       \right]
\,.
\label{W2}
\end{eqnarray}
After the radial integration this integral becomes,
\begin{eqnarray}
 W_2 &=&  i\frac{(4\pi)^\frac{D}{2}}{H^D}
        \frac{\Gamma\left(\frac{D}{2}\right)}{\Gamma(D)}
                2 \int_{1}^{a} \frac{da'}{a'}
           \left(1-\frac{a'}{a}\right)^3\ln(a')
               \left[2\ln\left(1\!-\!\frac{a'}{a}\right)
                    - \ln\left(\frac{a'}{a}\right)
                    - \frac{8}{3}
               \right]
      + {\cal O}(D\!-\!4)
\label{W2:2}
\\
&=&  i\frac{(4\pi)^\frac{D}{2}}{H^D}
        \frac{\Gamma\left(\frac{D}{2}\right)}{\Gamma(D)}
                 2\int_{\frac{1}{a}}^{1} \frac{dv}{v}
           \left(1-v\right)^3\left[\ln(v)+\ln(a)\right]
               \left[2\ln\left(1-v\right) - \ln(v) - \frac{8}{3}\right]
      + {\cal O}(D\!-\!4)
\,.\qquad\;
\nonumber
\end{eqnarray}
When integrated this gives,
\begin{eqnarray}
 W_2 &=&  i\frac{(4\pi)^\frac{D}{2}}{H^D}
        \frac{\Gamma\left(\frac{D}{2}\right)}{\Gamma(D)}
 \Bigg\{
         \frac{1}{3}\ln^3(a)
         - \frac{8}{3}\ln^2(a)
             + \left(\frac{21}{2} - \frac{2\pi^2}{3}
                    - \frac{6}{a}+\frac{3}{2a^2}
                    - \frac{2}{9a^3}
                \right)\ln(a)
\nonumber\\
 && \hskip 1.5cm
        +\, \left(-\frac{895}{54} + \frac{11\pi^2}{9}
             + \frac{187}{9a}-\frac{89}{18a^2}
             + \frac{20}{27a^3}\right)
             + \left(\frac{85}{9}
                    - \frac{12}{a}+\frac{3}{a^2}
                    - \frac{4}{9a^3}
                \right)\ln\left(1\!-\!\frac{1}{a}\right)
\nonumber\\
 && \hskip 1.5cm
      -\, \frac{22}{3}{\rm Li}_2\left(\frac{1}{a}\right)
         - 4{\rm Li}_3\left(\frac{1}{a}\right)
         + 4\zeta(3)
      \bigg]
   + {\cal O}(D-4)
\,.
\label{W2:3}
\end{eqnarray}

\bigskip

 We shall now act with d'Alembertians on the basic integrals
(\ref{Y1a}), (\ref{Y2:total}) and~(\ref{Y3a}). For simplicity
we neglect the terms that are suppressed as $\ln^2(a)/a$,
or more since they vanish in the limit when $a\rightarrow \infty$.
Since $Y_n$ are functions of the scale factor only,
the d'Alembertian operator simplifies to,
\begin{equation}
  \frac{\Box}{H^2} \rightarrow -\frac{1}{a^DH^2}\partial_0 a^{D-2} \partial_0
                   = -a^{2-D}\frac{d}{da} a^D \frac{d}{da}
\,.
\label{dAlembertian:time}
\end{equation}
 Considering that the d'Alembertian does not increase the power of $a$,
\begin{eqnarray}
  \frac{\Box}{H^2}\left(\frac{\ln^2(a)}{a}\right)
       &=& \frac{(D-2)\ln^2(a)}{a}- \frac{2(D-3)\ln(a)}{a}  -\frac{2}{a}
\nonumber\\
  \frac{\Box}{H^2}\left(\frac{\ln(a)}{a}\right)
       &=& \frac{(D-2)\ln(a)}{a}- \frac{D-3}{a}
\nonumber\\
  \frac{\Box}{H^2}\left(\frac{1}{a}\right)
       &=&  \frac{D-2}{a}
\,,
\label{dAlembertian:time:2}
\end{eqnarray}
we can use the asymptotic form
of the basic integrals~(\ref{Y1}), (\ref{Y2:total}) and~(\ref{Y3a}))
to calculate the late time behavior.
In the limit when $a\rightarrow \infty$, the integrals become,
\begin{eqnarray}
  Y_1 &=& i\frac{(4\pi)^\frac{D}{2}}{H^D}
           \frac{\Gamma\left(\frac{D}{2}\right)}{\Gamma\left(D\right)}
    \bigg\{\!
           \left(\ln(a) \!-\! \frac{11}{6}\right)
+(D\!-\!4)\left(\!\frac{49}{36} \!-\! \frac{\pi^2}{6}\right)
+\!(D\!-\!4)^2\left(-\frac{251}{216}+\zeta(3)\right)
\nonumber\\
   && \hskip 2.5cm
        +\, {\cal O}\left(\!\frac{\ln(a)}{a}\right)
        + {\cal O}\left((D-4)^3\right)
     \bigg\}
\label{Y1b}
\\
 Y_2 &=& i\frac{(4\pi)^\frac{D}{2}}{H^D}\frac{\Gamma\left(\frac{D}{2}\right)}
                                               {\Gamma(D)}
 \bigg\{
      \bigg(\ln^2(a) - \frac{16}{3}\ln(a)
          +  \frac{21}{2} - \frac{2\pi^2}{3}
          + {\cal O}\left(\frac{\ln(a)}{a}\right)
      \bigg)
\label{Y2:total:finite}
\\
 &+& (D\!-\!4)\bigg[\left(\frac{40}{9}-\frac{\pi^2}{2}\right)\ln(a)
                   + \left(- \frac{313}{18}+\frac{43\pi^2}{36}
                   + 6\zeta(3)\right)
                   + {\cal O}\left(\frac{\ln(a)}{a}\right)
              \bigg]
            + {\cal O}\left((D-4)^2\right)
  \bigg\}
\nonumber
\\
 Y_3 &=& i\frac{(4\pi)^\frac{D}{2}}{H^D}\frac{\Gamma\left(\frac{D}{2}\right)}
                                               {\Gamma(D)}
      \bigg\{
           \ln^3(a)
         - 8 \ln^2(a)
        +  \left(\frac{104}{3} - 2\pi^2 \right)\ln(a)
\label{Y3b}
\\
     && \hskip 2.5cm
       +\, \left(
               -  \frac{2495}{36} + \frac{16\pi^2}{3} + 12\zeta(3)
            \right)
       + {\cal O}\left(\frac{\ln^2(a)}{a}\right)
       + {\cal O}(D-4)
       \bigg\}
\,.
\nonumber
\end{eqnarray}

The asymptotic forms of the $W_n (n=1,2)$ integrals~(\ref{W1:3}),
(\ref{W2:3}) are,
\begin{eqnarray}
 W_1 &=&  i\frac{(4\pi)^\frac{D}{2}}{H^D}
        \frac{\Gamma\left(\frac{D}{2}\right)}{\Gamma(D)}
 \bigg\{
         \left[\frac{1}{2}\ln^2(a)-\frac{11}{6}\ln(a)
             + \frac{85}{36}
         \right]
         \nonumber\\
 &&+\, (D\!-\!4)
      \bigg[
         \left(\frac{49}{36}-\frac{\pi^2}{6}\right)\ln(a)
         - \frac{395}{108}+\frac{11\pi^2}{36}
         + \zeta(3)
      \bigg]
  + {\cal O}\left((D-4)^2\right)
\bigg\}
\,.
\label{W1:asym}
\end{eqnarray}
and
\begin{eqnarray}
 W_2 &=&  i\frac{(4\pi)^\frac{D}{2}}{H^D}
        \frac{\Gamma\left(\frac{D}{2}\right)}{\Gamma(D)}
 \bigg\{
         \frac{1}{3}\ln^3(a)
         - \frac{8}{3}\ln^2(a)
          + \left(\frac{21}{2} - \frac{2\pi^2}{3}
                \right)\ln(a)
        + \left(-\frac{895}{54} + \frac{11\pi^2}{9}
         + 4\zeta(3)
          \right)
\nonumber\\
 &&  +\, {\cal O}(D-4)
 \bigg\}
\,.
\label{W2:asym}
\end{eqnarray}

The asymptotic forms for the corresponding (nonvanishing) d'Alembertians are
\begin{eqnarray}
  \frac{\Box}{H^2}Y_1
       &=&  i\frac{(4\pi)^\frac{D}{2}}{H^D}
           \frac{\Gamma\left(\frac{D}{2}\right)}{\Gamma\left(D\right)}
              \left[-(D-1) + {\cal O}(1/a) + {\cal O}\left((D-4)^3)\right)\right]
\label{dAlembertian:Y1}
\\
  \frac{\Box}{H^2}Y_2
       &=&  i\frac{(4\pi)^\frac{D}{2}}{H^D}
           \frac{\Gamma\left(\frac{D}{2}\right)}{\Gamma\left(D\right)}
                 \bigg\{\Big[-6\ln(a)+14\Big]
                    + (D\!-\!4) \left[-2\ln(a) - 8 + \frac{3\pi^2}{2}\right]
\nonumber\\
&&\hskip 2.9cm
                                  + {\cal O}\left(\frac{\ln(a)}{a}\right)
                                  + {\cal O}\left((D\!-\!4)^2\right)
                              \bigg\}
\qquad
\label{dAlembertian:Y2:1}
\\
  \left(\frac{\Box}{H^2}\right)^2\!\!Y_2 \!
       &=&  \! i\frac{(4\pi)^\frac{D}{2}}{H^D}
           \frac{\Gamma\left(\frac{D}{2}\right)}{\Gamma\left(D\right)}
                \left\{ 18
                     + 12(D\!-\!4)
                   + {\cal O}\left(\frac{\ln(a)}{a}\right)
                   + {\cal O}\left((D\!-\!4)^2\right)
                \right\}
\qquad\;
\label{dAlembertian:Y2:2}
\\
  \frac{\Box}{H^2}Y_3
       \!&=&\! i\frac{(4\pi)^\frac{D}{2}}{H^D}
           \frac{\Gamma\left(\frac{D}{2}\right)}{\Gamma\left(D\right)}
                \bigg[ \!- \!9\ln^2(a) + 42\ln(a)
                                  - 88 + 6\pi^2
                         \!+\! {\cal O}\left(\!\frac{\ln^2(a)}{a}\right)
                                  \!+\! {\cal O}\left(D\!-\!4\right)\!
                              \bigg]\quad
\label{dAlembertian:Y3:1}
\\
  \left(\frac{\Box}{H^2}\right)^2Y_3
       &=&  i\frac{(4\pi)^\frac{D}{2}}{H^D}
           \frac{\Gamma\left(\frac{D}{2}\right)}{\Gamma\left(D\right)}
                 \left[54 \ln(a) - 108
                                  + {\cal O}(\ln^2(a)/a)
                                  + {\cal O}\left(D-4\right)
                               \right]
\label{dAlembertian:Y3:2}
\\
  \left(\frac{\Box}{H^2}\right)^3Y_3
       &=&  i\frac{(4\pi)^\frac{D}{2}}{H^D}
           \frac{\Gamma\left(\frac{D}{2}\right)}{\Gamma\left(D\right)}
                 \left[- 162
                     + {\cal O}(\ln^2(a)/a)
                     + {\cal O}\left(D-4\right)
                 \right]
\label{dAlembertian:Y3:3}
\,,
\end{eqnarray}
and similarly for $W_n$'s.

\medskip

\subsection*{B3. The $\Xi$ integrals}
\label{The Xi integrals}

 We are now ready to calculate intermediate integrals that we need
for evaluation of the one vertex integrals. They are of the general form,
\begin{equation}
\Xi_\alpha =  \int d^D x' {a'}^D \left[
                          \left(\frac{y_{++}}{4}\right)^{-\alpha}
                        -  \left(\frac{y_{+-}}{4}\right)^{-\alpha}
                    \right]
\label{xi:alpha}
\end{equation}
with $\alpha = \frac{3D}{2}-6, \frac{3D}{2}-5, \frac{3D}{2}-4,3-\frac{3D}{2},
                D-2, D-3, D-4, \frac{D}{2}-1$ and $\frac{D}{2}-2$.
All of $\Xi_\alpha$ integrals can be represented in terms of
the basic integrals $Y_n$ given in Eqs.~(\ref{Y1b}--\ref{Y3b}) and the
associated d'Alembertians~(\ref{dAlembertian:Y1}--\ref{dAlembertian:Y3:3}).

Let us begin with $\alpha = \frac{3D}{2}-6$ which, when
expanded in powers of $D-4$, yields
\begin{eqnarray}
\Xi_{\frac{3D}{2}-6}
    &=&  \int d^D x' {a'}^D \left[
                          \left(\frac{y_{++}}{4}\right)^{-\frac{3}{2}(D-4)}
                        -  \left(\frac{y_{+-}}{4}\right)^{-\frac{3}{2}(D-4)}
                    \right]
\nonumber\\
   &=& -\frac32(D-4)Y_1 + \frac{9}{8}(D-4)^2Y_2
    - \frac{9}{16}(D-4)^3Y_3
    + {\cal O}\left((D-4)^4\right)
\nonumber\\
    &=&  i\frac{(4\pi)^\frac{D}{2}}{H^D}
           \frac{\Gamma\left(\frac{D}{2}\right)}{\Gamma\left(D\right)}
  \bigg\{
     (D\!-\!4)\left[
                     -\frac32\ln(a) + \frac{11}{4}
                 \right]
     + (D\!-\!4)^2\left[
                     \frac98 \ln^2(a) - 6\ln(a)
                      + \frac{469}{48} - \frac{\pi^2}{2}
                 \right]
\nonumber\\
   &&  +\, (D\!-\!4)^3\bigg[
                     - \frac{9}{16} \ln^3(a) + \frac{9}{2} \ln^2(a)
                     +\left(-\frac{29}{2}+\frac{9\pi^2}{16}
                       \right) \ln(a)
\nonumber\\
     &&\hskip 2cm       +\,\left(
                           \frac{12191}{576}
                          -  \frac{53\pi^2}{32}
                          -  \frac{3}{2}\zeta(3)
                       \right)
                 \bigg]
     + {\cal O}\left((D-4)^4\right)
  \bigg\}
\,.
\qquad
\label{xi:3D/2-6}
\end{eqnarray}

Next integral we need corresponds to $\alpha ={\frac{D}{2}-2}$,
\begin{eqnarray}
\Xi_{\frac{D}{2}-2}
    &=&  \int d^D x' {a'}^D \left[
                          \left(\frac{y_{++}}{4}\right)^{-\frac{D-4}{2}}
                        -  \left(\frac{y_{+-}}{4}\right)^{-\frac{D-4}{2}}
                    \right]
\nonumber\\
   &=& -\frac12(D-4)Y_1 + \frac{1}{8}(D-4)^2Y_2
    - \frac{1}{48}(D-4)^3Y_3
    + {\cal O}\left((D-4)^4\right)
\nonumber\\
    &=&   i\frac{(4\pi)^\frac{D}{2}}{H^D}
           \frac{\Gamma\left(\frac{D}{2}\right)}{\Gamma\left(D\right)}
  \bigg\{
      (D\!-\!4)\left[
                     -\frac12\ln(a) + \frac{11}{12}
                 \right]
     + (D\!-\!4)^2\left[
                     \frac18 \ln^2(a) - \frac23\ln(a)
                      + \frac{91}{144}
                 \right]
\nonumber\\
   &&  +\, (D\!-\!4)^3\bigg[
                     - \frac{1}{48} \ln^3(a) + \frac{1}{6} \ln^2(a)
                     +\left(-\frac{1}{6}-\frac{\pi^2}{48}
                       \right) \ln(a)
\nonumber\\
     &&\hskip 2cm       +\,\left(
                          - \frac{257}{1728}
                          +  \frac{11\pi^2}{288}
                       \right)
                 \bigg]
     + {\cal O}\left((D-4)^4\right)
  \bigg\}
\,.
\qquad
\label{xi:D/2-2}
\end{eqnarray}

 The following integral is $\Xi_{\frac{3D}{2}-5}$, which we can rewrite
with the help of Eq.~(\ref{A1:DAlembertian:3D2m5}) as,
\begin{eqnarray}
\Xi_{\frac{3D}{2}-5}
  &=& \left[\frac{2}{3(D-4)(D-5)}\frac{\Box}{H^2} + \frac{D-10}{2(D-5)}\right]
                            \Xi_{\frac{3D}{2}-6}
\nonumber\\
    &=&   i\frac{(4\pi)^\frac{D}{2}}{H^D}
           \frac{\Gamma\left(\frac{D}{2}\right)}{\Gamma\left(D\right)}
  \bigg\{
      -3  -\frac{25}{4} (D\!-\!4)
     + (D\!-\!4)^2\left[
                    - \frac{85}{16} - \frac{3\pi^2}{8}
                  \right]
\nonumber\\
   &&  +\, (D\!-\!4)^3\bigg[
                     - \frac{27}{16} \ln^3(a) + \frac{189}{16} \ln^2(a)
                     +\left(-\frac{69}{2}+\frac{27\pi^2}{16}
                       \right) \ln(a)
\nonumber\\
     &&\hskip 1.6cm       +\,\left(
                           \frac{8377}{192}
                          -  \frac{151\pi^2}{32}
                          -  \frac{9}{2}\zeta(3)
                       \right)
                 \bigg]
     + {\cal O}\left((D-4)^4\right)
 \bigg\}
\,.
\qquad
\label{xi:3D/2-5}
\end{eqnarray}
Next integral we calculate is $\alpha = \frac{D}{2}-1$. With the help
of Eq.~(\ref{A1:DAlembertian:D2m1}), we arrive at,
\begin{eqnarray}
\Xi_{\frac{D}{2}-1}
    &=& \left[-\frac{2}{D-4}\frac{\Box}{H^2} + \frac{D+2}{2}\right]
        \Xi_{\frac{D}{2}-2}
\nonumber\\
    &=&   i\frac{(4\pi)^\frac{D}{2}}{H^D}
           \frac{\Gamma\left(\frac{D}{2}\right)}{\Gamma\left(D\right)}
  \bigg\{
     -3  - \frac74(D\!-\!4)
     + (D\!-\!4)^2\left[
                     \frac{11}{16} - \frac{\pi^2}{8}
                 \right]
\nonumber\\
   &&  +\, (D\!-\!4)^3\bigg[
                     - \frac{1}{16} \ln^3(a) + \frac{7}{16} \ln^2(a)
                     +\left(-\frac{1}{6}-\frac{\pi^2}{16}
                       \right) \ln(a)
\nonumber\\
     &&\hskip 2cm       +\,\left(
                          - \frac{89}{192}
                          +  \frac{7\pi^2}{96}
                       \right)
                 \bigg]
     + {\cal O}\left((D-4)^4\right)
  \bigg\}
\,.
\qquad
\label{xi:D/2-1}
\end{eqnarray}

 In order to evaluate the integral $\alpha = \frac{3D}{2}-4$, one
needs to subtract the divergent contribution according to
Eq.~(\ref{A1:DAlembertian:3D2m4}). The result is,
\begin{eqnarray}
\Xi_{\frac{3D}{2}-4}
  &=& \left[\frac{2}{(3D-10)(D-4)}\frac{\Box}{H^2} + \frac{D-8}{2(D-4)}\right]
                           \Xi_{\frac{3D}{2}-5}
\\
  &+& \frac{2}{(3D\!-\!10)(D\!-\!4)}
               \left[-\frac{\Box}{H^2} + \frac{D(D\!-\!2)}{4}\right]
                           \Xi_{\frac{D}{2}-1}
    + i\frac{1}{H^D}\frac{2}{(3D\!-\!10)(D\!-\!4)}
       \frac{(4\pi)^\frac{D}{2}}{\Gamma\left(\frac{D}{2}\!-\!1\right)}
\nonumber\\
    &=&   i\frac{(4\pi)^\frac{D}{2}}{H^D}
           \frac{\Gamma\left(\frac{D}{2}\right)}{\Gamma\left(D\right)}
  \bigg\{
      12 + (D\!-\!4)\left[
                    4 + \frac{\pi^2}{2}
                  \right]
\nonumber\\
   &&  +\, (D\!-\!4)^2\bigg[
                      \frac{13}{4} \ln^3(a) - \frac{65}{8} \ln^2(a)
                     +\left(\frac{61}{6}-\frac{7\pi^2}{2}
                       \right) \ln(a)
                      +\left(
                          - \frac{19}{4}
                          +  \frac{13\pi^2}{3}
                          +  \zeta(3)
                       \right)
                 \bigg]
\nonumber\\
&&
     +\, \frac{\Gamma(D)}{\Gamma\left(\frac{D}{2}\right)
                           \Gamma\left(\frac{D}{2}\!-\!1\right)}
         \frac{2}{(3D\!-\!10)(D\!-\!4)}
     + {\cal O}\left((D-4)^3\right)
\bigg\}
\,.
\qquad
\label{xi:3D/2-4}
\end{eqnarray}
Furthermore, for $\alpha = \frac{3D}{2}-3$ we make use of
Eq.~(\ref{A1:DAlembertian:3D2m3}) to obtain,
\begin{eqnarray}
\Xi_{\frac{3D}{2}-3}
  &=& \left[\frac{2}{(3D-8)(D-3)}\frac{\Box}{H^2} + \frac{D-6}{2(D-3)}\right]
                           \Xi_{\frac{3D}{2}-4}
\nonumber\\
    &=&   i\frac{(4\pi)^\frac{D}{2}}{H^D}
           \frac{\Gamma\left(\frac{D}{2}\right)}{\Gamma\left(D\right)}
  \bigg\{
      - 12 + (D\!-\!4)\left[
                    14 - \frac{\pi^2}{2}
                  \right]
\nonumber\\
   &&  +\, (D\!-\!4)^2\bigg[
                     -\frac{13}{4} \ln^3(a) - \frac{13}{2} \ln^2(a)
                     + \left(\frac{107}{24} + \frac{7\pi^2}{2}
                       \right) \ln(a)
\nonumber\\
     &&\hskip 1.6cm       +\,\left(
                          - \frac{115}{8}
                          +  \frac{5\pi^2}{3}
                          - 9 \zeta(3)
                       \right)
                 \bigg]
\nonumber\\
 &&   +\, \frac{\Gamma(D)}{\Gamma\left(\frac{D}{2}\right)
                           \Gamma\left(\frac{D}{2}\!-\!1\right)}
      \frac{D-6}{(D-3)(3D\!-\!10)(D\!-\!4)}
    + {\cal O}\left((D-4)^3\right)
\bigg\}
\,.
\qquad
\label{xi:3D/2-3}
\end{eqnarray}

The next class of integrals begins with $\alpha = D-4$.
The integral is of the form,
\begin{eqnarray}
\Xi_{D-4}
   &=& -(D-4)Y_1 + \frac{1}{2}(D-4)^2Y_2
    - \frac{1}{6}(D-4)^3Y_3
    + {\cal O}\left((D-4)^4\right)
\nonumber\\
    &=&  i\frac{(4\pi)^\frac{D}{2}}{H^D}
           \frac{\Gamma\left(\frac{D}{2}\right)}{\Gamma\left(D\right)}
  \bigg\{
     (D\!-\!4)\left[
                     -\ln(a) + \frac{11}{6}
                 \right]
     + (D\!-\!4)^2\left[
                     \frac12 \ln^2(a) - \frac83\ln(a)
                      + \frac{35}{9} - \frac{\pi^2}{6}
                 \right]
\nonumber\\
   &&  +\, (D\!-\!4)^3\bigg[
                     - \frac{1}{6} \ln^3(a)
                     + \frac{4}{3} \ln^2(a)
                     +\left(-\frac{32}{9}+\frac{\pi^2}{12}
                       \right) \ln(a)
\nonumber\\
     &&\hskip 2cm
      +\,\left(
                           \frac{217}{54}
                          -  \frac{7\pi^2}{24}
                       \right)
                 \bigg]
     + {\cal O}\left((D-4)^4\right)
  \bigg\}
\,.
\qquad
\label{xi:D-4}
\end{eqnarray}
With the help of Eq.~(\ref{A1:DAlembertian:Dm3})
we calculate the next integral,
\begin{eqnarray}
\Xi_{D-3}
  &=& \left[\frac{2}{(D-4)(D-6)}\frac{\Box}{H^2} - \frac{6}{D-6}\right]
                                     \Xi_{D-4}
\nonumber\\
    &=&   i\frac{(4\pi)^\frac{D}{2}}{H^D}
           \frac{\Gamma\left(\frac{D}{2}\right)}{\Gamma\left(D\right)}
  \bigg\{
      -3  -4 (D\!-\!4)
     + (D\!-\!4)^2\left[
                    - 1 - \frac{\pi^2}{4}
                  \right]
                  \!\!\!\!
\label{xi:D-3}
\\
   &+&\! (D\!-\!4)^3\bigg[
                     \!-\! \frac{1}{2} \ln^3(a) + \frac{7}{2} \ln^2(a)
                     +\left(-8+\frac{\pi^2}{4}
                       \right) \ln(a)
       +\left(
                           8
                          -  \frac{11\pi^2}{12}
                       \right)
                 \bigg]
     + {\cal O}\left((D-4)^4\right)
 \bigg\}
.
\nonumber
\end{eqnarray}
Next integral corresponds to $\alpha = D=2$, and can be calculated
by the means of Eq.~(\ref{A1:DAlembertian:Dm2}). The result is,
\begin{eqnarray}
\Xi_{D-2}
  &=& \left[\frac{2}{(D-3)(D-4)}\frac{\Box}{H^2} - \frac{4}{D-4}\right]
              \Xi_{D-3}
\nonumber
\\
  &+& \frac{2}{(D\!-\!3)(D\!-\!4)}
               \left[-\frac{\Box}{H^2} + \frac{D(D\!-\!2)}{4}\right]
              \Xi_{\frac{D}{2}-1}
    + \frac{2}{(D\!-\!3)(D\!-\!4)}
       \frac{(4\pi)^\frac{D}{2}}{\Gamma\left(\frac{D}{2}\!-\!1\right)}
            \frac{i}{H^D}
\nonumber\\
    &=&   i\frac{(4\pi)^\frac{D}{2}}{H^D}
           \frac{\Gamma\left(\frac{D}{2}\right)}{\Gamma\left(D\right)}
  \bigg\{
      12  +  (D\!-\!4)\left[
                          4 + \frac{\pi^2}{2}
                      \right]
\nonumber\\
   &&  +\, (D\!-\!4)^2\bigg[
                           \frac{7}{4} \ln^3(a) - \frac{35}{8} \ln^2(a)
                    + \left(-\frac16 - \frac{5\pi^2}{4}
                       \right) \ln(a)
                    +\left(
                           \frac{25}{12}
                          +  \frac{53\pi^2}{24}
                    \right)
                 \bigg]
\nonumber\\
    &+& \frac{\Gamma(D)}{\Gamma\left(\frac{D}{2}\right)
                           \Gamma\left(\frac{D}{2}\!-\!1\right)}
       \frac{2}{(D\!-\!3)(D\!-\!4)}
      + {\cal O}\left((D-4)^3\right)
  \bigg\}
\,.
\qquad
\label{xi:D-2}
\end{eqnarray}
The final integral corresponds to $\alpha = D-1$ and it evaluates to,
\begin{eqnarray}
\Xi_{D-1}
  &=& \left[\frac{2}{(D-2)^2}\frac{\Box}{H^2} - \frac{2}{D-2}\right]
              \Xi_{D-2}
\nonumber
\\
    &=&   i\frac{(4\pi)^\frac{D}{2}}{H^D}
           \frac{\Gamma\left(\frac{D}{2}\right)}{\Gamma\left(D\right)}
  \bigg\{
      - 12  +  (D\!-\!4)\left[
                          2 - \frac{\pi^2}{2}
                      \right]
\nonumber\\
   &&  +\, (D\!-\!4)^2\bigg[
                           - \frac{7}{4} \ln^3(a) - \frac{7}{2} \ln^2(a)
                    + \left(\frac{193}{24} + \frac{5\pi^2}{4}
                       \right) \ln(a)
 + \left(
                           \frac{37}{24}
                          -  \frac{\pi^2}{12}
                       \right)
                 \bigg]
\nonumber\\
  &&-\, \frac{\Gamma(D)}{\Gamma\left(\frac{D}{2}\right)
                           \Gamma\left(\frac{D}{2}\!-\!1\right)}
      \frac{4}{(D\!-\!2)(D\!-\!3)(D\!-\!4)}
     + {\cal O}\left((D-4)^3\right)
 \bigg\}
\,.
\qquad
\label{xi:D-1}
\end{eqnarray}

\bigskip

\subsection*{B4. The $\Lambda$ integrals}
\label{The Lambda integrals}

 There is one more class of integrals to be evaluated
which are based on the basic integrals~(\ref{Wn})
that are needed to evaluate the one vertex integral contributing to
the central diagram in Fig.~1. The general form of these integrals is,
\begin{equation}
\Lambda_\alpha =  \int d^D x' {a'}^D \ln(a')\left[
                          \left(\frac{y_{++}}{4}\right)^{-\alpha}
                     -  \left(\frac{y_{+-}}{4}\right)^{-\alpha}
                    \right]
\label{Lambda:alpha}
\end{equation}
with $\alpha = D-1, D-2, D-3, D-4, \frac{D}{2}-1$ and $\frac{D}{2}-2$.
All of these integrals can be represented in terms of
the $W_n$ functions given in Eqs.~(\ref{W1:3}) and (\ref{W2:3}). We give
these integrals to order $(D-4)^2$, which is also what we need for the
evaluation of the integral~(\ref{diagram:figure 8:2}).

When expanded in powers of $D-4$,
the $\Lambda_{\alpha = D-4}$ integral can be written as,
yields
\begin{eqnarray}
\Lambda_{D-4}
    &=&  \int d^D x' {a'}^D \ln(a')\left[
                          \left(\frac{y_{++}}{4}\right)^{-(D-4)}
                        -  \left(\frac{y_{+-}}{4}\right)^{-(D-4)}
                    \right]
\nonumber\\
   &=& -(D-4)W_1 + \frac{1}{2}(D-4)^2W_2
    + {\cal O}\left((D-4)^3\right)
\nonumber\\
    &=&  i\frac{(4\pi)^\frac{D}{2}}{H^D}
           \frac{\Gamma\left(\frac{D}{2}\right)}{\Gamma\left(D\right)}
  \bigg\{
     (D\!-\!4)\left[
                    -\frac12\ln^2(a) +\frac{11}{6}\ln(a) - \frac{85}{36}
                 \right]
\label{Lambda:D-4}
\\
 &&    +\, (D\!-\!4)^2\left[
                     \frac16 \ln^3(a)
                     -\frac43\ln^2(a)
                    + \left(\frac{35}{9} - \frac{\pi^2}{6}\right)\ln(a)
                    - \frac{125}{27} + \frac{11\pi^2}{36}
                    +\zeta(3)
                 \right]
\nonumber\\
 &&    + {\cal O}\left((D-4)^3\right)
  \bigg\}
\,.
\qquad
\nonumber
\end{eqnarray}
Similarly, for other integrals we need we have,
\begin{eqnarray}
\Lambda_{\frac{D}{2}-2}
   &=& -\frac{D-4}{2}W_1 + \frac{(D-4)^2}{8}W_2
    + {\cal O}\left((D-4)^3\right)
\nonumber\\
    &=&  i\frac{(4\pi)^\frac{D}{2}}{H^D}
           \frac{\Gamma\left(\frac{D}{2}\right)}{\Gamma\left(D\right)}
  \bigg\{
     (D\!-\!4)\left[
                    -\frac14\ln^2(a) +\frac{11}{12}\ln(a) - \frac{85}{72}
                 \right]
\label{Lambda:D2-2}
\\
 &&    +\, (D\!-\!4)^2\left[
                     \frac{1}{24} \ln^3(a)
                     -\frac13\ln^2(a)
                    + \frac{91}{144}\ln(a)
                    - \frac{35}{144}
                 \right]
     + {\cal O}\left((D-4)^3\right)
  \bigg\}
\,,
\qquad
\nonumber
\end{eqnarray}
\begin{eqnarray}
\Lambda_{D\!-\!3}
   &=& \left[\frac{2}{(D\!-\!4)(D\!-\!6)}\frac{\Box}{H^2}
          - \frac{6}{(D\!-\!6)}\right]\Lambda_{D\!-\!4}
    + {\cal O}\left((D-4)^3\right)
\nonumber\\
    &=&  i\frac{(4\pi)^\frac{D}{2}}{H^D}
           \frac{\Gamma\left(\frac{D}{2}\right)}{\Gamma\left(D\right)}
  \bigg\{
     \left[-3\ln(a)+\frac92\right]
     + (D\!-\!4)\left[
                    - 4\ln(a) + 6 - \frac{\pi^2}{2}
                 \right]
\label{Lambda:D-3}
\\
 &&    +\, (D\!-\!4)^2\left[
                     \frac{1}{2} \ln^3(a)
                    - \frac{7}{2} \ln^2(a)
                    + \left(7 - \frac{\pi^2}{2}\right)\ln(a)
                    - 7 + \frac{\pi^2}{2} + 3\zeta(3)
                 \right]
     + {\cal O}\left((D-4)^3\right)
  \bigg\}
\,,
\nonumber
\end{eqnarray}
\begin{eqnarray}
\Lambda_{\frac{D}{2}-1}
   &=& \left[-\frac{2}{D\!-\!4}\frac{\Box}{H^2} + \frac{D\!+\!2}{4}\right]\Lambda_{\frac{D}{2}-2}
\nonumber\\
    &=&  i\frac{(4\pi)^\frac{D}{2}}{H^D}
           \frac{\Gamma\left(\frac{D}{2}\right)}{\Gamma\left(D\right)}
  \bigg\{-3\ln(a)+\frac92
       + (D\!-\!4)\left[
                    -\frac74\ln(a) + \frac{3}{4}
                 \right]
\label{Lambda:D2-1}
\\
 && \hskip 2.4cm   +\, (D\!-\!4)^2\left[
                     \frac{1}{8}\ln^3(a)
                     -\frac78\ln^2(a)
                    + \frac{49}{48}\ln(a)
                    - \frac{1}{18}
                 \right]
     + {\cal O}\left((D-4)^3\right)
  \bigg\}
\,,
\qquad
\nonumber
\end{eqnarray}
\begin{eqnarray}
\Lambda_{D\!-\!2}
   &=& \left[\frac{2}{(D\!-\!3)(D\!-\!4)}\frac{\Box}{H^2}
          - \frac{4}{(D\!-\!4)}\right]\Lambda_{D\!-\!3}
     + \frac{2}{(D\!-\!3)(D\!-\!4)}
        \left[-\frac{\Box}{H^2}
          + \frac{D(D\!-\!2)}{4}\right]\Lambda_{\frac{D}{2}\!-\!1}
\nonumber\\
&&    +\, i\frac{(4\pi)^\frac{D}{2}}{H^D}
                               \frac{1}{\Gamma\left(\frac{D}{2}\!-\!1\right)}
            \frac{2}{(D\!-\!3)(D\!-\!4)}\ln(a)
      + {\cal O}\left((D-4)^3\right)
\nonumber\\
    &=&  i\frac{(4\pi)^\frac{D}{2}}{H^D}
           \frac{\Gamma\left(\frac{D}{2}\right)}{\Gamma\left(D\right)}
  \bigg\{ \left[ 12 \ln(a) - 12 + 2\pi^2\right]
\nonumber\\
&&    +\, (D\!-\!4)\left[
                    -\frac32\ln^3(a)
                    + \frac{15}{4}\ln^2(a)
                    +\left(\frac{1}{3} + 2\pi^2\right)\ln(a)
                     - \frac{43}{72} + \pi^2 - 12\zeta(3)
                 \right]
\nonumber
\\
 &&    +\, (D\!-\!4)^2\left[
                    -\frac{1}{24}\ln^3(a)
                    + \frac{61}{12}\ln^2(a)
                    - \frac{2323}{144}\ln(a)
                    + \frac{2533}{144} - 2\pi^2
                 \right]
\nonumber
\\
  &&             +\, \frac{\Gamma(D)}{\Gamma\left(\frac{D}{2}\right)
                               \Gamma\left(\frac{D}{2}\!-\!1\right)}
                      \frac{2}{(D\!-\!3)(D\!-\!4)}\ln(a)
  \bigg\}
              + {\cal O}\left((D-4)^3\right)
\label{Lambda:D-2}
\,,
\qquad
\end{eqnarray}
\begin{eqnarray}
\Lambda_{D\!-\!1}
   &=& \left[\frac{2}{(D\!-\!2)^2}\frac{\Box}{H^2}
          - \frac{2}{D\!-\!2}\right]\Lambda_{D\!-\!2}
      + {\cal O}\left((D-4)^3\right)
\nonumber\\
    &=&  i\frac{(4\pi)^\frac{D}{2}}{H^D}
           \frac{\Gamma\left(\frac{D}{2}\right)}{\Gamma\left(D\right)}
  \bigg\{\left[ - 12\ln(a) - 6 - 2\pi^2\right]
\nonumber\\
&&    +\, (D\!-\!4)\left[
                    \frac32\ln^3(a)
                    + 3\ln^2(a)
                    +\left(-\frac{13}{12} - 2\pi^2\right)\ln(a)
                    + \frac{169}{72} - 3\pi^2 + 12\zeta(3)
                 \right]
\label{Lambda:D-1}
\\
 &&    +\, (D\!-\!4)^2\left[
                    -\frac{17}{24}\ln^3(a)
                    - \frac{361}{48}\ln^2(a)
                    + \left(\frac{169}{144} + \pi^2\right)\ln(a)
                    + \frac{233}{288} + 4\pi^2 - 6\zeta(3)
                 \right]
\nonumber\\
 &&  - \frac{\Gamma(D)}{\Gamma\left(\frac{D}{2}\right)^2}
            \frac{2}{(D\!-\!3)(D\!-\!4)}\ln(a)
      - \frac{\Gamma(D)}{\Gamma\left(\frac{D}{2}\right)^2}
            \frac{2(D\!-\!1)}{(D\!-\!2)(D\!-\!3)(D\!-\!4)}
       +  {\cal O}\left((D-4)^3\right)
  \bigg\}
\,.
\qquad
\nonumber
\end{eqnarray}

\bigskip

\subsection*{B5. The $\Omega$ integrals}
\label{The Omega integrals}

 We can now use these results to calculate the following class of integrals
we need, which are of the form,
\begin{eqnarray}
\lefteqn{  \Omega_{\alpha,\beta} = \int d^Dx' {a'}^D
         \left[\partial_\mu\left(\frac{y_{++}}{4}\right)^{-\alpha}
                \partial_\rho\left(\frac{y_{++}}{4}\right)^{-\beta}
            -  \partial_\mu\left(\frac{y_{+-}}{4}\right)^{-\alpha}
                \partial_\rho\left(\frac{y_{+-}}{4}\right)^{-\beta}
         \right] }
\label{Zeta:alpha:beta}
\\
  & & = \left[\frac{\alpha\beta}{(\alpha+\beta)(\alpha+\beta+1)}
        \nabla_{\mu}\nabla_\rho
   + \frac{\alpha\beta H^2g_{\mu\rho}}{D-2(\alpha+\beta+1)}
       \left(
           - \frac{1}{(\alpha+\beta)(\alpha+\beta+1)}\frac{\Box}{H^2}
           + 1
      \right)
       \right] \Xi_{\alpha+\beta}
\,, \quad
\nonumber
\end{eqnarray}
where we made use of Eq.~(\ref{DyalphaDybeta})
and assumed $\alpha+\beta+1\neq \frac{D}{2}$.
These integrals can be evaluated with a help of
 Eqs.~(\ref{Dy1-D2Dy2-D}--\ref{Dy1-D2Dy-1})
and the $\Xi_\gamma$ integrals~(\ref{xi:3D/2-6}--\ref{xi:D-2}).
We need the integrals corresponding to the pairs,
$(\alpha,\beta) = (\frac{D}{2}-1,D-2),
(\frac{D}{2}-1,D-2), (\frac{D}{2}-1,D-3), (\frac{D}{2}-1,D-4),
(\frac{D}{2}-1,D-5), (\frac{D}{2}-1,D-6),
 (\frac{D}{2}-2,D-2), (\frac{D}{2}-2,D-3), (\frac{D}{2}-2,D-4),
 (\frac{D}{2}-5,D-5),
(\frac{D}{2}-1,\frac{D}{2}-2),
(\frac{D}{2}-1,\frac{D}{2}-3),
(\frac{D}{2}-2,\frac{D}{2}-2)$.

 Note that when $\nabla_\mu \nabla_\rho $ acts on a function of time
 (or equivalently of $a$, which is a scalar function),
one obtains,
\begin{eqnarray}
   \frac{\nabla_\mu \nabla_\rho}{H^2} &=&
                         - g_{\mu\rho}a\partial_a
                         + \left(a^2\delta_\mu^0\delta_\rho^0\right)
                             a^2\partial_a^2
\label{doubleDerivative:time}
\\
    \frac{\Box}{H^2} &=& g^{\mu\rho} \frac{\nabla_\mu \nabla_\rho}{H^2}
                      =   - a^{2-D}\partial_a a^D\partial_a
\,.
\label{dAlembertian:time:3}
\end{eqnarray}
Making use of these relations and of Eq.~(\ref{DyalphaDybeta})
(and assuming that $\alpha+\beta+1\neq \frac{D}{2}$),
Eq.~(\ref{Zeta:alpha:beta}) reduces to,
\begin{eqnarray}
  \Omega_{\alpha,\beta}
  &=& H^2g_{\mu\rho}
      \left[-\frac{\alpha\beta}{(\alpha\!+\!\beta)(\alpha\!+\!\beta\!+\!1)}
        a\partial_a
   + \frac{\alpha\beta}{D-2(\alpha\!+\!\beta\!+\!1)}
       \left(
         \frac{a^{2-D}\partial_aa^D\partial_a}
              {(\alpha\!+\!\beta)(\alpha\!+\!\beta\!+\!1)}
           + 1
      \right)
    \right] \Xi_{\alpha+\beta}
\,,
\nonumber\\
  &+& H^2\left(a^2\delta_\mu^0\delta_\rho^0\right)
         \frac{\alpha\beta}{(\alpha+\beta)(\alpha+\beta+1)}
        a^2\partial^2_a
         \Xi_{\alpha+\beta}
\,.
\label{Zeta:alpha:beta:2}
\end{eqnarray}
Note that the last term breaks the Lorentz symmetry. The $\Omega_{\alpha,\beta}$
integrals are symmetric under $\alpha \rightleftarrows \beta$.

Since the one vertex integrals are at least linearly divergent in $D-4$,
to extract the finite contribution, we only need terms up to linear order
in the $D-4$ expansion.

 We begin with the integrals of the form $\alpha = \frac{D}{2}-n, \beta = D-n$
($\alpha + \beta =  \frac{3D}{2}-(m+n); m=1,2,3,4;  n=2,3,4,5$).
We get,
\begin{eqnarray}
\Omega_{\frac{D}{2}-1,D-2}
 &=&   ig_{\mu\rho}\frac{(4\pi)^\frac{D}{2}}{H^{D-2}}
        \frac{\Gamma\left(\frac{D}{2}\right)}{\Gamma\left(D\right)}
  \bigg\{ 6
             + (D-4)\left(-4 + \frac{\pi^2}{4} \right)
\label{Omega:D2-1:D-2}
\\
  && \hskip 3.5cm 
    -\,  \frac{\Gamma\left(D\right)}
    {\Gamma\left(\frac{D}{2}\right)\Gamma\left(\frac{D}{2}\!-\!1\right)}
    \frac{(D-2)(D-6)}{4(D\!-\!3)(3D\!-\!10)(D\!-\!4)}
  \bigg\}
   + {\cal O}\left((D-4)^2\right)
\nonumber\\
\Omega_{\frac{D}{2}-1,D-3}
 &=&   ig_{\mu\rho}\frac{(4\pi)^\frac{D}{2}}{H^{D-2}}\frac{\Gamma\left(\frac{D}{2}\right)}{\Gamma\left(D\right)}
  \bigg\{ - 6
          + (D-4)\left( - 5 - \frac{\pi^2}{4}\right)
\label{Omega:D2-1:D-3}
\\
  && \hskip 3.5cm
    -\,\frac{\Gamma\left(D\right)}
       {\Gamma\left(\frac{D}{2}\right)\Gamma\left(\frac{D}{2}\!-\!1\right)}
    \frac{D-2}{2(3D\!-\!10)(D\!-\!4)}
  \bigg\}
  + {\cal O}\left((D-4)^2\right)
\nonumber\\
\Omega_{\frac{D}{2}-1,D-4}
 &=&   ig_{\mu\rho}\frac{(4\pi)^\frac{D}{2}}{H^{D-2}}
           \frac{\Gamma\left(\frac{D}{2}\right)}{\Gamma\left(D\right)}
          \left[\frac32
             + \frac{31}{8}(D-4)
           \right]
    + {\cal O}\left((D-4)^2\right)
\label{Omega:D2-1:D-4}
\\
\Omega_{\frac{D}{2}-1,D-5}
 &=&   ig_{\mu\rho}\frac{(4\pi)^\frac{D}{2}}{H^{D-2}}
           \frac{\Gamma\left(\frac{D}{2}\right)}{\Gamma\left(D\right)}
          \left[\frac12
             + \frac{7}{8}(D-4)
           \right]
\nonumber\\
 &+&  i(a^2\delta_\mu^0\delta_\rho^0)\frac{(4\pi)^\frac{D}{2}}{H^{D-2}}
           \frac{\Gamma\left(\frac{D}{2}\right)}{\Gamma\left(D\right)}
          \left[-1
             +(D-4)\left(\frac{3}{2}\ln(a) - \frac72\right)
           \right]
    + {\cal O}\left((D-4)^2\right)
\qquad
\label{Omega:D2-1:D-5}
\end{eqnarray}
\begin{eqnarray}
\Omega_{\frac{D}{2}-2,D-2}
 &=&   ig_{\mu\rho}\frac{(4\pi)^\frac{D}{2}}{H^{D-2}}
           \frac{\Gamma\left(\frac{D}{2}\right)}{\Gamma\left(D\right)}
  \bigg\{\!-6(D\!-\!4) +(D\!-\!4)^2\left(1-\frac{\pi^2}{4}\right)
  \nonumber\\
  && \hskip 3cm
    -\, \frac{\Gamma\left(D\right)}
           {\Gamma\left(\frac{D}{2}\right)\Gamma\left(\frac{D}{2}\!-\!1\right)}
    \frac{D-2}{2(3D\!-\!10)(D\!-\!3)}
  \bigg\}
     + {\cal O}\left((D\!-\!4)^3\right)
\label{Omega:D2-2:D-2}
\\
\Omega_{\frac{D}{2}-2,D-3}
 &=&  \! ig_{\mu\rho}\frac{(4\pi)^\frac{D}{2}}{H^{D-2}}
           \frac{\Gamma\left(\frac{D}{2}\right)}{\Gamma\left(D\right)}
  \bigg\{\!\frac34+\frac{37}{16}(D\!-\!4)
     + (D\!-\!4)^2\left(\frac{185}{64}\!+\!\frac{3\pi^2}{32}\right)
  \bigg\}
     + {\cal O}\left((D\!-\!4)^3\right)
\quad
\label{Omega:D2-2:D-3}
\\
\Omega_{\frac{D}{2}-2,D-4} &=&
ig_{\mu\rho}\frac{(4\pi)^\frac{D}{2}}{H^{D-2}}
           \frac{\Gamma\left(\frac{D}{2}\right)}{\Gamma\left(D\right)}
  \bigg\{\!-\frac14(D\!-\!4)^2\bigg\}
 + ia^2\delta^0_\mu\delta^0_\rho\frac{(4\pi)^\frac{D}{2}}{H^{D-2}}
            \frac{\Gamma\left(\frac{D}{2}\right)}{\Gamma\left(D\right)}
  \bigg\{\frac12(D\!-\!4)^2\bigg\}
  \nonumber\\
  && \hskip 3cm
    +\,{\cal O}\left((D\!-\!4)^3\right)
\label{Omega:D2-2:D-4}
\end{eqnarray}
\begin{eqnarray}
\Omega_{\frac{D}{2}-3,D-2}
 &=&   ig_{\mu\rho}\frac{(4\pi)^\frac{D}{2}}{H^{D-2}}
           \frac{\Gamma\left(\frac{D}{2}\right)}{\Gamma\left(D\right)}
  \bigg\{\!-\frac{3}{D\!-\!4}-\frac{25}{4}
        + (D\!-\!4)\left(\!-\!\frac{73}{16}\!-\!\frac{3\pi^2}{8}\right)
  \nonumber\\
  &&      +\,   (D\!-\!4)^2\bigg[-\frac{27}{16}\ln^3(a)
                   + \frac{135}{32}\ln^2(a)
                   +\left(-\frac{33}{8} + \frac{27\pi^2}{16}\right)\ln(a)
\nonumber\\
&&\hskip 2cm
                   +\, \frac{1009}{192}\!-\!\frac{35\pi^2}{16}
                   -\frac92\zeta(3)\bigg]
    + {\cal O}\left((D\!-\!4)^3\right)
  \bigg\}
\label{Omega:D2-3:D-2}
\\
\Omega_{\frac{D}{2}-3,D-3}
 &=&   ig_{\mu\rho}\frac{(4\pi)^\frac{D}{2}}{H^{D-2}}
           \frac{\Gamma\left(\frac{D}{2}\right)}{\Gamma\left(D\right)}
  \bigg\{\frac12+\frac{11}{8}(D\!-\!4)
           + (D\!-\!4)^2\bigg(\frac{133}{96} + \frac{\pi^2}{16}\bigg)
  \bigg\}
\nonumber\\
 &+&  i(a^2\delta_\mu^0\delta_\rho^0)\frac{(4\pi)^\frac{D}{2}}{H^{D-2}}
           \frac{\Gamma\left(\frac{D}{2}\right)}{\Gamma\left(D\right)}
          \bigg[ - 1 + (D\!-\!4)\Big(\frac32\ln(a) - \frac{9}{2}\Big)
\label{Omega:D2-3:D-3}
\\
&& \hskip 2cm
            +\,(D\!-\!4)^2\bigg(-\frac98\ln^2(a) + \frac{27}{4}\ln(a)
              - \frac{67}{6} + \frac{3\pi^2}{8}\bigg)
           \bigg]
            + {\cal O}\left((D\!-\!4)^3\right)
\nonumber
\\
\Omega_{\frac{D}{2}-4,D-2}
 &=&   ig_{\mu\rho}\frac{(4\pi)^\frac{D}{2}}{H^{D-2}}
           \frac{\Gamma\left(\frac{D}{2}\right)}{\Gamma\left(D\right)}
  \bigg\{2 + 5(D\!-\!4)
         +\,(D\!-\!4)^2\bigg(\frac{31}{6} - \frac{\pi^2}{16}\bigg)
  \bigg\}
\nonumber\\
 &+&  i(a^2\delta_\mu^0\delta_\rho^0)\frac{(4\pi)^\frac{D}{2}}{H^{D-2}}
           \frac{\Gamma\left(\frac{D}{2}\right)}{\Gamma\left(D\right)}
          \Big[-4
             +(D-4)\Big(6\ln(a)-17\Big)
 \label{Omega:D2-4:D-2}
 \\
 && \hskip 2cm
           +\, (D\!-\!4)^2\bigg(-\frac92\ln^2(a) + \frac{51}{2}\ln(a)
              - \frac{253}{6} + \frac{3\pi^2}{2}\bigg)
           \Big]
    + {\cal O}\left((D\!-\!4)^3\right)
\nonumber
\,.
\end{eqnarray}
The following integrals are also useful,
\begin{eqnarray}
\Omega_{-1,D-2}
 &=&   ig_{\mu\rho}\frac{(4\pi)^\frac{D}{2}}{H^{D-2}}
           \frac{\Gamma\left(\frac{D}{2}\right)}{\Gamma\left(D\right)}
  \bigg\{-\frac{6}{D\!-\!4} - 11
        + (D\!-\!4)\left(-6-\frac{\pi^2}{2}\right)
\label{Omega:-1:D-2}\\
  &&      +\, (D\!-\!4)^2\left(-\ln^3(a) + \frac52\ln^2(a)
        + \Big(2+\frac{\pi^2}{2}\Big)\ln(a) - 2 - \frac{4\pi^2}{3}\right)
  \bigg\}
    + {\cal O}\left((D\!-\!4)^3\right)
\nonumber
\\
\Omega_{-1,D-3}
 &=&   ig_{\mu\rho}\frac{(4\pi)^\frac{D}{2}}{H^{D-2}}
           \frac{\Gamma\left(\frac{D}{2}\right)}{\Gamma\left(D\right)}
  \left[\frac12  + \frac76(D\!-\!4)
         + (D\!-\!4)^2\left(\frac56+\frac{\pi^2}{24}\right)
  \right]
\nonumber\\
  &+& i\delta_\mu^0\delta_\rho^0\frac{(4\pi)^\frac{D}{2}}{H^{D-2}}
           \frac{\Gamma\left(\frac{D}{2}\right)}{\Gamma\left(D\right)}
  \bigg[-1 + (D\!-\!4)\left(\ln(a)-\frac{11}{3}\right)
\nonumber\\
 &&\hskip 2cm
      + \, (D\!-\!4)^2\left(-\frac12\ln^2(a)+\frac{11}{3}\ln(a)
                      -\frac{56}{9}+\frac{\pi^2}{12}\right)
       \bigg]
    + {\cal O}\left((D\!-\!4)^3\right)
\label{Omega:-1:D-3}
\\
\Omega_{-2,D-2}
 &=&   ig_{\mu\rho}\frac{(4\pi)^\frac{D}{2}}{H^{D-2}}
           \frac{\Gamma\left(\frac{D}{2}\right)}{\Gamma\left(D\right)}
  \left[2  + \frac{11}{3}(D\!-\!4) + (D\!-\!4)^2\left(2+\frac{\pi^2}{3}\right)\right]
\nonumber\\
  &+& i\delta_\mu^0\delta_\rho^0\frac{(4\pi)^\frac{D}{2}}{H^{D-2}}
           \frac{\Gamma\left(\frac{D}{2}\right)}{\Gamma\left(D\right)}
  \bigg[-4 +(D\!-\!4)\left(4\ln(a)-\frac{38}{3}\right)
\nonumber\\
&& \hskip 2cm
  +\, (D\!-\!4)^2\left(-2\ln^2(a)+\frac{38}{3}\ln(a)-\frac{176}{9}+\frac{\pi^2}{3}\right)
  \bigg]
    + {\cal O}\left((D\!-\!4)^3\right)
\label{Omega:-2:D-2}
\,.
\end{eqnarray}

 Next class of the $\Omega$ integrals is of the form,
$\alpha = \frac{D}{2}-m,\beta = \frac{D}{2}-n$
$(\alpha+\beta=D-(n+m); m,n=1,2,3)$. Since the coefficient of these
integrals is not singular when $D=4$, actually
we need only the ${\cal O}((D-4)^0)$ contribution. The integrals are,
\begin{eqnarray}
\Omega_{\frac{D}{2}-1,\frac{D}{2}-1}
 &=&   ig_{\mu\rho}\frac{(4\pi)^\frac{D}{2}}{H^{D-2}}
          \frac{\Gamma\left(\frac{D}{2}\right)}{\Gamma\left(D\right)}
  \bigg\{  \left[-6
             + (D-4)\left(-5 - \frac{\pi^2}{4}\right)
           \right]
\label{Omega:D2-1:D2-1}
\\
  && \hskip 3.5cm
    -\,  \frac{\Gamma\left(D\right)}
    {\Gamma\left(\frac{D}{2}\right)\Gamma\left(\frac{D}{2}\!-\!1\right)}
    \frac{(D-2)}{2(D\!-\!3)(D\!-\!4)}
    + {\cal O}\left((D-4)^2\right)
  \bigg\}
\,.
\nonumber\\
\Omega_{\frac{D}{2}-1,\frac{D}{2}-2}
 &=&   ig_{\mu\rho}\frac{(4\pi)^\frac{D}{2}}{H^{D-2}}
           \frac{\Gamma\left(\frac{D}{2}\right)}{\Gamma\left(D\right)}
          \left[\frac32 + \frac{11}{4}(D-4)
           \right]
    + {\cal O}\left((D-4)^2\right)
\label{Omega:D2-1:D2-2}
\\
\Omega_{\frac{D}{2}-1,\frac{D}{2}-3}
 &=&   ig_{\mu\rho}\frac{(4\pi)^\frac{D}{2}}{H^{D-2}}
           \frac{\Gamma\left(\frac{D}{2}\right)}{\Gamma\left(D\right)}
          \left[\frac12 + \frac{2}{3}(D-4)
           \right]
\label{Omega:D2-1:D2-3}
\\
 &+& i(a^2\delta_\mu^0\delta_\rho^0)\frac{(4\pi)^\frac{D}{2}}{H^{D-2}}
           \frac{\Gamma\left(\frac{D}{2}\right)}{\Gamma\left(D\right)}
          \left[-1 + (D-4)\left(\ln(a)-\frac{8}{3}\right)
           \right]
    + {\cal O}\left((D-4)^2\right)
\nonumber
\\
\Omega_{\frac{D}{2}-2,\frac{D}{2}-2}
 &=& 0     + {\cal O}\left((D-4)^2\right)
\label{Omega:D2-2:D2-2}
\\
\Omega_{-1,\frac{D}{2}-1}
 &=&   ig_{\mu\rho}\frac{(4\pi)^\frac{D}{2}}{H^{D-2}}
           \frac{\Gamma\left(\frac{D}{2}\right)}{\Gamma\left(D\right)}
          \left[\frac12 + \frac{11}{24}(D-4)
           \right]
\label{Omega:-1:D2-1}
\\
 &+& i(a^2\delta_\mu^0\delta_\rho^0)\frac{(4\pi)^\frac{D}{2}}{H^{D-2}}
           \frac{\Gamma\left(\frac{D}{2}\right)}{\Gamma\left(D\right)}
          \left[-1 + (D-4)\left(\frac12\ln(a)-\frac{11}{6}\right)
           \right]
    + {\cal O}\left((D-4)^2\right)
\nonumber
\,.
\end{eqnarray}

\bigskip

\subsection*{B6. The radial integrals}
\label{The radial integrals}

 The radial integrals are of the form
\begin{equation}
 J_n = \int_0^1 dx x^{D-2} [\ln(1-x^2)]^n
\,
\end{equation}
and when evaluated result in
\begin{eqnarray}
 J_0 &=& \frac{1}{D-1}
 \label{J0}
\\
 J_1 &=& \frac{1}{D-1}\left[-\psi\left(\frac{D+1}{2}\right)+\psi(1)\right]
\nonumber\\
      &=& \frac{1}{D-1}\left[\left(\ln(4)-\frac 83\right)
       + (D-4)\left(\frac{20}{9}-\frac{\pi^2}{4}\right)
       +{\cal O}((D-4)^2)  \right]
 \label{J1}
\\
 J_2 &=& \frac{1}{D\!-\!1}
       \left[
            \frac{\pi^2}{6}
         + \left(\psi\left(\frac{D+1}{2}\right)-\psi(1)\right)^2
         - \psi^\prime\left(\frac{D+1}{2}\right)
       \right]
\nonumber\\
 &=& \frac{1}{D\!-\!1}
       \left[
          \frac{104}{9} - \frac{\pi^2}{3} - \frac{16}{3}\ln(4)
           + \ln^2(4)
           + {\cal O}\left(D-4\right)
        \right]
 \label{J2}
\\
 J_3 &=& \frac{1}{D\!-\!1}
       \left[
      -\frac{640}{9} + \frac{8\pi^2}{3}-\pi^2 \ln(4)
      + \frac{104}{3}\ln(4) - 8\ln^2(4)
      + \ln^3(4) + 12\zeta(3)
        \right]
   \nonumber\\
     && +\, {\cal O}\left(D-4\right)
\,.
 \label{J3}
\end{eqnarray}
%

\section*{\large \bf
           Appendix C: The finite two leg integral}
\label{The finite two leg integral}

  In this appendix we evaluate the two leg finite
  integral~(\ref{I:fin}). The relevant part of the integral is of the form,
\begin{eqnarray}
 {\cal I}_{\mu\rho}=\sum_{\pm\pm}(\pm)(\pm)\int d^4x'{a'}^4\int
 d^4x''{a''}^4\left(\frac{\partial}{\partial x^\mu}\frac{1}{y_{+\pm}}\right)
 \left(\frac{\partial}{\partial
 x^\rho}\frac{1}{y''_{+\pm}}\right)\ln\left(\frac{y'_{\pm\pm}}{4}\right)
\label{AppC:Imn}
  \end{eqnarray}
The integral can be naturally split into the spatial and timelike
parts, each of which we evaluate separately. The spatial part of
(\ref{AppC:Imn}) can be written as,
\begin{eqnarray}
 {\cal I}_{ij}=\frac{4\delta_{ij}}{3a^2H^4}\sum_{\pm\pm}(\pm)(\pm)
 \int \frac{d^4x'{a'}^3}{(\Delta x_{+\pm})^4}\int
\frac{ d^4x''{a''}^3}{(\Delta x''_{+\pm})^4}
 \Delta\vec x\cdot\Delta \vec x\,''
 \ln\left(\frac{a'a''H^2}{4}{\Delta x'}^2_{\pm\pm}\right)
\label{AppC:Iij}
  \end{eqnarray}
It is natural to introduce the coordinates $r=\|\vec x-\vec x'\|$,
$r''=\|\vec x-\vec x\,''\|$, in terms of which $\|\vec x\,'-\vec
x\,''\|^2=r^2+{r''}^2-2rr''\cos(\theta)$, where
 $\theta=\angle(\vec x-\vec x\,',\vec x-\vec x\,'')$. The angular
 integral over $\theta$ can be easily performed. The result is,
\begin{eqnarray}
 {\cal I}_{ij}&=&\frac{16\pi^2}{3}\frac{\delta_{ij}}{a^2H^4}
 \sum_{\pm\pm}(\pm)(\pm)
 \int_{\eta_0}^\eta d\eta' {a'}^3\int_{\eta_0}^{\eta'}d\eta'' {a''}^3
 \int_{-\infty}^{\infty}\frac{dr r}{(r^2-\Delta\eta^2_{+\pm})^2}
 \int_{0}^{\infty}
\frac{ dr'' r''}{({r''}^2-{\Delta\eta''}^2_{+\pm})^2} \nonumber\\
&\times& \Bigg\{
 \Big({r^2}+{r''}^2-{\Delta\eta'}^2\Big)
 \Big[\big(r+r''\big)^2-{\Delta\eta'}^2\Big]
 \Bigg(\ln\bigg[\frac{H^2}{4}\Big((r+r'')^2-{\Delta\eta'}^2\Big)
           \bigg]-1\Bigg)
\nonumber\\
&& -\,\frac12\Big[\big(r+r''\big)^2-{\Delta\eta'}^2\Big]^2
 \Bigg(\ln\bigg[\frac{H^2}{4}\Big((r+r'')^2-{\Delta\eta'}^2\Big)
          \bigg]-\frac12\Bigg)
\Bigg\}\,, \label{AppC:Iij:2}
  \end{eqnarray}
where we also used the symmetry of the resulting integrand under
the exchange $r\rightarrow -r$ to extend the limit of integration
of the $r$ integral from $[0,\infty)$ to $(-\infty,\infty)$.
Finally we made use of the fact that temporal integrals extend
from $\eta_0=-1/H$ to $\eta$, and that the symmetry allows us to
constrain the $\eta''$ integral up to $\eta'$, thus gaining a
factor 2. With this we have achieved an important time ordering,
$\eta''\leq\eta'\leq \eta$, which we will find useful below. Next
we split the integral~(\ref{AppC:Iij:2}) into two integrals ${\cal
I}_{ij}={\cal I}_{ij}^{(1)}+{\cal I}_{ij}^{(2)}$ as follows,
\begin{eqnarray}
 {\cal I}_{ij}^{(1)}&=&\frac{16\pi^2}{3}\frac{\delta_{ij}}{a^2H^4}
 \int_{\eta_0}^\eta d\eta' {a'}^3\int_{\eta_0}^{\eta'}d\eta'' {a''}^3
 \sum_{x''=\pm}(\pm)\int_{0}^{\infty}
\frac{ dr''r''}{({r''}^2-{\Delta\eta''}^2_{+\pm})^2} J^{(1)}
 \label{AppC:Iij:3:1}
\\
{\cal I}_{ij}^{(2)}&=&\frac{16\pi^2}{3}\frac{\delta_{ij}}{a^2H^4}
 \int_{\eta_0}^\eta d\eta' {a'}^3\int_{\eta_0}^{\eta'}d\eta'' {a''}^3
  \sum_{x''=\pm}(\pm)\int_{0}^{\infty}
\frac{dr''{r''}^2}{({r''}^2-{\Delta\eta''}^2_{+\pm})^2} J^{(2)}
\,,
\label{AppC:Iij:3:2}
  \end{eqnarray}
where
\begin{eqnarray}
 J^{(1)} \!\!\!&=&\!\!\!
  \sum_{x'=\pm}(\pm)\!\!\int_{-\infty}^{\infty}\!\frac{dr \, r}{(r^2\!-\!
    \Delta\eta^2_{+\pm})^2}
\bigg\{\!
 \Big[(r\!+\!r'')^2\!-\!{\Delta\eta'}^2_{\pm\pm}\Big]^2
 \bigg(\frac12\ln\!\bigg[\frac{H^2}{4}\Big((r\!+\!r'')^2\!-\!
     {\Delta\eta'}^2_{\pm\pm}\Big) \bigg]\!-\!\frac34\bigg)\!
\bigg\} \qquad\label{AppC:J(1)}
\\
J^{(2)}\!\!\! &=&\!\!
\!\sum_{x'=\pm}\!(\pm)\!\!\int_{-\infty}^{\infty}\!\frac{dr \,
r^2}{(r^2\!-\!\Delta\eta^2_{+\pm})^2} \Bigg\{
 \Big[(r\!+\!r'')^2\!-\!{\Delta\eta'}^2_{\pm\pm}\Big]
 \bigg(\!\!-\!2\ln\!\bigg[\frac{H^2}{4}\Big((r\!+\!r'')^2\!-\!
    {\Delta\eta'}^2_{\pm\pm}\Big) \bigg]\!+2\!\bigg)\!
\bigg\}.\qquad\; \label{AppC:J(2)}
\end{eqnarray}
Before we begin evaluation of these integrals, we recall the
$i\epsilon$ prescription,
\begin{eqnarray}
 \Delta x_{++}^2 &=& \|\Delta \vec x\|^2 - (|\Delta\eta|-i\epsilon)^2
        \rightarrow \|\Delta \vec x\|^2 - (\Delta\eta-i\epsilon)^2
 \nonumber\\
 \Delta x_{+-}^2 &=& \|\Delta \vec x\|^2 - (\Delta\eta+i\epsilon)^2
 \,,\qquad
 \Delta x_{-+}^2 = \|\Delta \vec x\|^2 - (\Delta\eta-i\epsilon)^2
\nonumber\\
 \Delta x_{--}^2 &=& \|\Delta \vec x\|^2 - (|\Delta\eta|+i\epsilon)^2
 \rightarrow \|\Delta \vec x\|^2 - (\Delta\eta+i\epsilon)^2\,,
\label{AppC:ieps prescription}
\end{eqnarray}
where the implication follows as a result of the time ordering
$\eta\geq\eta'\geq \eta''$ in Eqs.~(\ref{AppC:Iij:3:1}--\ref{AppC:Iij:3:2}).
Analogous prescriptions hold for the
other two intervals. Based on this, we arrive at the $i\epsilon$
prescriptions as given in Table~I. Note that $\Delta
\eta''$ and $\Delta\eta'$ have identical $i\epsilon$ prescriptions
and that they are completely specified by the {\it signature}  of the
$x''$ vertex. On the other hand, the $i\epsilon$ prescriptions of
$\Delta\eta$ are completely given in terms of the signature of
the $x'$ vertex.
%
%
\begin{table}\label{Table:one}
\vbox{\tabskip=0pt \offinterlineskip
\def\tablerule{\noalign{\hrule}}
\halign to390pt {\strut#& \vrule#\tabskip=1em plus2em&
\hfil#\hfil& \vrule#& 
\hfil#\hfil& \vrule#& 
\hfil#\hfil& \vrule#& 
\hfil#\hfil& \vrule#& 
\hfil#\hfil& \vrule#& \hfil#\hfil&
\vrule#\tabskip=0pt\cr
\tablerule
\omit&height4pt&\omit&&\omit&&\omit&&\omit&&\omit&&\omit
&\cr
&&  $x$ &&  $x'$ &&  $x''$ && $\Delta\eta$ && $\Delta\eta''$ && $\Delta\eta'$ 
& \cr
\omit&height4pt&\omit&&\omit&&\omit&&\omit&&\omit&&\omit&\cr
\tablerule
\omit&height2pt&\omit&&\omit&&\omit&&\omit&&\omit&&\omit&\cr
\tablerule
\omit&height2pt&\omit&&\omit&&\omit&&\omit&&\omit&&\omit&\cr
&& $+$ && $+$ && $+$ && $-i\epsilon$ && $-i\epsilon$ && $-i\epsilon$
& \cr
\omit&height2pt&\omit&&\omit&&\omit&&\omit&&\omit&&\omit&\cr
\tablerule
\omit&height2pt&\omit&&\omit&&\omit&&\omit&&\omit&&\omit&\cr
&& $+$ && $+$ && $-$ && $-i\epsilon$ && $+i\epsilon$ && $+i\epsilon$
& \cr
\omit&height2pt&\omit&&\omit&&\omit&&\omit&&\omit&&\omit&\cr
\tablerule
\omit&height2pt&\omit&&\omit&&\omit&&\omit&&\omit&&\omit&\cr
&& $+$ && $-$ && $+$ && $+i\epsilon$ && $-i\epsilon$ && $-i\epsilon$
& \cr
\omit&height2pt&\omit&&\omit&&\omit&&\omit&&\omit&&\omit&\cr
\tablerule
\omit&height2pt&\omit&&\omit&&\omit&&\omit&&\omit&&\omit&\cr
&& $+$ && $-$ && $-$ && $+i\epsilon$ && $+i\epsilon$ && $+i\epsilon$
& \cr
\omit&height2pt&\omit&&\omit&&\omit&&\omit&&\omit&&\omit&\cr
\tablerule
\omit&height2pt&\omit&&\omit&&\omit&&\omit&&\omit&&\omit&\cr
\tablerule}}

\caption{The $i\epsilon$ prescriptions of the intervals 
$\Delta \eta=\eta-\eta'$, $\Delta \eta''=\eta-\eta''$ and 
$\Delta \eta'=\eta'-\eta''$ as a function of the signature $\pm$ of the 
vertices $x$, $x'$ and $x''$.}

\end{table}

We shall now use this fact to perform the $r$
integrals~(\ref{AppC:J(1)}--\ref{AppC:J(2)}). The first integral
$J^{(1)}$ can be performed by making use of the Dirac identity,
\begin{eqnarray}
 \frac{r}{[r^2-(\Delta\eta-i\epsilon)^2]^2}-\frac{r}{[r^2-(\Delta\eta+i\epsilon)^2]^2}
 =\frac{1}{4\Delta\eta}2\pi i
  \Big[\frac{\partial}{\partial r}\delta(r+\Delta\eta)
  +\frac{\partial}{\partial r}\delta(r-\Delta\eta)\Big]
\,.
  \label{AppC:Dirac identity:1}
\end{eqnarray}
The derivative $\partial/\partial r$ can be moved by partial
integration to the integrand, such that the result of integration
of~(\ref{AppC:J(1)}) is,
\begin{eqnarray}
 J^{(1)} &=& - \frac{\pi i}{\Delta \eta}
 \Bigg\{
  (r''+\Delta\eta)\bigg[\Big((r''+\Delta\eta)^2-{\Delta\eta'}^2\Big)\bigg]
  \bigg[\ln\bigg(\frac{H^2}{4}\Big[(r''+\Delta\eta)^2-{\Delta\eta'}^2\Big]\bigg)-1\bigg]
\nonumber\\
 &&\hskip 1cm +\,(r''-\Delta\eta)\bigg[\Big((r''-\Delta\eta)^2-{\Delta\eta'}^2\Big)\bigg]
  \bigg[\ln\bigg(\frac{H^2}{4}\Big[(r''-\Delta\eta)^2-{\Delta\eta'}^2\Big]\bigg)-1\bigg]
 \Bigg\}\,.\qquad
 \label{AppC:J(1):2}
\end{eqnarray}
Analogously, to evaluate $J^{(2)}$ we can use the following Dirac
identity,
\begin{eqnarray}
\frac{r^2}{[r^2 - (\Delta\eta-i\epsilon)^2]^2}-\frac{r^2}{[r^2 -
(\Delta\eta+i\epsilon)^2]^2}
 &=&\frac{\pi i}{2}
  \bigg\{\Big[\frac{\partial}{\partial r}\delta(r-\Delta\eta)
  -\frac{\partial}{\partial r}\delta(r+\Delta\eta)\Big]
\nonumber\\
&&\hskip 0.5cm
  -\,\frac{1}{\Delta\eta}
    \Big[\delta(r-\Delta\eta)+\delta(r+\Delta\eta)\Big]
  \bigg\}
  \,,
  \label{AppC:Dirac identity:2}
\end{eqnarray}
with whose help one can evaluate $J^{(2)}$. The result is,
\begin{eqnarray}
 J^{(2)} &=& \frac{\pi i}{2}
 \Bigg\{\!
  4(r''\!+\!\Delta\eta)
  \ln\bigg(\!\frac{H^2}{4}\Big[(r''\!+\!\Delta\eta)^2 \!-\! 
      {\Delta\eta'}^2\Big]\bigg) \!-\! 4(r'' \!-\! \Delta\eta)
  \ln\bigg(\!\frac{H^2}{4}\Big[(r'' \!-\! \Delta\eta)^2 \!-\!
      {\Delta\eta'}^2\Big]\!\bigg) \nonumber\\
 &&+\,\frac{2}{\Delta\eta}\Big[(r''+\Delta\eta)^2-{\Delta\eta'}^2\Big]
  \bigg[\ln\bigg(\frac{H^2}{4}\Big[(r''+\Delta\eta)^2-{\Delta\eta'}^2\Big]\bigg)-1\bigg]
\nonumber\\
&&+\,\frac{2}{\Delta\eta}\Big[(r''-\Delta\eta)^2-{\Delta\eta'}^2\Big]
  \bigg[\ln\bigg(\frac{H^2}{4}\Big[(r''-\Delta\eta)^2-{\Delta\eta'}^2\Big]\bigg)-1\bigg]
   \Bigg\}\,.\qquad
 \label{AppC:J(2):2}
\end{eqnarray}
The symmetry of the integrals $J^{(1)}$ and $J^{(2)}$ under
$r''\rightarrow -r''$ allows one to extend the $r''$ integration
to $-\infty$. Upon inserting these two integrals into
Eqs.~(\ref{AppC:Iij:3:1}--\ref{AppC:Iij:3:2}) and summing them, we
obtain,
\begin{eqnarray}
 {\cal I}_{ij} &=& -i
 \frac{16\pi^3}{3}\frac{\delta_{ij}}{a^2H^4}
 \int_{\eta_0}^\eta d\eta' {a'}^3\int_{\eta_0}^{\eta'} d\eta'' {a''}^3
\sum_{x''=\pm}(\pm)\int_{-\infty}^\infty\frac{dr''r''}{({r''}^2-{\Delta\eta''}_{+\pm}^2)^2}
\nonumber\\
&\times& \Bigg\{
\Big[\Delta\eta^2-{r''}^2-{\Delta\eta'}_{+\pm}^2\Big]
\ln\bigg(\frac{H^2}{4}\Big[(r''+\Delta\eta)^2-{\Delta\eta'}_{+\pm}^2\Big]\bigg)
 -\Big[(r''+\Delta\eta)^2-{\Delta\eta'}_{+\pm}^2\Big]\Bigg\}
 \,.\qquad
\label{AppC:Iij:3}
\end{eqnarray}
It is convenient to rewrite this expression in terms of two radial
integrals as follows,
\begin{eqnarray}
 {\cal I}_{ij} &=& -i
 \frac{16\pi^3}{3}\frac{\delta_{ij}}{a^2H^4}
 \int_{\eta_0}^\eta d\eta' {a'}^3\int_{\eta_0}^{\eta'} d\eta'' {a''}^3
\Big[L^{(1)}+L^{(2)}\Big] \,, \label{AppC:Iij:4}
\end{eqnarray}
where
\begin{eqnarray}
 L^{(1)} \!&=& \!\sum_{x''=\pm}(\pm)
 \int_{-\infty}^\infty\!dr''\frac{r''}{({r''}^2-{\Delta\eta''}_{+\pm}^2)^2}
 \Big[(r''\!+\!\Delta\eta)^2\!-{\Delta\eta'}_{+\pm}^2\Big]
\nonumber\\
&&\hskip6cm\times\,
\bigg\{\ln\bigg(\frac{H^2}{4}\Big[(r''\!+\!\Delta\eta)^2\!-\!{\Delta\eta'}_{+\pm}^2\Big]\bigg)
 \!-\!1\bigg\}\qquad
\label{AppC:L(1)}
\\
L^{(2)} \!&=&\!-2\sum_{x''=\pm}(\pm) \int_{-\infty}^\infty
dr''\frac{{r''}^2}{({r''}^2-{\Delta\eta''}_{+\pm}^2)^2}
\Big[r''+\Delta\eta\Big]
\ln\bigg(\frac{H^2}{4}\Big[(r''+\Delta\eta)^2-{\Delta\eta'}_{+\pm}^2\Big]\bigg)
\,. \qquad\label{AppC:L(2)}
\end{eqnarray}
$L^{(1)}$ can be partially integrated to give,
\begin{eqnarray}
 L^{(1)} \!&=& \!\sum_{x''=\pm}(\pm)
 \int_{-\infty}^\infty\!\frac{dr''}{{r''}^2-{\Delta\eta''}_{+\pm}^2}
 \Big[r''\!+{\Delta\eta}\Big]
\ln\bigg(\frac{H^2}{4}\Big[(r''\!+\!\Delta\eta)^2\!-\!{\Delta\eta'}_{+\pm}^2\Big]\bigg)
\,. \qquad \label{AppC:L(1):2}
\end{eqnarray}
The first term in $L^{(2)}$ can be rewritten as
\begin{equation}
-2\frac{{r''}^2}{({r''}^2-{\Delta\eta''}^2)^2} =
-\frac{1}{{r''}^2-{\Delta\eta''}^2}
-\frac12\frac{1}{({r''}-{\Delta\eta''})^2}
-\frac12\frac{1}{({r''}+{\Delta\eta''})^2} \,,
\end{equation}
resulting in four simple integrals,
$L^{(1)}+L^{(2)}=L_A+L_B+L_C+L_D$, where
\begin{eqnarray}
 L_A &=& -\frac12\sum_{x''=\pm}(\pm)
 \int_{-\infty}^\infty\frac{dr''}{r''-\Delta\eta''}_{+\pm}
\ln\bigg(\frac{H^2}{4}\Big[(r''+\Delta\eta)^2-{\Delta\eta'}_{+\pm}^2\Big]\bigg)
\label{AppC:LA}\\
L_B &=& -\frac12\sum_{x''=\pm}(\pm)
 \int_{-\infty}^\infty\frac{dr''}{r''+\Delta\eta''_{+\pm}}
\ln\bigg(\frac{H^2}{4}\Big[(r''+\Delta\eta)^2-{\Delta\eta'_{+\pm}}^2\Big]\bigg)
\label{AppC:LB}\\
L_C &=& -\frac12\sum_{x''=\pm}(\pm)
 \int_{-\infty}^\infty\frac{dr''(\Delta\eta+\Delta\eta'')}{(r''-\Delta\eta''_{+\pm})^2}
\ln\bigg(\frac{H^2}{4}\Big[(r''+\Delta\eta)^2-{\Delta\eta'}_{+\pm}^2\Big]\bigg)
\label{AppC:LC}\\
L_D &=& \frac12\sum_{x''=\pm}(\pm)
 \int_{-\infty}^\infty\frac{dr''\Delta\eta'}{(r''+\Delta\eta''_{+\pm})^2}
\ln\bigg(\frac{H^2}{4}\Big[(r''+\Delta\eta)^2-{\Delta\eta'}_{+\pm}^2\Big]\bigg)
\label{AppC:LD}
\end{eqnarray}
where we made use of
$r''+\Delta\eta=r''-\Delta\eta''+\Delta\eta+\Delta\eta''
=r''+\Delta\eta''-\Delta\eta'$. All of these integrals are simple
to evaluate by contour integration. For example, upon writing the
$i\epsilon$ prescription explicitly ({\it e.g.} from
Table~I), $L_A$ can be written as
 \begin{eqnarray}
 L_A &=& -\frac12
 \int_{-\infty}^\infty dr''\Bigg\{
 \frac{\ln\Big[r''+\Delta\eta-\Delta\eta'+i\epsilon\Big]}{r''-\Delta\eta''+i\epsilon}
+\frac{\ln\Big[r''+\Delta\eta''-i\epsilon\Big]}{r''-\Delta\eta''+i\epsilon}
\nonumber\\
&&\hskip2.3cm
-\,\frac{\ln\Big[r''+\Delta\eta-\Delta\eta'-i\epsilon\Big]}{r''-\Delta\eta''-i\epsilon}
-\frac{\ln\Big[r''+\Delta\eta''+i\epsilon\Big]}{r''-\Delta\eta''-i\epsilon}
\Bigg\} \,. \label{AppC:LA:2}
\end{eqnarray}
This is straightforward to evaluate,
 \begin{eqnarray}
 L_A = 2\pi i \Big[\ln(2\Delta\eta'')\Big]
\,. \label{AppC:LA:fin}
\end{eqnarray}
Similarly, the remaining integrals in
Eqs.~(\ref{AppC:LB}--\ref{AppC:LD}) evaluate to,
 \begin{eqnarray}
 L_B &=& 2\pi i \Big[-\ln(2\Delta\eta')\Big]
\label{AppC:LB:fin}\\
L_C &=& 2\pi i
\Big[\frac12+\frac12\frac{\Delta\eta}{\Delta\eta''}\Big]
\label{AppC:LC:fin}\\
L_D &=& 2\pi i \Big[-\frac12\Big]
 \,. \label{AppC:LD:fin}
\end{eqnarray}
Upon combining these integrals, we can write the spatial integral
${\cal I}_{ij}$~(\ref{AppC:Iij:4}) as follows,
\begin{eqnarray}
 {\cal I}_{ij} &=& -
 \frac{32\pi^4}{3}\frac{\delta_{ij}}{a^2H^4}
 \int_{\eta_0}^\eta d\eta' {a'}^3\int_{\eta_0}^{\eta'} d\eta'' {a''}^3
\bigg[\ln\bigg(\frac{\Delta\eta'}{\Delta\eta''}\bigg)
-\frac12\frac{\Delta\eta}{\Delta\eta''}\bigg]
\,,
\label{AppC:Iij:5}
\end{eqnarray}
Upon changing the variables to $v=a'/a$ and $w=a''/a$, this
reduces to
\begin{eqnarray}
 {\cal I}_{ij} &=& -
 \frac{32\pi^4}{3}\frac{a^2\delta_{ij}}{H^6}
 \int_{1/a}^a dv v \int_{1/a}^{v} d w w
\bigg[\ln\bigg(1-\frac{w}{v}\bigg)-\ln(1-w)
-\frac12\frac{w(1-v)}{v(1-w)}\bigg]
 \,, \label{AppC:Iij:6}
\end{eqnarray}
These integrals are straightforward to do. The result is,
\begin{eqnarray}
 {\cal I}_{ij} &=& 
 \frac{\pi^4}{H^6}a^2\delta_{ij}
\bigg[1-\frac{8}{3a^3}+\frac{5}{3a^4}-\frac{4}{3a^4}\ln(a)\bigg]
 \,. \label{AppC:Iij:fin}
\end{eqnarray}
This is the final result for the spatial part of the finite two
vertex integral.

 The de Sitter invariant form of the integrand in
Eq.~(\ref{AppC:Imn}) indicates that the leading order contribution
to the integral ${\cal I}_{\mu\nu}$ should be de Sitter invariant,
such that when combined with the $00$ component of ${\cal
I}_{\mu\nu}$ one expects to get, ${\cal
I}_{\mu\nu}=\frac{\pi^4}{H^6}g_{\mu\nu}$ plus terms suppressed by
powers of the scale factor $a$. A detailed evaluation confirms this
expectation. Albeit more cumbersome, the evaluation of ${\cal
I}_{00}$ closely resembles the evaluation of ${\cal I}_{ij}$, and
here we present only the major steps.

 From Eq.~(\ref{AppC:Imn}) it follows
\begin{eqnarray}
 {\cal I}_{00}&=&\frac{1}{H^2}\sum_{\pm\pm}(\pm)(\pm)
 \int d^4x'{a'}^3\bigg[\frac{1}{\Delta x^2_{+\pm}}
          +\frac{2\eta\Delta\eta}{(\Delta x_{+\pm})^4}\bigg]
\int d^4x''{a''}^3\bigg[\frac{1}{\Delta {x''}^2_{+\pm}}
          +\frac{2\eta\Delta\eta''}{(\Delta x''_{+\pm})^4}\bigg]
          \nonumber\\
&&\times\ln\left(\frac{a'a''H^2}{4}{\Delta x'}^2_{\pm\pm}\right)
\,. \label{AppC:I00}
  \end{eqnarray}
Upon performing the $r$ integral one obtains,
\begin{eqnarray}
 {\cal I}_{00}&=&i\frac{16\pi^3}{H^2}\sum_{x''=\pm}(\pm)
 \int_{\eta_0}^\eta d\eta'{a'}^3\int_{\eta_0}^{\eta'}d\eta''{a''}^3
 \int_{-\infty}^\infty dr''
 \bigg[\frac{r''}{{r''}^2-{\Delta\eta''}^2_{+\pm}}
 +\frac{2\eta r''\Delta\eta''}{({r''}^2-\Delta{\eta''}^2_{+\pm})^2}
 \bigg]
 \nonumber\\
&&\times\Bigg\{
-\frac12\Big[({r''}+\Delta\eta)^2-\Delta{\eta'}^2_{+\pm}\Big]
\bigg[\ln\left(\frac{a'a''H^2}{4}\right) +\ln\left(
({r''}+\Delta\eta)^2-\Delta{\eta'}^2_{+\pm}\right)-1\bigg]
\nonumber\\
&&\hskip 0.5cm
-\,\eta\Big[{r''}+\Delta\eta\Big]
\bigg[\ln\left(\frac{a'a''H^2}{4}\right) +\ln\left(
({r''}+\Delta\eta)^2-\Delta{\eta'}^2_{+\pm}\right)\bigg] \Bigg\}
 \,. \label{AppC:I00:2}
  \end{eqnarray}
Performing the $r''$ integral requires more work,
\begin{eqnarray}
 {\cal I}_{00}&=&-\frac{32\pi^4}{H^2}
 \int_{\eta_0}^\eta d\eta'{a'}^3\int_{\eta_0}^{\eta'}d\eta''{a''}^3
\bigg\{\bigg[\Delta\eta\Delta\eta''\ln\Big[a'a''H^2(\Delta\eta'')^2\Big]
-\Delta\eta(\Delta\eta''+\Delta\eta')\bigg] \nonumber\\
&& \hskip 5cm
+\,\eta\bigg[(\Delta\eta+\Delta\eta'')\ln\Big[a'a''H^2(\Delta\eta'')^2\Big]
             -2\Delta\eta'\bigg]
\nonumber\\
&&\hskip 5cm
+\,\eta^2\bigg[\ln\Big[a'a''H^2\Delta\eta''\Delta\eta'\Big]+1
+\frac12\frac{\Delta\eta}{\Delta\eta''}\bigg] \bigg\}
 \,. \label{AppC:I00:3}
  \end{eqnarray}
By making the substitutions $v=a'/a$ and $w=a''/a$, this can be
reduced to a set of relatively simple integrals. The result of
integration is,
\begin{eqnarray}
 {\cal I}_{00}\!&=&\!\frac{\pi^4a^2}{H^6}
\bigg\{\!\!-\!\Big(1\!-\!\frac{1}{a}\Big)\Big(1+\frac{17}{a}-\frac{23}{a^2}+\frac{17}{a^3}\Big)
+\frac{4}{a^2}\Big(4\!-\!\frac{1}{a^2}\Big)\ln(a)
-\frac{32}{a^2}\Big(1\!-\!\frac{1}{a}\Big)^2\ln\Big(1\!-\!\frac{1}{a}\Big)
\bigg\}
 \,.\qquad\; \label{AppC:I00:fin}
  \end{eqnarray}
When the two parts ${\cal I}_{ij}$ in Eq.~(\ref{AppC:Iij:fin}) and
${\cal I}_{00}$ in Eq.~(\ref{AppC:I00:fin}) are combined, the
leading order contribution is de Sitter invariant,
\begin{eqnarray}
 {\cal I}_{\mu\nu}&=&\frac{\pi^4}{H^6}g_{\mu\nu}
 \, \label{AppC:Imn:fin}
  \end{eqnarray}
plus terms suppressed as powers of scale factors, in accordance
with the expectation.

 From this result we see that the finite integral ${\cal I}_{\rm
fin}$ does not contribute a leading logarithm to the two loop
photon field strength bilinear.


\begin{thebibliography}{999}

\bibitem{Prokopec:2006ue}
  T.~Prokopec, N.~C.~Tsamis and R.~P.~Woodard,
  ``Two loop scalar bilinears for inflationary SQED,''
  Class.\ Quant.\ Grav.\  {\bf 24} (2007) 201
  [arXiv:gr-qc/0607094].

\bibitem{Prokopec:2007ak}
  T.~Prokopec, N.~C.~Tsamis and R.~P.~Woodard,
  ``Stochastic Inflationary Scalar Electrodynamics,''
  arXiv:0707.0847 [gr-qc], Annals Phys.,\ in press (2007).

\bibitem{Starobinsky:1986fx}
  A.~A.~Starobinsky,
``Stochastic de Sitter (inflationary) stage in the early universe,''
{\it  In De Vega, H.J. (Ed.), Sanchez, N. (Ed.):
Field Theory, Quantum Gravity and Strings, 107-126}.

\bibitem{Sasaki:1987gy}
  M.~Sasaki, Y.~Nambu and K.~i.~Nakao,
  ``Classical Behavior Of A Scalar Field In The Inflationary Universe,''
  Nucl.\ Phys.\  B {\bf 308} (1988) 868.

\bibitem{Starobinsky:1994bd}
  A.~A.~Starobinsky and J.~Yokoyama,
  ``Equilibrium state of a selfinteracting scalar field in the De Sitter
  background,''
  Phys.\ Rev.\  D {\bf 50} (1994) 6357
  [arXiv:astro-ph/9407016].

\bibitem{Martin:2005ir}
  J.~Martin and M.~Musso,
  ``Solving stochastic inflation for arbitrary potentials,''
  Phys.\ Rev.\  D {\bf 73} (2006) 043516
  [arXiv:hep-th/0511214].

\bibitem{Nambu:1988je}
  Y.~Nambu and M.~Sasaki,
  ``Stochastic Approach To Chaotic Inflation And The Distribution Of
  Universes,''
  Phys.\ Lett.\  B {\bf 219} (1989) 240.

\bibitem{Linde:1993xx}
  A.~D.~Linde, D.~A.~Linde and A.~Mezhlumian,
  ``From the Big Bang theory to the theory of a stationary universe,''
  Phys.\ Rev.\  D {\bf 49} (1994) 1783
  [arXiv:gr-qc/9306035].

\bibitem{Morikawa:1989xz}
  M.~Morikawa,
  ``Dissipation and fluctuation of quantum fields in expanding universes,''
  Phys.\ Rev.\  D {\bf 42} (1990) 1027.

\bibitem{Habib:1992ci}
  S.~Habib,
  ``Stochastic inflation: The Quantum phase space approach,''
  Phys.\ Rev.\  D {\bf 46} (1992) 2408
  [arXiv:gr-qc/9208006].

\bibitem{Rigopoulos:2005ae}
  G.~I.~Rigopoulos, E.~P.~S.~Shellard and B.~J.~W.~van Tent,
  ``Large non-Gaussianity in multiple-field inflation,''
  Phys.\ Rev.\  D {\bf 73} (2006) 083522
  [arXiv:astro-ph/0506704].

\bibitem{Rigopoulos:2005xx}
  G.~I.~Rigopoulos, E.~P.~S.~Shellard and B.~J.~W.~van Tent,
  ``Non-linear perturbations in multiple-field inflation,''
  Phys.\ Rev.\  D {\bf 73} (2006) 083521
  [arXiv:astro-ph/0504508].

\bibitem{Rigopoulos:2004ba}
  G.~I.~Rigopoulos, E.~P.~S.~Shellard and B.~J.~W.~van Tent,
  ``A simple route to non-Gaussianity in inflation,''
  Phys.\ Rev.\  D {\bf 72} (2005) 083507
  [arXiv:astro-ph/0410486].

\bibitem{Rigopoulos:2004gr}
  G.~I.~Rigopoulos and E.~P.~S.~Shellard,
  ``Non-linear inflationary perturbations,''
  JCAP {\bf 0510} (2005) 006
  [arXiv:astro-ph/0405185].

\bibitem{Riotto:2008mv}
  A.~Riotto and M.~S.~Sloth,
  ``On Resumming Inflationary Perturbations beyond One-loop,''
  arXiv:0801.1845 [hep-ph].

\bibitem{Rigopoulos:2008kt}
  K.~Enqvist, S.~Nurmi, D.~Podolsky and G.~I.~Rigopoulos,
  ``On the divergences of inflationary superhorizon perturbation,''
  [arXiv:0802.0395].

\bibitem{Tsamis:1996qq}
  N.~C.~Tsamis and R.~P.~Woodard,
  ``Quantum Gravity Slows Inflation,''
  Nucl.\ Phys.\  B {\bf 474}, 235 (1996)
  [arXiv:hep-ph/9602315].

\bibitem{Tsamis:1996qm}
  N.~C.~Tsamis and R.~P.~Woodard,
  ``The quantum gravitational back-reaction on inflation,''
  Annals Phys.\  {\bf 253} (1997) 1
  [arXiv:hep-ph/9602316].


\bibitem{Tsamis:2005hd}
  N.~C.~Tsamis and R.~P.~Woodard,
  ``Stochastic quantum gravitational inflation,''
  Nucl.\ Phys.\  B {\bf 724} (2005) 295
  [arXiv:gr-qc/0505115].

\bibitem{Davis:2000zp}
  A.~C.~Davis, K.~Dimopoulos, T.~Prokopec and O.~Tornkvist,
  ``Primordial spectrum of gauge fields from inflation,''
  Phys.\ Lett.\  B {\bf 501} (2001) 165
  [Phys.\ Rev.\ Focus {\bf 10} (2002) STORY9]
  [arXiv:astro-ph/0007214].

\bibitem{Prokopec:2002jn}
  T.~Prokopec, O.~T\"ornkvist and R.~P.~Woodard,
  ``Photon mass from inflation,''
  Phys.\ Rev.\ Lett.\  {\bf 89} (2002) 101301
  [arXiv:astro-ph/0205331].

\bibitem{Prokopec:2002uw}
  T.~Prokopec, O.~T\"ornkvist and R.~P.~Woodard,
  ``One loop vacuum polarization in a locally de Sitter background,''
  Annals Phys.\  {\bf 303} (2003) 251
  [arXiv:gr-qc/0205130];

\bibitem{Prokopec:2003bx}
  T.~Prokopec and R.~P.~Woodard,
  ``Vacuum polarization and photon mass in inflation,''
  Am.\ J.\ Phys.\  {\bf 72} (2004) 60
  [arXiv:astro-ph/0303358];

\bibitem{Prokopec:2003iu}
  T.~Prokopec and R.~P.~Woodard,
  ``Dynamics of super-horizon photons during inflation with vacuum
  polarization,''
  Annals Phys.\  {\bf 312} (2004) 1
  [arXiv:gr-qc/0310056].

\bibitem{Prokopec:2003tm}
  T.~Prokopec and E.~Puchwein,
  ``Photon mass generation during inflation: de Sitter invariant case,''
  JCAP {\bf 0404} (2004) 007
  [arXiv:astro-ph/0312274].

\bibitem{Dimopoulos:2001wx}
  K.~Dimopoulos, T.~Prokopec, O.~Tornkvist and A.~C.~Davis,
  ``Natural magnetogenesis from inflation,''
  Phys.\ Rev.\  D {\bf 65} (2002) 063505
  [arXiv:astro-ph/0108093].


\bibitem{ProkopecPuchwein:2004}
  T.~Prokopec and E.~Puchwein,
  ``Nearly minimal magnetogenesis,''
  Phys.\ Rev.\ D {\bf 70} (2004) 043004
  [arXiv:astro-ph/0403335].


\bibitem{Prokopec:2003qd}
  T.~Prokopec and R.~P.~Woodard,
  ``Production of massless fermions during inflation,''
  JHEP {\bf 0310} (2003) 059
  [arXiv:astro-ph/0309593].

\bibitem{Garbrecht:2006jm}
  B.~Garbrecht and T.~Prokopec,
  ``Fermion mass generation in de Sitter space,''
  Phys.\ Rev.\  D {\bf 73} (2006) 064036
  [arXiv:gr-qc/0602011].

\bibitem{Miao:2006pn}
  S.~P.~Miao and R.~P.~Woodard,
  ``Leading log solution for inflationary Yukawa,''
  Phys.\ Rev.\  D {\bf 74} (2006) 044019
  [arXiv:gr-qc/0602110].

\bibitem{Duffy:2005ue}
  L.~D.~Duffy and R.~P.~Woodard,
  ``Yukawa scalar self-mass on a conformally flat background,''
  Phys.\ Rev.\  D {\bf 72} (2005) 024023
  [arXiv:hep-ph/0505156].

\bibitem{Onemli:2002hr}
  V.~K.~Onemli and R.~P.~Woodard,
  ``Super-acceleration from massless, minimally coupled phi**4,''
  Class.\ Quant.\  Grav. {\bf 19} (2002) 4607
  [arXiv:gr-qc/0204065].

\bibitem{Onemli:2004mb}
  V.~K.~Onemli and R.~P.~Woodard,
  ``Quantum effects can render $w < -1$ on cosmological scales,''
  Phys.\ Rev.\  D {\bf 70} (2004) 107301
  [arXiv:gr-qc/0406098].

\bibitem{Wise} 
  S.~D.~H.~Hsu, A.~Jenkins and M.~B.~Wise,
  ``Gradient Instability for $w < -1$,''
  Phys.\ Lett.\ B {\bf 597} (2004) 270
  [arXiv:astro-ph/0406043].

\bibitem{BOW}
  T.~Brunier, V.~K.~Onemli and R.~P.~Woodard,
  ``Two loop scalar self-mass during inflation,''
  Class.\ Quant.\ Grav.\ {\bf 22} (2005) 59
  [arXiv:gr-qc/0408080].

\bibitem{KO}
  E.~O.~Kahya and V.~K.~Onemli,
  ``Quantum Stability of a $w < -1$ Phase of Cosmic Acceleration,''
  Phys.\ Rev.\ D {\bf 76} (2007) 043512
  [arXiv:gr-qc/0612026].

\bibitem{Miao:2005am}
  S.~P.~Miao and R.~P.~Woodard,
  ``The fermion self-energy during inflation,''
  Class.\ Quant.\ Grav.\  {\bf 23} (2006) 1721
  [arXiv:gr-qc/0511140].

\bibitem{Miao:2006gj}
  S.~P.~Miao and R.~P.~Woodard,
  ``Gravitons enhance fermions during inflation,''
  Phys.\ Rev.\  D {\bf 74} (2006) 024021
  [arXiv:gr-qc/0603135].

\bibitem{Kahya:2007bc}
  E.~O.~Kahya and R.~P.~Woodard,
  ``Quantum Gravity Corrections to the One Loop Scalar Self-Mass during
  Inflation,''
  arXiv:0709.0536 [gr-qc].

\bibitem{KW}
  E.~O.~Kahya and R.~P.~Woodard,
  ``Scalar Field Equations from Quantum Gravity during Inflation,''
  arXiv:0710.5282 [gr-qc].
  

\bibitem{Abramo:2001dc}
  L.~R.~Abramo and R.~P.~Woodard,
  ``No one loop back-reaction in chaotic inflation,''
  Phys.\ Rev.\  D {\bf 65} (2002) 063515
  [arXiv:astro-ph/0109272].

\bibitem{GB} 
  G.~Geshnizjani and R.~Brandenberger,
  ``Back reaction and local cosmological expansion rate,''
  Phys.\ Rev.\ D {\bf 66} (2002) 123507
  [arXiv:gr-qc/0204074].

\bibitem{Weinberg:2005vy}
  S.~Weinberg,
  ``Quantum contributions to cosmological correlations,''
  Phys.\ Rev.\  D {\bf 72} (2005) 043514
  [arXiv:hep-th/0506236].

\bibitem{Weinberg:2006ac}
  S.~Weinberg,
  ``Quantum contributions to cosmological correlations. II: Can these
  corrections become large?,''
  Phys.\ Rev.\  D {\bf 74} (2006) 023508
  [arXiv:hep-th/0605244].

\bibitem{Boyanovsky:2005px}
  D.~Boyanovsky, H.~J.~de Vega and N.~G.~Sanchez,
  ``Quantum corrections to the inflaton potential and the power spectra  from
  Phys.\ Rev.\  D {\bf 72} (2005) 103006
  [arXiv:astro-ph/0507596].

\bibitem{LU}
  B.~Losic and W.~G.~Unruh,
  ``On leading order gravitational backreactions in de Sitter spacetime,''
  Phys.\ Rev.\ D {\bf 74} (2006) 023511
  [arXiv:gr-qc/0604122]

\bibitem{WNF} 
  C.~H.~Wu, K.~N.~Ng and L.~H.~Ford, 
  ``Possible constraints on the Duration of Inflationary Expansion
    from Quantum Stress Tensor Fluctuations,''
  Phys.\ Rev.\ D {\bf 75} (2007) 103502
  [arXiv:gr-qc/0608002].

\bibitem{Sloth:2006nu}
  M.~S.~Sloth,
  ``On the one loop corrections to inflation. II: The consistency relation,''
  Nucl.\ Phys.\  B {\bf 775} (2007) 78
  [arXiv:hep-th/0612138].

\bibitem{Bilandzic:2007nb}
  A.~Biland\v{z}i\'c and T.~Prokopec,
  ``Quantum radiative corrections to slow-roll inflation,''
  arXiv:\-0704.1905 [astro-ph].

\bibitem{Garriga:2007zk}
  J.~Garriga and T.~Tanaka,
  ``Can infrared gravitons screen Lambda?,''
  Phys. Rev.\ D {\bf 77} (2008) 024021
  [arXiv:0706.0295 [hep-th]].

\bibitem{van der Meulen:2007ah}
  M.~van der Meulen and J.~Smit,
  ``Classical approximation to quantum cosmological correlations,''
  arXiv:0707.0842 [hep-th].

\bibitem{TW}
  N.~C.~Tsamis and R.~P.~Woodard,
  ``Reply to `Can infrared gravitons screen Lambda?',''
  arXiv:0708.2004 [hep-th].

\bibitem{JS} 
  J.~Schwinger,
  ``Brownian motion of a quantum oscillator,''
  J.\ Math.\ Phys.\ {\bf 2} (1961) 407.

\bibitem{KTM} 
  K.~T.~Mahanthappa, 
  ``Multiple production of photons in quantum electrodynamics,''
  Phys.\ Rev.\ {\bf 126} (1962) 329.

\bibitem{BM} 
  P.~M.~Bakshi and K.~T.~Mahanthappa, 
  ``Expectation value formalism in quantum field theory. 1,''
  J.\ Math.\ Phys. {\bf 4} (1963) 1; 
  ``Expectation value formalism in quantum field theory. 2,''
  J.\ Math.\ Phys. {\bf 4} (1963) 12.

\bibitem{LVK} 
  L.~V.~Keldysh, 
  ``Diagram technique for nonequilibrium processes,''
  Sov.\ Phys.\ JETP\ {\bf 20} (1965) 1018.

\bibitem{Chernikov:1968zm}
  N.~A.~Chernikov and E.~A.~Tagirov,
  ``Quantum theory of scalar fields in de Sitter space-time,''
  Annales Poincare Phys.\ Theor.\  A {\bf 9} (1968) 109.

\bibitem{Allen:1987tz}
  B.~Allen and A.~Folacci,
  ``The massless minimally coupled scalar field in de Sitter space,''
  Phys.\ Rev.\  D {\bf 35} (1987) 3771.

\bibitem{Janssen:2007ht}
  T.~Janssen and T.~Prokopec,
  ``A graviton propagator for inflation,''
  arXiv:0707.3919 [gr-qc].

\bibitem{Allen:1985wd}
  B.~Allen and T.~Jacobson,
  ``Vector Two Point Functions In Maximally Symmetric Spaces,''
  Commun.\ Math.\ Phys.\  {\bf 103} (1986) 669.

\bibitem{Tsamis:2006gj}
  N.~C.~Tsamis and R.~P.~Woodard,
  ``A maximally symmetric vector propagator,''
  J.\ Math.\ Phys.\  {\bf 48} (2007) 052306
  [arXiv:gr-qc/0608069].

\bibitem{Weinberg}
  S.~Weinberg, {\it The Quantum Theory of Fields}, Vol. II
  (Cambridge University Press, 1996), pp. 115-118.

\bibitem{Itzykson}
  C.~Itzykson and J.~B.~Zuber, {\it Quantum Field Theory},
  (McGraw-Hill, New York, 1980) pp. 399-402, 684.

\bibitem{Tsamis:1994ca}
  N.~C.~Tsamis and R.~P.~Woodard
  ``Strong infrared effects in quantum gravity,''
  Annals Phys.\ {\bf 238} (1995) 1.

\bibitem{KW0}
  E.~O.~Kahya and R.~P.~Woodard,
  ``Charged scalar self-mass during inflation,''
  Phys.\ Rev.\ D {\bf 72} (2005) 104001
  [arXiv:gr-qc/0508015].

\bibitem{WMAP}
  D.~N.~Spergel {\it et al.}, 
  ``Wilkinson Microwave Anisotropy Probe (WMAP) three year results:
    implications for cosmology''
  Astrophys.\ J.\ Suppl.\ {\bf 170} (2007) 377
  [arxiv:astro-ph/0603449].

\end{thebibliography}
\end{document}